\documentclass[a4paper]{article}

\usepackage{amsmath}
\usepackage{amsfonts}
\usepackage{amssymb}
\usepackage{latexsym}
\usepackage{stmaryrd}
\usepackage{array}
\usepackage{exscale}
\newcommand{\nc}{\newcommand}
\newcommand{\ol}{\overline}
\newcommand{\ul}{\underline}
\newcommand{\es}{\emptyset}
\newcommand{\sm}{\setminus}
\newcommand{\ve}{\varepsilon}
\newcommand{\vp}{\varphi}

\newcommand{\bc}{\bigcup}

\newcommand{\Lra}{\Leftrightarrow}

\newcommand{\Ra}{\Rightarrow}

\newcommand{\ra}{\rightarrow}

\newcommand{\lra}{\leftrightarrow}

\newcommand{\sse}{\subseteq}

\newcommand{\fa}{\forall}
\newcommand{\ex}{\exists}
\newcommand{\mr}{\mathrm}
\newcommand{\mc}{\mathcal}
\newcommand{\mf}{\mathfrak}

\newcommand{\DMO}{\DeclareMathOperator}
\newcommand{\DST}{\displaystyle}

\newcommand{\NN}{\mathbb{N}}
\newcommand{\NNZ}{\NN_0}
\newcommand{\ZZ}{\mathbb{Z}}

\newcommand{\FF}{\mathbb{F}}

\newcommand{\PP}{\mathbb{P}}

\mathchardef\breakingcomma\mathcode`\,
{\catcode`,=\active
  \gdef,{\breakingcomma\discretionary{}{}{}}
}
\newcommand{\mathlist}[1]{$\mathcode`\,=\string"8000 #1$}
\newcommand{\aru}{\ar @{-}} 
\newcommand{\und}{{\:\wedge\:}} 
\newcommand{\mb}{{\:|\:}} 
\newcommand{\set}[1]{\{ #1 \}}
\newcommand{\setb}[1]{\big \{ \, #1 \, \big \}}

\DeclareMathOperator{\id}{id}

\nc{\simlvi}[1]{\!\sim_{#1}}
\DeclareMathOperator{\rstr}{|} 
\DeclareMathOperator*{\addbcup}{{\stackrel{\text{\raisebox{-4.2ex}[-0ex][-0ex]{\Large$\cdot$}}}{\bigcup}}} 
\nc{\apprel}[3]{{#1}(#2)_{(#3)}} 
\newcommand{\nni}{\NNZ \cup \{+\infty\}} 
\newcommand{\tb}[2]{\set{#1, \dots, #2}} 
\providecommand{\abs}[1]{\lvert #1 \rvert} 
\newcommand{\trans}[1]{#1^{\hspace{0.05em}\mr{t}}} 

\DeclareMathOperator{\enachbarn}{N}

\DeclareMathOperator{\length}{lgth} 

\DeclareMathOperator{\pot}{\PP} 
\newcommand{\ceil}[1]{\lceil #1 \rceil}
\DeclareMathOperator{\teilt}{\mid} 
\DeclareMathOperator{\auto}{Aut} 
\DeclareMathOperator{\ug}{ug} 
\newcommand{\Va}{\mc{V\hspace{-0.1em}A}}

\newcommand{\Lit}{\mc{LIT}}
\newcommand{\Cl}{\mc{CL}}
\newcommand{\Cls}{\mc{CLS}}

\newcommand{\Pcls}[1]{#1\mbox{--}\Cls}

\newcommand{\Musat}{\mc{M\hspace{0.8pt}U}} 
\newcommand{\Musati}[1]{\Musat_{\!#1}} 
\newcommand{\Smusat}{\mc{S}\Musat} 
\newcommand{\Smusati}[1]{\Smusat_{\!#1}}

\nc{\Clsoo}{\Cls^{1,1}} 
\DeclareMathOperator{\lit}{lit}
\DeclareMathOperator{\var}{var}

\DMO{\dos}{ds} 
\DMO{\mdos}{mds} 
\newcommand{\Clash}{\mc{HIT}} 

\newcommand{\Uclash}{\mc{U}\Clash} 
\newcommand{\Uclashi}[1]{\Uclash_{\!\!#1}}

\DeclareMathOperator{\dpl}{DP} 
\newcommand{\dpi}[1]{\dpl_{\!#1}}
\DMO{\hts}{hs} 
\DeclareMathOperator{\ldeg}{ld} 
\DeclareMathOperator{\vdeg}{vd} 
\DeclareMathOperator{\minvdeg}{\mu\!\vdeg} 
\DMO{\varmvd}{\var_{\minvdeg}} 
\DMO{\nfc}{fc} 
\DMO{\maxnfc}{\nu\!\nfc} 
\nc{\svbf}{\mc{VB}} 
\nc{\svbfs}{\mc{VB}^*} 
\DMO{\potp}{pp} 
\DMO{\potprec}{NM} 
\DMO{\minnonmer}{VDM} 
\DMO{\minnonmerh}{VDH} 
\DMO{\maxsmar}{FCM} 
\DMO{\maxsmarh}{FCH} 
\DMO{\varsing}{\var_s} 
\DMO{\varosing}{\var_{1s}} 
\DMO{\varnosing}{\var_{\neg1s}} 
\DMO{\nsv}{\mathit{n}_s} 
\DMO{\nosv}{\mathit{n}_{1s}}
\DMO{\nnosv}{\mathit{n}_{\neg1s}}
\nc{\Musatns}{\Musat'} 
\nc{\Musatnsi}[1]{\Musati{#1}'}
\nc{\Smusatns}{\Smusat'} 
\nc{\Smusatnsi}[1]{\Smusati{#1}'}
\nc{\Uclashns}{\Uclash'} 
\nc{\Uclashnsi}[1]{\Uclashi{#1}'}
\nc{\tsdp}{\xrightarrow{\text{sDP}}}
\nc{\tsdps}{\tsdp_{\!*}}
\nc{\tosdp}{\xrightarrow{\text{1sDP}}}
\nc{\tosdps}{\tosdp_{\!*}}
\DMO{\sdp}{sDP} 
\DMO{\osdp}{sDP_1} 
\nc{\cflmusat}{\mc{CF}\Musat} 
\nc{\cflmusati}[1]{\mc{CF}\Musati{#1}}
\nc{\cflimusat}{\mc{CFI}\Musat} 
\DMO{\sNF}{sNF} 
\DMO{\eqp}{eqp} 
\DMO{\sgp}{sp} 
\DMO{\singind}{si} 
\DMO{\osingind}{si_1} 
\DMO{\shyp}{svh} 
\DMO{\sdph}{ssh} 
\DMO{\msdph}{mss} 
\DMO{\osdph}{ssh_1} 
\DMO{\mosdph}{mss_1} 
\DMO{\mps}{mps} 
\DMO{\purec}{puc} 
\DMO{\doping}{D}
\usepackage{theorem} 
\usepackage[driverfallback=hypertex]{hyperref} 
\newtheorem{defi}{Definition}[section]
\newtheorem{lem}[defi]{Lemma}
\newtheorem{thm}[defi]{Theorem}
\newtheorem{corol}[defi]{Corollary}

\newtheorem{examp}[defi]{Example}

\theorembodyfont{\rmfamily}

\theorembodyfont{}
\newenvironment{prf}{\noindent\textbf{Proof:}\;}{\par\noindent\ignorespacesafterend}

\newcommand{\Qed}{\hfill $\square$}

\nc{\bm}{\boldmath}
\nc{\bmm}[1]{\mbox{\bm$\DST #1$}}
\nc{\mi}[1]{\bmm{\mathrm{(#1):}} \quad}

\usepackage{enumerate}
\usepackage[active]{srcltx}
\usepackage[all]{xy}
\usepackage{cancel}

\usepackage{xr}

\providecommand{\keywords}[1]{\textbf{\textit{Keywords}} #1}

\newcommand{\Bmusat}{2\mbox{--}\!\Musat}
\newcommand{\Bmusats}{\Bmusat^*}
\newcommand{\Bmusati}[1]{\Bmusat_{\!#1}}
\newcommand{\Bmusatds}{\Bmusati{\delta=1}^*}
\newcommand{\Bclss}{\Pcls{2}^*}

\newcommand{\Bmusatdp}[1]{\Bmusat_{\!#1}^+}
\newcommand{\DiCo}{\mc{DC}}

\newcommand{\Musatnos}{\Musat^+}

\newcommand{\Bcl}{\Cl_2}
\newcommand{\Blit}{\Lit_{\!2}}

\DeclareMathOperator{\idg}{idg}
\DeclareMathOperator{\sidg}{sidg}
\DeclareMathOperator{\ig}{ig}
\DeclareMathOperator{\img}{img}

\DeclareMathOperator{\1dp}{1sDP}

\DeclareMathOperator{\isos}{iso}

\DeclareMathOperator{\pfirst}{first}
\DeclareMathOperator{\plast}{last}
\DeclareMathOperator{\smooth}{sm}
\DeclareMathOperator{\llin}{LL}
\DeclareMathOperator{\dgtomg}{mg}
\DeclareMathOperator{\gtomg}{mg}
\DeclareMathOperator{\gtodg}{dg}
\DeclareMathOperator{\dgtog}{ug}
\DeclareMathOperator{\mgtog}{ug}
\DeclareMathOperator{\cycleg}{CG}
\DeclareMathOperator{\dcycleg}{DCG}
\DeclareMathOperator{\scs}{sc}
\DeclareMathOperator{\bcs}{bc}
\DeclareMathOperator{\overlaps}{ovl}
\DeclareMathOperator{\interiors}{int}

\DeclareMathOperator{\scg}{scg}

\newcommand{\Bpt}[1]{\mc{B}_{#1}}
\newcommand{\Hp}{\operatorname{U}^2}
\newcommand{\Hl}{\operatorname{U}^1}
\newcommand{\Shl}{\operatorname{U}^0}
\newcommand{\Ub}{\operatorname{U}^0}
\newcommand{\GUb}[1]{\operatorname{U}^{#1}}
\DeclareMathOperator{\canonCLS}{canon}

\newcommand{\lexorder}{\preceq}
\newcommand{\slexorder}{\prec}
\newcommand{\brsim}{\sim}
\DeclareMathOperator{\allbr}{\mf B}
\DeclareMathOperator{\brop}{br}

\begin{document}

\title{The classification of minimally unsatisfiable 2-CNFs
  --- a fundamental study
}

\author{
  \href{https://orcid.org/0000-0002-9575-4758}{Hoda Abbasizanjani}\\
  Population Data Science\\
  Swansea University Medical School\\
  Swansea, SA2 8PP, UK
  \and
  \href{http://cs.swan.ac.uk/~csoliver}{Oliver Kullmann}\\
  Computer Science Department\\
  Swansea University\\
  Swansea, SA1 8EN, UK
}

\maketitle

\begin{abstract}
  Conjunctive normal forms, where every clause has length at most two, are called 2-CNFs.
  They have efficient algorithms for many interesting problems.
  We study minimally unsatisfiable 2-CNFs, short 2-MUs, that is, unsatisfiable 2-CNFs where removing any clause destroys unsatisfiability.
  The main result of this article is their full classification (up to isomorphism).
  Characterisations of 2-MUs have only been known for the nonsingular case (where every variable occurs positively and negatively at least twice), and the cases with a unit-clause.
  We now characterise all 2-MUs.
  The main tool is the implication digraph, and we show that for 2-MUs they are ``weak double cycles'' (WDCs), big cycles of small cycles (with possible overlaps).
  Combining logical and graph-theoretical methods, we prove that WDCs have at most one skew-symmetry (a self-inverse fixed-point free anti-symmetry, reversing the direction of arcs).
  It follows that the isomorphisms between 2-MUs are exactly the isomorphisms between their implication digraphs, thus reducing the classification of 2-MUs to the classification of a nice class of digraphs.

  We obtain a variety of applications for 2-MUs $F$ of deficiency $k$, the difference of the number of clauses of $F$ and the number $n$ of variables of $F$.
  The smoothing (removal of linear vertices) of skew-symmetric WDCs corresponds exactly to the canonical normalform $F'$ of $F$ obtained by 1-singular DP-reduction, reducing variables occurring exactly twice (a restricted form of DP-reduction, or ``variable elimination'').
  The isomorphism types of these normalforms $F'$, i.e., the homeomorphism types of skew-symmetric WDCs, are in one-to-one correspondence with binary bracelets (or ``turnover necklaces'') of length $k$.
  The automorphism group of $F$ is a subgroup of the Dihedral group with $4 k$ elements.
  The isomorphism problem restricted to 2-MUs $F$ is decidable in quadratic time.
  The number of isomorphism types of 2-MUs for fixed $k$ is $\Theta(n^{3k-1})$.
  And finally we obtain more precise information on the contradictory cycles of a 2-CNF.

  The article is addressed to both the logic and the graph theory communities, and provides complete proofs throughout.
  Including foundational results on digraphs with skew-symmetries and classification up to isomorphism, it may serve as a self-contained, fundamental study.
\end{abstract}

\keywords{Minimal unsatisfiability, 2-CNF, implication digraph, skew-sym\-met\-ry}

\tableofcontents

\section{Introduction}
\label{sec:intro}

A CNF is a propositional formula as a conjunction of disjunctions of literals, and a 2-CNF has at most two literals per disjunction (i.e., clause).
A CNF is \emph{minimally unsatisfiable} (MU) iff it is unsatisfiable and removing any clause yields a satisfiable formula.
A simple example is the 2-CNF MU (short 2-MU) $F = (v) \land (\neg v)$ with two clauses (of length one) and one variable.

We study the \emph{isomorphism problem} for 2-MUs, i.e., for given 2-MUs $F, F'$, decide whether $F$ is isomorphic to $F'$.
The isomorphism problem for subclasses of MUs has been considered in \cite{KleineBuening2000SubclassesMU,KleineBueningXu2005HomomorphismsMu,KullmannZhao2012ConfluenceJ,AbbasizanjaniKullmann2016MUconference} and in the Handbook chapter \cite{Kullmann2007HandbuchMU2021}:
\begin{itemize}
\item It is shown in \cite{KleineBueningXu2005HomomorphismsMu} that the isomorphism problem for MUs with fixed deficiency, the difference between the number of clauses and variables, is GI-complete (graph-isomorphism complete).
\item Typical GI-complete problems are graph isomorphism, CNF isomorphism, and 2-CNF isomorphism.
\item Furthermore even the class of Horn MUs (which is a subset of deficiency one) is still GI-complete.
\end{itemize}
We give the first example of a class of restricted but still rich MUs, namely 2-MUs, where we can obtain a very clear picture of the possible isomorphism types, which includes polytime isomorphism decision.
That picture of 2-MUs is that they are one big cycle of small cycles.
The simplest variables in any MU are \emph{1-singular variables}, occurring positively and negatively exactly once.
The subclass of 2-MUs without 1-singular variables corresponds exactly to the class of binary strings called ``bracelets'' (\cite{GilbertRiordan1961Sequences}).
This shows that there are exponentially many isomorphism types of 2-MUs (depending on the number of variables).

\subsection{Generating all 2-MUs}
\label{sec:introgenerating}

The starting point of the investigations of this article are the most basic 2-MUs, the \emph{nonsingular} 2-MUs.
A nonsingular MU $F$ is characterised by the property that every variable occurs positively and negatively at least twice -- or, in other words, they do not contain \emph{singular variables}, which are variables which occur in at least one sign only once.
The (non-trivial) characterisation of nonsingular 2-MUs was first obtained in the technical report \cite{KBZ2002b}, and then with a more general proof in \cite{AbbasizanjaniKullmann2016MUconference}.
A nonsingular 2-MU $F$ with $n(F) \ge 2$ (the number of variables occurring in $F$) is isomorphic to $\Bpt{n(F)}$, a cycle of equivalences with a final negation, given as
\begin{eqnarray*}
  \bmm{\Bpt n} & := & v_1 \lra v_2 \lra v_3 \lra \dots \lra v_{n-1} \lra v_n \lra \neg v_1 \\
  & = & (\neg v_1 \vee v_2) \und (v_1 \vee \neg v_2) \und \dots \und (\neg v_{n-1} \vee v_n) \und (v_{n-1} \vee \neg v_n) \und \\
    && (v_1 \vee v_n) \und (\neg v_1 \vee \neg v_n)
\end{eqnarray*}
for $n \ge 2$.
The simplest case is $\Bpt 2 = v_1 \lra v_2 \lra \neg v_1 = (\neg v_1 \vee v_2)  \und (v_1 \vee \neg v_2) \und (v_1 \vee v_2) \und (\neg v_1 \vee \neg v_2)$.
Note that $\Bpt n$ has $n$ variables $v_1, \ldots, v_n$ and $2 n$ clauses (each of the $n$ equivalences yields two clauses).
Thus the deficiency is $\delta(\Bpt n) = 2 n - n = n$.
For general MUs holds, that removing singular variables via DP-reduction (also known as ``variable elimination'') maintains minimal unsatisfiability and deficiency (\cite{KullmannZhao2012ConfluenceJ}).
So singular DP-reduction for a 2-MU of deficiency $k \ge 2$ yields some 2-MU isomorphic to $\Bpt k$.

To refine this, \emph{1-singular DP-reduction} is considered, i.e., DP-reduction for (only) 1-singular variables.
1-singular DP-reduction for any MU $F$ is confluent (\cite{KullmannZhao2012ConfluenceJ}), yielding the non-1-singular normalform of $F$.
We obtain the basis for this article: to generate all 2-MUs of deficiency $k$,
\begin{enumerate}
\item[(i)] start with $\Bpt k$ (all variables occur in both signs exactly twice),
\item[(ii)] first reverse non-1-singular DP-reductions (introducing variables which occur in one sign once and in the other sign twice),
\item[(iii)] and then reverse 1-singular DP-reduction (introducing variables which occur in both signs exactly once).
\end{enumerate}
Via this generation process, it is not hard to prove the fundamental observation on literal degrees (how often a literal occurs) in 2-MUs: Every literal occurs at most twice.
Thus we obtain that there are just three possible variable-degrees (how often a variable, positively and negatively, occurs), and that these variable-degrees determine the occurrences of the two signs (polarities) up to sign-symmetry:
\begin{itemize}
\item degree-4-variables, which necessarily occur in both signs (exactly) twice;
\item degree-3-variables, which occur in one sign once, in the other twice;
\item degree-2-variables, which occur in both signs once.
\end{itemize}
The starting point, the $\Bpt k$, contain only degree-4-variables.
Reverse non-1-singular DP-reductions introduces degree-3-variables (replacing one degree-4-variable with two degree-3-variables), and reverse 1-singular DP-reduction introduces degree-2-variables.

\subsection{Implication digraphs of 2-MUs}
\label{sec:introimpldigraphs}

Now how do the clause-sets generated in this way ``look''?
Graph theory can answer this question.
As it turns out, 2-MUs correspond closely to a nice class of digraphs, called \emph{weak double cycles} (WDCs; studied in \cite{SeymourThomassen1987GeradeGerichteteGraphen}).
For this, the concept of the \emph{implication digraph} of a 2-CNF $F$ is needed (introduced in \cite{AspvallPlassTarjan1979Binary}; for an overview see \cite[Section 5.4.3]{CramaHammer2011BooleanFunctions}).
For a 2-CNF $F$, we denote the implication digraph by \bmm{\idg(F)}, where we always assume that $F$ does not contain the empty clause.
The vertex-set of $\idg(F)$ consists of the $2 n(F)$ literals $v_1, \dots, v_n, \ol{v_1}, \dots, \ol{v_n}$ of $F$, now using complementation instead of negation, i.e., $\ol x = \neg x$.
A clause $(x \vee y)$ in $F$ (mathematically written as $\set{x,y} \in F$) yields the
\begin{center}
  \emph{two} arcs $\ol x \ra y$, $\ol y \ra x$
\end{center}
in the implication digraph; these arcs become one in case $x=y$.
$F$ is unsatisfiable iff the implication digraph $\idg(F)$ contains a contradictory directed closed walk (a closed walk containing a literal and its complement), as first noticed in \cite{AspvallPlassTarjan1979Binary} (there as the statement, that $F$ is unsatisfiable iff there is a strongly connected component containing complementary literals).

We note that if the variable $v$ occurs positively $a$ times in the 2-MU $F$, and negatively $b$ times, then the literal $v$ has indegree $a$ and outdegree $b$ in $\idg(F)$, while the literal $\ol v$ has indegree $b$ and outdegree $a$ in $\idg(F)$.
So the degree of variable $v$ in $F$ (which is $a + b$) equals the degrees of \emph{both} associated literals $v, \ol v$ in $\idg(F)$.

A basic observation now is that the reversal of singular DP-reduction for 2-MUs corresponds to the following two graph-theoretical operations:
\begin{itemize}
\item \emph{Splitting a vertex} replaces a vertex $x$ by two new vertices $u, v$, with an arc from $u$ to $v$: $u$ collects the ingoing arcs of $x$, and $v$ the outgoing arcs.

  If $x$ has indegree $a$ and outdegree $b$, then $u$ has indegree $a$ and outdegree $1$, while $v$ has indegree $1$ and outdegree $b$.
\item \emph{Splitting an arc} adds a midpoint (a new vertex) to an arc.

  The new vertex has in- and outdegree $1$, while in- and out-degrees of the two original vertices do not change.
\end{itemize}
Performing reverse non-1-singular DP-reduction for a 2-MU $F$ corresponds to splitting vertices, while the reverse of 1-singular DP-reduction corresponds to splitting arcs.
WDCs are obtained from ``double $m$-cycles'' by splitting of vertices and arcs.
These double $m$-cycles are undirected cycles of length $m$ converted to digraphs (with $2m$ arcs), and these are the implication digraphs of $\Bpt{\frac m2}$ for even $m$ (see below for an example).

\subsection{Skew-symmetry}
\label{sec:introskew}

Now not all WDCs correspond to clause-sets (at all) --- digraphs in general do not allow negation (complementation), and this further ingredient is needed.
The corresponding concept indeed exists in the literature on digraphs under the name ``skew-symmetry''.

The implication digraph of a 2-CNF $F$, in its labelled form, with the literals of $F$ as vertices, allows \emph{exact} reconstruction of $F$, since the arcs faithfully encode the clauses, while the vertices-as-literals reveal full information on the complement-relation between vertices.
Forgetting the labels, the complement-relation between vertices (a literal $x$ is mapped to $\ol x$) is provided \emph{explicitly} (and additionally) by a \emph{skew-symmetry}:
\begin{center}
  A permutation of the vertices,\\
  which is an ``anti-automorphism'', reversing the direction of arcs.
\end{center}

Digraphs might have no skew-symmetry (then they don't correspond to 2-CNFs at all), or they might have many (then they correspond to several 2-CNFs).
Digraphs with \emph{given} skew-symmetry are basically the same as 2-CNFs.

We show that WDCs have at most one skew-symmetry.
That is, there is at most one way to add complementation of the vertices to a WDC and obtain a 2-CNF.
The main technical result of this article follows easily: The isomorphisms between 2-MUs are \emph{exactly} the isomorphisms between their implication digraphs.
So we reduced determining isomorphisms/automorphisms of 2-MUs to a purely graph-theoretical problem between (nice) digraphs.
It follows that the automorphisms of a 2-MU $F$ with deficiency $k \ge 2$ form a subgroup of the Dihedral group with $4k$ elements, and this allows efficient enumeration and counting of isomorphism types of 2-MUs.

\subsection{The running example}
\label{sec:introruningexample}

The running example of this article is based on $\Bpt 3$, which as logical formula is $v_1 \lra v_2 \lra v_3 \lra \neg v_1$.
It has the following implication digraph ($3$ variables, thus $6 = 2 \cdot 3$ vertices, and $6$ clauses, yielding $12 = 2 \cdot 6$ arcs):
  \begin{displaymath}
    \idg(\Bpt 3) =\hspace{-0.5em}
      \xymatrix @C=4.2em @R=1.7em{
        v_1 \ar@/^1pc/[rr] \ar@/_1pc/[dd] & K_1 & v_2 \ar@/_1pc/[rr] \ar@/^1pc/[ll] & K_2 & v_3 \ar@/^1pc/[dd] \ar@/_1pc/[ll]\\
        -K_3 && {\makebox[0pt]{$\set{\ol{v_1},v_2}, \set{v_1, \ol{v_2}}, \set{\ol{v_2},v_3}, \set{v_2,\ol{v_3}}, \set{v_1,v_3}, \set{\ol{v_1}, \ol{v_3}}$}} && K_3\\
     \ol{v_3} \ar@/^1pc/[rr] \ar@/_1pc/[uu] & -K_2 & \ol{v_2} \ar@/_1pc/[rr] \ar@/^1pc/[ll] & -K_1 & \ol{v_1} \ar@/_1pc/[ll] \ar@/^1pc/[uu]
    }
  \end{displaymath}
$\idg(\Bpt 3)$ is a double $6$-cycle and so has six \emph{small) cycles} (of length two), the cycles $K_1, K_2, K_3$, and their contrapositions $-K_1, -K_2, -K_3$.
The contraposition of an arc $(x,y)$ is the arc $\ol{(x,y)} := (\ol y ,\ol x)$, and the contraposition of a cycle contraposes all arcs (we don't use the notation $\ol {K_i}$ here for typographical reasons).
We note here, that the contraposition of each small cycle is its ``antipodal'' cycle, on the ``opposite side'' of the digraph.

$\idg(\Bpt 3)$ has also two \emph{big cycles}, namely $K_4: v_1 \ra v_2 \ra v_3 \ra \ol{v_1} \ra \ol{v_2} \ra \ol{v_3} \ra v_1$ and its contraposition $-K_4$, and these two cycles are exactly the contradictory cycles.
We emphasise that in this article, when applied to a digraph, ``cycle'' always means ``directed cycle''; when we want to speak of the undirected cycles in a digraph, then we always explicitly convert the digraph to the underlying (undirected) graph.

In general the implication digraph of $\Bpt n$ is a double $2n$-cycle with $2n$ small cycles (non-contradictory), and two big cycles (contradictory), so that together $\idg(\Bpt n)$ has exactly $2n + 2$ cycles (recall, a cycle of a digraph is always directed).

To display the unlabelled $\idg(\Bpt 3)$ with complementation, the three pairs of complementary literals (this is all what is needed to know about complementation) in $\idg(\Bpt 3)$ are shown below by three different types $\bullet, \circ, \times$ of vertices (note their antipodal positions).
Furthermore we show the abstract implication digraph (the unlabelled $\idg(\Bpt 3)$), which has lost the information on the complementation.
The final abstraction for a 2-MU $F$ is the homeomorphism type of the implication graph (undirected) of $F$, which here is a cycle of 6 (small) cycles connected by single vertices (note that this is a multigraph, with parallel edges):
\begin{displaymath}
  \xymatrix @C=0.5em @R=0.8em {
    \bullet \ar@/^1pc/[rr] \ar@/_1pc/[dd] & K_1 & \circ \ar@/_1pc/[rr] \ar@/^1pc/[ll] & K_2 & \times \ar@/^1pc/[dd] \ar@/_1pc/[ll]\\
    -K_3 &&&& K_3\\
    \times \ar@/^1pc/[rr] \ar@/_1pc/[uu] & -K_2 & \circ \ar@/_1pc/[rr] \ar@/^1pc/[ll] & -K_1 & \bullet \ar@/_1pc/[ll] \ar@/^1pc/[uu]
  }
  \hspace{1.4em}
  \xymatrix @C=1.1em @R=1.5em {
    \bullet \ar@/^1pc/[rr] \ar@/_1pc/[dd] && \bullet \ar@/_1pc/[rr] \ar@/^1pc/[ll] && \bullet \ar@/^1pc/[dd] \ar@/_1pc/[ll]\\
    &&&& \\
    \bullet \ar@/^1pc/[rr] \ar@/_1pc/[uu] && \bullet \ar@/_1pc/[rr] \ar@/^1pc/[ll] && \bullet \ar@/_1pc/[ll] \ar@/^1pc/[uu]
  }
  \hspace{1.6em}
  \xymatrix @C=1.1em @R=1.5em {
    \bullet \aru@/^1pc/[rr] \aru@/_1pc/[dd] && \bullet \aru@/_1pc/[rr] \aru@/^1pc/[ll] && \bullet \aru@/^1pc/[dd] \aru@/_1pc/[ll]\\
    &&&& \\
    \bullet \aru@/^1pc/[rr] \aru@/_1pc/[uu] && \bullet \aru@/_1pc/[rr] \aru@/^1pc/[ll] && \bullet \aru@/_1pc/[ll] \aru@/^1pc/[uu]
  }
\end{displaymath}

We use natural numbers for variables and non-zero integers for literals (as in the DIMACS file format for CNFs), e.g., the clause $\set{-1,2}$ stands for the usual clause $\set{\ol{v_1}, v_2}$.
The following implication digraph, which is a WDC, is our running example, obtained from $\idg(\Bpt 3)$ by splitting vertices $2,-2$ via replacing them with arcs $(4,2), (-2,-4)$ (using new variable $4$), and then splitting these two arcs (using variable $6$) as well as arcs $(-3,1), (-1,3)$ (using variable $5$):
\begin{equation}
  \label{eq:running}
  \xymatrix @C=1.8em @R=1.2em {
    &&&& 4 \ar[d] &&&& \\
    & 1 \ar@/^1.4pc/[rrru] \ar@/_2pc/[dddd] && K_1 & 6 \ar[d] & K_2 && 3 \ar@/_1.4pc/[lllu] \ar@/^2pc/[dddd] & \\
    &&&& 2 \ar@/^1.4pc/[lllu] \ar@/_1.4pc/[rrru] &&&& \\
    & -K_3 & 5 \ar@/_0.4pc/[luu] &&&& -5 \ar@/^0.4pc/[ruu] & K_3 &\\
    &&&& -2 \ar[d] &&&& \\
    & -3 \ar@/^1.4pc/[rrru] \ar@/_0.4pc/[ruu] && -K_2 & -6 \ar[d] & -K_1 && -1 \ar@/_1.4pc/[lllu] \ar@/^0.4pc/[luu] & \\
    &&&& -4 \ar@/^1.4pc/[lllu] \ar@/_1.4pc/[rrru] &&&&
  }
\end{equation}
As mentioned before, the implication digraph together with complementation of vertices is essentially the same as the original clause-set.
The underlying clause-set of the above implication digraph is (where the order of clauses follows a contradictory cycle, starting at vertex $1$):
\begin{multline*}
  F = \set{\set{-1,4}, \set{-4,6}, \set{-6,2}, \set{-2,3}, \set{-3,-1}, \set{1,-2},\\
    \cancel{\set{2,-6}, } \cancel{\set{6,-4}, } \set{4,-3}, \set{3,5}, \set{-5,1}} \in \Bmusat.
\end{multline*}
Note that two duplicated clauses are cancelled.
Key structural elements of $\idg(F)$ are:
\begin{itemize}
\item The two big (contradictory) cycles are $1,4,6,2,3,-1,-2,-6,-4,-3,5,1$ and $1,-3,-2,-6,-4,-1,-5,3,4,6,2,1$.
\item It might be instructive for the reader to check that indeed $F$ is minimally unsatisfiable, by checking that the removal of any clause indeed disruptes both contradictory cycles; for example for the clause $\set{-1,4}$ we have the arc $1,4$ in the first cycle and the contraposition $-4,-1$ in the second cycle.
\item Note that the big cycles alternatingly use the outer and the inner arcs of the small cycles --- in this way the splitting of vertices, here splitting vertex $2$ into $4, 2$ (keeping vertex $2$) can be done ``vertically'', since the frontier vertices of the overlap in this way have always either the cycle-arcs both incoming or both outgoing (that is, we have the incoming arcs on one side, and the outgoing arcs on the other).
\item As before, there are six small cycles ($K_1, K_2, K_3$ and their contrapositions $-K_1, -K_2, -K_3$, as in $\idg(\Bpt 3)$).
\item There are four linear vertices (with exactly one ingoing and one outgoing arc), namely $5,-5,6,-6$, corresponding to the two degree-2-variables $5, 6$.
\item Two degree-3-variables $2, 4$, corresponding to four degree-3-vertices.
\item Two degree-4-variables $1, 3$, corresponding to four degree-4-vertices.
\end{itemize}
In order to understand better the structure of $F$, we consider its non-1-singular normalform, denoted by $\bmm{\1dp(F)}$, with its implication digraph obtained by removing all the linear vertices. Together with the homeomorphism type of $F$ these graphs are:
\begin{equation}
  \label{eq:runninghomeo}
  \xymatrix @C=1.4em @R=0.3em {
    && 4 \ar[dd] && \\
    1 \ar@/^1pc/[rru] \ar@/_1.5pc/[dddd] & K_1 && K_2 & 3 \ar@/_1pc/[llu] \ar@/^1.5pc/[dddd] \\
    && 2 \ar@/^1pc/[llu] \ar@/_1pc/[rru] && \\
    -K_3&&&& K_3 \\
    && -2 \ar[dd] && \\
    -3 \ar@/_1.5pc/[uuuu] \ar@/^1pc/[rru] & -K_2 && -K_1 & -1 \ar@/^1.5pc/[uuuu] \ar@/_1pc/[llu] \\
    && -4 \ar@/^1pc/[llu] \ar@/_1pc/[rru] &&
  } \qquad
  \xymatrix @C=1.9em @R=0.6em {
    && \bullet \aru[dd] && \\
    \bullet \aru@/^1pc/[rru] \aru@/_1pc/[dddd] &&&& \bullet \aru@/_1pc/[llu] \aru@/^1pc/[dddd] \\
    && \bullet \aru@/^1pc/[llu] \aru@/_1pc/[rru] && \\
    &&&& \\
    && \bullet \aru[dd] && \\
    \bullet \aru@/_1pc/[uuuu] \aru@/^1pc/[rru] &&&& \bullet \aru@/^1pc/[uuuu] \aru@/_1pc/[llu] \\
    && \bullet \aru@/^1pc/[llu] \aru@/_1pc/[rru] &&
  }
\end{equation}
The underlying clause-set (again in the order of a contradictory cycle) is:
\begin{multline*}
  \1dp(F) = \set{\set{-1,4}, \set{-4,2}, \set{-2,3}, \set{-3,-1}, \set{1,-2},\\
    \cancel{\set{2,-4}, } \set{4,-3}, \set{3,1}} \in \Bmusat.
\end{multline*}

\subsection{Overview on the literature}
\label{sec:introoverview}

CNFs (conjunctive normal forms, conjunctions of disjunctions of literals) and DNFs (disjunctive normal forms, disjunctions of conjunctions of literals) have long been studied in propositional logic.
Restrictions to the lengths of the ``clauses'' (in CNFs the disjunctions) resp.\ the ``terms'' (in DNFs the conjunctions) were studied especially with the advent of automated theorem proving in the middle of the 20th century.
2-CNFs (all clauses have length at most two) were also called ``Krom formulas'' in the context of first-order logic.
The first explicit proof of polytime SAT decision for 2-CNFs (via resolution closure, in the context of first-order logic) seems to be in \cite{Krom1967Binary}.
Another proof for propositional logic was pointed out in the seminal paper \cite{Cook1971CTP} using the Davis-Putnam procedure in \cite{DP60}; note that this is not the usual splitting algorithm, often referred to as ``DPLL'', but the elimination of one variable at a time, replacing the clauses containing this variable by all their resolvents on this variable --- for us this ``DP-reduction'' is of fundamental importance.
Later the bound was improved by the linear time algorithms of \cite{EIS76} and \cite{AspvallPlassTarjan1979Binary} (the latter even for quantified 2-CNFs).
For an overview on the dual form of (general) 2-DNFs and their underlying boolean functions, called ``quadratic functions'' (which are constant zero for unsatisfiable 2-CNFs resp.\ constant one for tautological 2-DNFs), see \cite[Chapter 5]{CramaHammer2011BooleanFunctions}.
Irredundant 2-CNFs (no clause can be removed without changing the underlying boolean function) are studied in \cite{Liberatore2008Redundanz}, mostly concentrating on satisfiable cases.

A classical connection of SAT to combinatorics is random satisfiability.
For random 2-CNFs with $c$ clauses and $n$ variables, the satisfiability threshold was proven for the critical density $\frac{c}n = 1$ by \cite{CR1992} and independently by \cite{Goe1996}.
That is, a random 2-CNF $F$ with $\frac{c}n > 1$ is unsatisfiable with high probability, while $F$ with $\frac{c}n < 1$ is satisfiable with high probability.
A more precise picture of phase transition and its scaling window for random 2-CNF was achieved in \cite{BBCKW2001}, and an overview is given in \cite{Ve2001}.

2-CNFs are close to renamable Horn formulas in the following sense:
Consider a satisfiable 2-CNF $F$ and a satisfying assignment $\vp$ for $F$.
Obtain $F'$ from $F$ by flipping all literals whose variable is set to true in $\vp$.
So $F'$ is isomorphic to $F$, and since each clause in $F'$ has at least one negative literal (and so at most one positive literal), $F'$ is a Horn formula.
That is, $F$ is a renamable Horn formula.
Regarding unsatisfiable cases, \cite{He74} established the basic fact that every unsatisfiable formula is renamable Horn iff it is refutable by unit-resolution (the resolution rule where at least one of the clauses involved is a unit-clause), and thus we see that an unsatisfiable 2-CNF without a unit-clause is not renamable Horn.
Below we look at the case with a unit-clause, which in the MU-case is renamable Horn.

We now turn to \emph{unsatisfiable} 2-CNFs.
\cite{AspvallPlassTarjan1979Binary} introduced the implication digraph and showed that a 2-CNF $F$ is unsatisfiable iff the implication digraph of $F$ has a strongly connected component containing a literal and its complement.
Every unsatisfiable 2-CNF has a variable $v$ such that via so-called input-resolution, i.e., a chain of resolution steps, one can derive $v$ and $\ol v$ (\cite[Lemma 5.6]{Ku99b}), where the length of each chain is at most the number of variables.
In the general framework of \cite{GwynneKullmann2012SlurJ}, these are those resolution trees $T$ with the Horton-Strahler number $\hts(T)$ at most $2$.
Considering resolution complexity, \cite{OppenheimMitchell2006Min2CNF} obtained a polytime algorithms for finding a smallest tree-like resolution refutation for 2-CNFs, while \cite{OppenheimMitchell2007Min2CNF} provided a polytime algorithm for finding a smallest general resolution refutation, both using implication digraphs of 2-CNFs.
Study of some incomplete refinements of resolution has been carried out in \cite{KBWS20162CNFJ}, namely so-called ``read-once'' resolution refutation and its variations, and the authors have investigated the complexity of finding such resolution refutations.

A different study of graphs related to 2-CNFs is the recent \cite{KarveHirani2020U2CNF} which is mainly interested in distinguishing satisfiability and unsatisfiability.
For a 2-CNF $F$ they obtain a graph by first applying some form of preprocessing of $F$ to remove clauses $C, D \in F$, with $C \ne D$ but $\var(C) = \var(D)$ (these yield the parallel edges).
This process destroys information on isomorphism types, and thus is not suitable for our investigations, but the obtained graph can distinguish satisfiable and unsatisfiable $F$ (\cite[Corollary 21]{KarveHirani2020U2CNF}).

In this article we are only interested in \emph{minimally} unsatisfiable 2-CNFs (i.e., 2-MUs).
Before considering the literature here, we mention that ``MUSs'', minimally unsatisfiable \emph{sub-}sets of 2-CNFs, have been studied in \cite{OppenheimMitchell2006Min2CNF}, showing how to compute some shortest MUS in polytime.

Running through all clauses and testing their irredundancy, the minimal unsatisfiability problem for 2-CNFs can be decided in quadratic time.
Just expressing the above special form of resolution refutations for 2-CNFs, \cite[Lemma 19]{Liberatore2008Redundanz} states a general pattern of 2-MUs.
The notation for the 2-MUs ``$\Bpt n$'' was introduced in \cite{AbbasizanjaniKullmann2016MUconference}, while they had been used before in the report \cite{KBZ2002b} and also in \cite{Lee2009MU3CNF} (called ``$F^{(2)}$'' there).
Regarding the number of clauses $c(F)$ for a 2-MU $F$, \cite{Liberatore2008Redundanz} and \cite{Lee2009MU3CNF} provide some bounds, while the sharp bound $c(F) \le 4 n(F) - 2$ (attained exactly for the $\Bpt n$) is given in \cite{AbbasizanjaniKullmann2016MUconference}.

The main complexity measure for MUs $F$ (and so for 2-MUs) is the deficiency $\delta(F) := c(F) - n(F)$, which was introduced in \cite{FrGe98}.
The basic fact is that for an MU $F$ we have $\delta(F) \ge 1$ (\cite{AhLi86}).
``Classification'' of MUs is concerned with determining all the isomorphism types of MUs with fixed deficiency $\delta \ge 1$, first for the easier nonsingular cases and then also for the singular cases.
This line of research started by studying the deficiency one case, also called $\Musati{\delta=1}$, and the earliest papers are \cite{DDK98} and \cite{Ku99dKo}.
For the class $\Musati{\delta=1}$ we only have the singular cases, as by \cite{DDK98} any MU $F$ with $\delta(F) = 1$ and $n(F) \ge 1$ has a variable occurring once positively and once negatively.
The classes of MUs with fixed deficiency $\delta \ge 2$ contain singular and nonsingular cases, however so far only classification of the nonsingular cases has been investigated in the literature (see the handbook chapter \cite{Kullmann2007HandbuchMU2021} for an overview).
Concerning the complexity of isomorphism decision, \cite{KleineBueningXu2005HomomorphismsMu} showed that the isomorphism problem for MUs of any fixed deficiency $\delta \ge 1$ is GI-complete.
Even for the special class of renamable Horn MUs, which is a sub-class of $\Musati{\delta=1}$ (first noted in \cite{DDK98}), the isomorphism problem is still GI-complete (\cite{KleineBueningXu2005HomomorphismsMu}).

Considering 2-MUs, only the nonsingular cases and those with a unit-clause have been characterised in the literature.
A 2-MU $F$ with a unit-clause has a unit-resolution refutation, since otherwise unit-clause propagation would yield a non-trivial autarky (a partial assignment satisfying some clauses and not touching the other).
Thus, as mentioned above, $F$ is renamable Horn, and so $\delta(F) = 1$.
In \cite[Lemma 5.1, Parts 1,2]{KBWS20162CNFJ} the isomorphism types of 2-MUs with a unit-clause are determined, leaving open the determination of (singular) 2-MUs of deficiency one without unit-clauses.
Now we come to the 2-MUs of higher deficiencies, which are necessarily 2-uniform (all clauses have length 2).
Classification of 2-MUs of a fixed deficiency $\delta \ge 2$ is split into nonsingular and singular cases.
Nonsingular cases have been characterised as the 2-MUs
\begin{displaymath}
\Bpt n = \set{\set{-1,2}, \set{1,-2}, \dots, \set{-(n-1),n}, \set{n-1,-n}, \set{1,n}, \set{-1,-n}},
\end{displaymath}
(recall the definition above) in the technical report \cite{KBZ2002b}.
Adding further details, \cite{AbbasizanjaniKullmann2016MUconference} provided a simplified proof of this characterisation using the ``positive implication digraph'', that is, only the implications between positive literals.

The present article is the full version of the preliminary report \cite{AbbasizanjaniKullmann20202MUa}, which appeared as a 27-page arXiv preprint in 2020 and contained the main results in condensed form.
The current version provides complete proofs, an expanded foundational development (including the full treatment of digraphs with skew-symmetries and smoothing of multigraphs), and several additional results and applications.
The PhD thesis of the first author \cite{HodaAbbasizanjani2021} contains related material and further background.

\subsection{Overview}
\label{sec:introover}

In this article we consider \emph{all} 2-MUs, allowing singular variables, and we obtain a very precise overview, including a polytime isomorphism decision.

After discussing basic terminology in Section \ref{sec:prelim}, we discuss the important notion of implication digraphs of 2-CNFs in Section \ref{sec:idg}, which gives an incomplete picture of 2-MUs (as an unlabelled digraph, without information on complementation).
This picture is completed by a skew-symmetry, where then the isomorphism type of a 2-MU is uniquely determined by the unlabelled implication digraph together with the skew-symmetry given by complementation.
In Section \ref{sec:nearlyunique} we focus on skew-symmetries, where the associated clause-set does not have a unit-clause; these are called ``unit-free'' skew-symmetries.
The basic Lemma \ref{lem:isoexactoneunitfree} shows that if there is exactly one unit-free skew-symmetry, then the isomorphisms between 2-CNFs and between their implication digraphs are exactly the same.
Continuing the investigation of skew-symmetries, all skew-symmetries of some very basic digraph classes are determined in Section \ref{sec:skewsymbasic}.

In Section \ref{sec:2-MU(1)} we characterise 2-MUs of deficiency one.
As already mentioned, all 2-MUs with a unit-clause have deficiency one.
In Theorem \ref{thn:classd1} we determine precise isomorphism types of the cases with a unit-clause, which were implicitly handled in \cite{KBWS20162CNFJ}, and also for the 2-uniform cases (which is new).

In Section \ref{sec:sDPred} we first discuss singular DP-reduction in general, including the specialisation of 1-singular DP-reduction.
Then we define the smoothing process for digraphs, that is, their underlying multigraphs, which is a new conceptual tool and is strongly related to 1-singular DP-reduction for 2-CNFs (though details related to unit-clauses differ).
The concept of smoothing is known in graph theory, but we exploit it in more details, showing new connections between graph theory and propositional logic.
Theorem \ref{thm:smooting2MUsk2} shows the precise correspondence for 2-MUs of deficiency at least $2$, while Theorem \ref{thm:homeodef1} determines the four homeomorphism types for deficiency $1$ (each of the Families I-IV yields exactly one type).

Then in Section \ref{sec:2MU-WDC} we discuss WDCs from a graph-theoretical point of view, and we characterise their isomorphism types.
Furthermore we show that the homeomorphism types of WDCs, obtained by the smoothing process, correspond exactly to binary bracelets (Theorem \ref{thm:binarybrhomeo}).

In Section \ref{sec:class2MUk} we classify 2-MUs of higher deficiency $k \ge 2$.
We first provide a generation process for the elements of $\Bmusati{\delta=k}$, which implies that the implication digraphs of 2-MUs with deficiency $k \ge 2$ are $2k$-WDCs.
A main result of this section is to show the uniqueness of skew-symmetry for WDCs in Theorem \ref{thm:wdcuniqueskewsym}.
Our second major result is Theorem \ref{thm:idgcompleteisovar}, showing that for 2-MUs $F,F'$ the set of isomorphisms between $F,F'$ is equal to the set of isomorphisms between $\idg(F), \idg(F')$.
That is, the isomorphism problem for 2-MUs has been ``completely faithfully'' transported to the realm of certain (simple) digraphs (the WDCs, via uniqueness of
skew-symmetries), where known combinatorial/graph-theoretical tools can
be applied.
We obtain a variety of applications.
\begin{itemize}
\item The automorphism groups of $F \in \Musati{\delta=k}$, $k \ge 2$, are subgroups of the Dihedral group with $4k$ elements (Corollary \ref{cor:automudih}).
\item The isomorphism problem for 2-MUs is decidable in quadratic time (Corollary \ref{cor:isodec2MU}).
\item The number of isomorphism types of $\Bmusati{\delta=k}$ is $\Theta(n^{3k-1})$ (Corollary \ref{cor:numisotypes}).
\item The smoothing of skew-symmetric WDCs corresponds to the canonical normalform of 2-MUs $F$ obtained by 1-singular DP-reduction, and so the isomorphism types of these normalforms, for deficiency $k$, are in one-to-one correspondence with binary bracelets of length $k$ (Corollary \ref{cor:binbracelet}).
\end{itemize}
We conclude in Section \ref{sec:conclusion} with the summary and a discussion of the main open questions.

\section{Preliminaries}
\label{sec:prelim}

The concepts defined here are all quite standard, and need to be consulted only to look up details (e.g., what exactly is a ``graph'') and notations, though the strict set-theoretical treatment of isomorphisms in Subsection \ref{sec:setsisos}, which are treated strictly as maps, not as some kind of ``morphism'', might be considered upfront.
Regarding sets, we use $\NN = \set{1,2,\dots}$ and $\NNZ = \NN \cup \set{0}$.
For $x, y \in \ZZ$ we use the divisor relation $x \teilt y :\Lra \ex d \in \ZZ : x \cdot d = y$.
And by $\ZZ_n$ for $n \in \NN$ we denote the cyclic group of order $n$ (represented by the natural numbers $\set{0,\dots,n-1}$ together with addition modulo $n$).

\subsection{Sets and isomorphisms}
\label{sec:setsisos}

A map $f: X \ra Y$ is a set of pairs, that is, $f = \set{(x,f(x)) : x \in X}$.
A special map is $\id_X := \set{(x,x) : x \in X}$, the identity (map) of $X$.
The composition of maps $f: X \ra Y$ and $g: Y \ra Z$ is the map $g \circ f: X \ra Z$ given by $g \circ f = \set{(x,g(f(x))) : x \in X}$.

Any form of isomorphism between mathematical objects is itself just a map (and thus a set of pairs).
We emphasise this fact, since we consider sets $\mc I$ of isomorphisms, which are sets of maps, and we thus may compare for example via ``$\mc I \sse \mc I'$'' sets of isomorphisms, where the structures of isomorphisms underlying $\mc I$ resp.\ $\mc I'$ are in general \emph{unrelated}.
An isomorphism is just a map, and knows itself nothing about the structures it relates --- these structures are additionally stated in statements like ``$f$ is an isomorphism from graph $G$ to graph $G'$'', where $f$ itself is just a (special) map from the vertex-set of $G$ to the vertex-set of $G'$.
For structures $\mc S, \mc S'$ of the same type we use $\isos(\mc S, \mc S')$ for the set of all isomorphisms from $\mc S$ to $\mc S'$.

For any structure $\mc S$ the automorphism group is the set of automorphisms of $\mc S$, that is isomorphisms $f: \mc S \ra \mc S$, which we denote by $\auto(\mc S) := \isos(\mc S, \mc S)$.
The set of automorphisms is automatically supplied with the composition of maps, which makes it a group; the identity element of this group is $\id_X = \set{(x,x) : x \in X}$, where $X$ is the underlying set of $\mc S$.

For any structures $\mc S, \mc S'$ we can obtain the set of isomorphisms from $\mc S$ to $\mc S'$ from the automorphisms of $\mc S$ by composition with a single isomorphism from $\mc S$ to $\mc S'$, that is:
\begin{lem}\label{lem:isosstructs}
  $\fa\, f \in \isos(\mc S, \mc S') : \isos(\mc S, \mc S') = \set{f \circ \alpha : \alpha \in \auto(\mc S)}$.
\end{lem}

\subsection{Overview on notions}
\label{Sec:overview}

Concerning logic, we define the following notions in Subsection \ref{sec:logic}:
\begin{itemize}
\item variables (the set of variables is $\Va$), literals (with complementation), clauses (clash-free), clause-sets (the set of clause-sets is $\Cls$)
\item empty clause $\bot$, empty clause-set $\top$, $\Pcls 2$ (all clauses have length at most two), $k$-uniform clause-sets (all clauses have length $k$)
\item $\var(F)$, $\lit(F)$ for clause-sets $F$ (sets of variables and literals of $F$)
\item $n(F) = \abs{\var(F)}$, $c(F) = \abs{F}$, $\delta(F) = c(F) - n(F)$
\item literal-degree $\ldeg_F(x)$, variable-degree $\vdeg_F(v)$
\item isomorphisms between clause-sets $F, G$, which are special bijections $f: \lit(F) \ra \lit(G)$ (the set of isomorphisms is $\isos(F,G)$)
\item the resolvent of two clauses, DP-reduction $\dpi{v}(F)$ on variable $v$
\item minimally unsatisfiable clause-sets (MUs; the set of all MUs is $\Musat$), $\Bmusat$
\item the starred variations $\Bclss, \Bmusats$ not allowing the empty clause.
\end{itemize}
Concerning graph theory we define (in Subsection \ref{sec:graph}):
\begin{itemize}
\item graphs, digraphs, multigraphs
\item isomorphisms between $G, G'$ for the three graph types, which are special bijections $f: V(G) \ra V(G')$ (the set of isomorphisms is $\isos(G,G')$)
\item reversing the arcs in a digraph by $\trans G$
\item converting (``promoting'') a graph to a digraph by $\gtodg(G)$, while the underlying graph of a digraph is $\dgtog(G)$
\item promoting a graph to a multigraph by $\gtomg(G)$, while the underlying graph of a multigraph is $\mgtog(G)$
\item converting a digraph to a multigraph also by $\dgtomg(G)$
\item in-degrees, out-degrees, degrees of vertices in digraphs
\item degrees $\deg_G(v)$ of vertices in graphs and multigraphs
\item linear vertices (of degree $2$)
\item cycle graphs (standardised $\cycleg_n$), cycle digraphs.
\end{itemize}

\subsection{Clause-sets}
\label{sec:logic}

The set of all variables is denoted by $\Va$, and we assume $\NN = \set{1,2,\dots} \sse \Va$ (as in the DIMACS format).
Literals are variables $v \in \Va$ and their complementations $\ol v$ ($\ol v = -v$ for $v \in \NN$), the underlying variable of a literal $x$ is $\var(x) \in \Va$.
The set of all literals is denoted by $\Lit$.
For a set $L$ of literals we denote by $\ol L := \set{\ol x : x \in L}$ the elementwise complementation.
Thus $\Lit = \Va \cup \ol{\Va}$ with $\Va \cap \ol{\Va} = \es$.

A clause is a finite set $C$ of literals, which we assume to be clash-free (i.e., non-tautological), that is, $C \cap \ol C = \es$.
A clause-set is a finite set of clauses, and we use $\Cls$ for the set of all clause-sets.
The empty clause-set is denoted by $\top := \es \in \Cls$ and the empty clause by $\bot := \es$.
By $\Pcls 2$ we denote the set of clause-sets $F \in \Cls$ such that for all clauses $C \in F$ holds $\abs C \le 2$.
A clause-set $F$ is uniform resp.\ $k$-uniform, if all clauses of $F$ have the same length resp.\ length $k$.
The set of variables in a clause $C$ is denoted by $\var(C) := \set{\var(x) : x \in C} = (C \cup \ol C) \cap \Va$.
The set of variables in $F$ is $\var(F) := \bc_{C \in F} \var(C) \subset \Va$, while $\lit(F) := \var(F) \cup \ol{\var(F)}$ is the set of all literals whose variable is in $\var(F)$.
For $F \in \Cls$ we use $\bmm{n(F)} := \abs{\var(F)} \in \NNZ$ for the number of variables and $\bmm{c(F)} := \abs{F} \in \NNZ$ for the number of clauses, while $\bmm{u(F)} := \abs{\set{C \in F : \abs C = 1}}$ is the number of unit-clauses.
The \textbf{deficiency} of $F$ is $\bmm{\delta(F)} := c(F) - n(F) \in \ZZ$.

For clause-sets $F, G \in \Cls$ an \textbf{isomorphism} $f: F \ra G$ is a map $f: \lit(F) \ra \lit(G)$, which is bijective and complement-preserving (i.e., $f(\ol x) = \ol{f(x)}$ for all $x \in \lit(F)$), such that $f(F) := \set{f(C) : C \in F} = G$.
The condition $f(F) = G$ here is equivalent to $c(F) = c(G)$ together with $\fa\, C \in F :f(C) = \set{f(x) : x \in C} \in G$ (using that $f$ is a bijection on the literals, together with the finite number of clauses).

Two clause-sets $F, G \in \Cls$ are \textbf{isomorphic}, denoted by \bmm{F \cong G}, if there exists an isomorphism $f: F \ra G$.
The set of isomorphisms $f: F \ra G$ (as maps $f: \lit(F) \ra \lit(G)$) is denoted by \bmm{\isos(F, G)} (so $\isos(F,G) \ne \es \Lra F \cong G$).

Two clauses $C, D$ are resolvable if they clash in exactly one variable $v \in \Va$, i.e., $C \cap \ol D = \set{v}$, in which case the resolvent on $v$ is the clause $(C \sm \set{v}) \cup (D \sm \set{\ol v})$.
The \textbf{DP-reduction} for $F \in \Cls$ and a variable $v$, denoted by $\bmm{\dpi v(F)} \in \Cls$, replaces all $C \in F$ with $v \in \var(C)$ by all their resolvents on $v$.
$\dpi v(F)$ is satisfiability-equivalent to $F$.
In fact, in this article we do not need to handle assignments, and thus we define that $F$ is unsatisfiable, if repeated applications of DP-reduction yields $\set{\bot}$; otherwise $F$ is satisfiable (and we obtain $\top$ from DP-reduction).

The set of minimally unsatisfiable clause-sets (unsatisfiable, while removal of any clause renders it satisfiable) is denoted by $\bmm{\Musat}$.
For some background on MUs, see \cite{Kullmann2007HandbuchMU2021} (though the present article is self-contained).
It is well-known that $\delta(F) \ge 1$ holds for $F \in \Musat$ (\cite{AhLi86}).
We use $\bmm{\Bmusat} := \Musat \cap \Pcls 2$, and the subsets $\Bmusati{\delta=k}$ resp.\ $\Bmusati{\delta\ge k}$ given by $F \in \Bmusat$ with $\delta(F)=k$ resp.\ $\delta(F) \ge k$.
Since here often the empty clause $\bot$ is just in the way, by an upper-index ``$*$'' we exclude it: $\Bclss := \set{F \in \Pcls 2 : \bot \notin F}$, $\Bmusats := \Bmusat \sm \set{\set{\bot}}$, and $\Bmusatds := \Bmusati{\delta=1} \sm \set{\set{\bot}}$.

\subsubsection{Degrees}
\label{sec:defsdegs}

Less standard for the study of propositional logic, but of great importance to us, are several notions related to ``degrees''.

For a literal $x$, the \textbf{literal-degree} $\bmm{\ldeg_F(x)} := \abs{\set{C \in F : x \in C}} \in \NNZ$ is the number of clauses of $F$ containing $x$, while the \textbf{variable-degree} of a variable $v$ is $\bmm{\vdeg_F(v)} := \ldeg_F(v) + \ldeg_F(\ol v) \in \NNZ$.
Note $0 \le \vdeg_F(v) \le c(F)$, while we have $v \in \var(F)$ iff $\vdeg_F(v) \ge 1$.

For $k \in \NNZ$ let $\bmm{\var_k(F)} := \set{v \in \var(F) : \vdeg_F(v) = k}$ be the set of degree-$k$-variables; note $\var_0(F) = \es$ (since $\var(F)$ only contains actually occurring variables).
And let $\bmm{n_k(F)} := \abs{\var_k(F)} \in \NNZ$ be the number of degree-$k$-variables (similarly to a common notation in graph theory).
So $n_0(F) = 0$ and $n(F) = \sum_{k=1}^{c(F)} n_k(F)$.

Similarly we use $\bmm{\lit_k(F)} := \set{x \in \lit(F) : \ldeg_F(x) = k}$ for the set of degree-$x$-literals; now $\lit_0(F) \ne \es$ is possible for non-MUs ($x \in \lit_0(F)$ iff $\ol x$ is a ``pure literal'', only occurring in one sign).
Furthermore we use $\bmm{n'_k(F)} := \abs{\lit_k(F)} \in \NNZ$ for the number of degree-$k$-literals.
So $2 n(F) = \sum_{k=0}^{c(F)} n'_k(F)$.

\subsection{Graphs, digraphs and multigraphs}
\label{sec:graph}

A \textbf{graph} resp.\ \textbf{digraph} $G$ is a pair $(V,E)$, where $V(G) := V$ is a finite set of vertices and $E(G) := E$ is the set of edges resp.\ arcs defined as two-element subsets $\set{a,b} \sse V$ resp.\ pairs $(a,b) \in V^2$ with $a \ne b$.
Note that we do not allow (self-)loops, and that there are no parallel edges resp.\ arcs (though there might be antiparallel arcs).
When making complexity statements about graphs or digraphs, we assume a standard representation by adjacency lists.

A (di)graph $G$ is a sub(di)graph of another (di)graph $G'$ if $V(G) \sse V(G')$ and $E(G) \sse E(G')$.
For a digraph $G$, the transposed digraph, obtained by reversing the direction of all arcs, is denoted by \bmm{\trans G} (the \textbf{transpose} of $G$).

For two digraphs $G_1, G_2$, an \textbf{isomorphism} from $G_1$ to $G_2$ is a bijection $f: V(G_1) \ra V(G_2)$ such that $f(E(G_1)) = \set{(f(a), f(b)): (a,b) \in E(G_1)} = E(G_2)$; if $G_1, G_2$ are graphs, then the condition is $f(E(G_1)) = \set{\set{f(a), f(b)}: \set{a,b} \in E(G_1)} = E(G_2)$.
If there is an isomorphism between $G_1$ and $G_2$, then we write \bmm{G_1 \cong G_2}.
By \bmm{\isos(G_1, G_2)} we denote the set of isomorphisms $f: G_1 \ra G_2$.
A digraph $G$ is called \textbf{self-converse} if $G \cong \trans G$.

A graph $G$ is promoted to a digraph by $\bmm{\gtodg(G)} := (V(G), \set{(a,b),(b,a): \set{a,b} \in E(G)})$, converting every edge $\set{a,b}$ into two arcs $(a,b), (b,a)$.
The conversion of a digraph $G$ to its underlying graph (forgetting directions, and contracting antiparallel arcs into one edge) is denoted by \bmm{\dgtog(G)}.
A map $f$ is an isomorphism from graph $G$ to graph $G'$ iff $f$ is an isomorphism from $\gtodg(G)$ to $\gtodg(G')$, that is,
\begin{displaymath}
  \isos(G, G') = \isos(\gtodg(G), \gtodg(G')).
\end{displaymath}
Every isomorphism $f$ from a digraph $G$ to a digraph $G'$ is also an isomorphism from $\dgtog(G)$ to $\dgtog(G')$, that is, $\isos(G, G') \sse \isos(\dgtog(G), \dgtog(G'))$.

For a set $V$ and $m \in \NNZ$, let $\binom Vm := \set{S \sse V : \abs S = m}$ be the set of $m$--element subsets of $V$, and furthermore $\binom V{a,b} := \binom Va \cup \binom Vb$.
A \textbf{multigraph} is a pair $(V,E)$ where $V$ is a set and $E: \binom V{1,2} \ra \NNZ$.
The set of neighbours of a vertex $v$ in a multigraph $G$ is $\enachbarn_G(v) := \set{w \in V(G) : E(G)(\set{v,w}) \ne 0} \sse V(G)$.
A submultigraph $G'$ of a multigraph $G$ has $V(G') \sse V(G)$ and $\fa\, \set{u,v} \in \binom{V(G')}{1,2} : E(G')(\set{u,v}) \le E(G)(\set{u,v})$ (that is, the multiplicities of all edges in $G'$ is at most their multiplicity in $G$).
A graph $G$ is promoted to a multigraph \bmm{\gtomg(G)} by using the same vertex-set $V(G)$, and using the characteristic function of $E(G) \sse \binom V2$, while the underlying graph \bmm{\mgtog(G)} of a multigraph just forgets the multiplicities of edges and discards loops.
A digraph $G$ is converted to a multigraph $\dgtomg(G)$ by forgetting the direction of arcs, while not contracting edges (so antiparallel arcs yield edges of multiplicity $2$).
An isomorphism from a multigraph $G$ to a multigraph $G'$ is a bijection $f: V(G) \ra V(G')$ with $\fa\, v, w \in V(G) : E(G')(\set{f(v),f(w)}) = E(G)(\set{v,w})$.
Every isomorphism $f: G \ra G'$ between multigraphs is also an isomorphism $f: \mgtog(G) \ra \mgtog(G')$ between the underlying graphs, i.e., $\isos(G, G') \sse \isos(\mgtog(G), \mgtog(G'))$.
A map $f$ is an isomorphism from a graph $G$ to a graph $G'$ iff $f$ is an isomorphism from $\gtomg(G)$ to $\gtomg(G')$, i.e., $\isos(G, G') = \isos(\gtomg(G), \gtomg(G'))$.
Every isomorphism $f: G \ra G'$ between digraphs is also an isomorphism $f: \dgtomg(G) \ra \dgtomg(G')$, i.e., $\isos(G, G') \sse \isos(\dgtomg(G), \dgtomg(G'))$.

The \textbf{in-degree} of a vertex $v \in V(G)$ of a digraph $G$ is the number of arcs going into $v$, the \textbf{out-degree} is the number of outgoing arcs, and the \textbf{degree} of $v$ is the sum of in- and out-degree.
If $G$ is a graph, then the degree of $v$ is the number of vertices $w$ adjacent to $v$ (that is $\abs{\set{w \in V(G) : \set{v,w} \in E(G)}} \in \NNZ$).
More generally, the degree of a vertex $v \in V(G)$ in a multigraph is $\deg_G(v) := \sum_{w \in V(G)} E(G)(\set{v,w}) \in \NNZ$, the number of adjacent edges.
A \textbf{linear vertex} in a multigraph $G$ is a vertex $v \in G$ of degree two, while a linear vertex in a digraph is a vertex of in- and out-degree one.
By $n_k(G) \in \NNZ$ for $k \in \NNZ$ we denote the number of degree-$k$-vertices in $G$.

A \textbf{walk} in any form of graph is a sequence of (adjacent) vertices and the connecting edges/arcs.
The standardised \textbf{path (di)(multi)graph} of length $n-1 \ge 0$ has vertices $1, \dots, n$ and edges/arcs $1 \ra 2 \dots \ra n$, while a \textbf{path} in a (di)(multi)graph is a substructure of the same type, which is isomorphic to a pathgraph (of that type).
So walks can repeat vertices (and edges/arcs), and they can be thought of as processes, while a path does not repeat vertices (and thus also not edges/arcs), and is a fixed (static) structure (and thus inherently can't repeat anything).
Given a walk or path $P$, by $\pfirst(P), \plast(P)$ we denote the (unique) first resp.\ last vertex of $P$.

\subsection{Cycles}
\label{sec:defcycles}

A \textbf{cycle graph} is a connected graph (every two vertices are connected by some walk), where every vertex is linear (so it has at least three vertices).
The standardised cycle graph \bmm{\cycleg_n} of length $n \ge 3$ has vertices $1, \dots, n$ and edges $1 \ra 2 \dots \ra n \ra 1$ (since ``C'' is also used for clauses, we don't use ``$C_n$'' here).
A \textbf{cycle multigraph} allows additionally for length $2$ (two vertices and two parallel edges) and length $1$ (one vertex with a loop).
A \textbf{cycle} in a (multi)graph $G$ is a (sub)multigraph which is isomorphic to some cycle (multi)graph.
A \textbf{cycle digraph} is a strongly connected digraph (from every vertex every other vertex is reachable by a (directed) path), where every vertex is linear (so it has at least two vertices).
We denote the standardised cycle digraph with $n \ge 2$ vertices by \bmm{\dcycleg_n}.
A \textbf{cycle} in a digraph is a subdigraph which is isomorphic to some cycle digraph (so cycles in a digraph are always directed).
More generally is the concept of a \textbf{closed walk} in a digraph $G$, which is a sequence $v_0, \dots, v_n$, $n \ge 0$, of vertices of $G$ such that $v_0 = v_n$, and for all $0 \le i < n$ we have $(v_i, v_{i+1}) \in E(G)$; modulo cyclic permutation of the order, the closed walks with $n \ge 1$, where for $0 \le i < j \le n$ we have $v_i = v_j \Lra i = 0 \land j=n$, correspond to the cycles.
\begin{examp}\label{exam:dgraph}
  The cycle graph $\cycleg_4$, the digraph $\gtodg(\cycleg_4)$ and the multigraph and underlying graph of this digraph are as follows:
  \begin{displaymath}
  \cycleg_4 = \xymatrix {
    1 \aru[r] & 2 \aru[d] \\
    4 \aru[u] & 3 \aru[l]
  } \quad
  \gtodg(\cycleg_4) = \xymatrix {
    1 \ar@/^/[r] \ar@/^/[d] & 2 \ar@/^/[d] \ar@/^/[l] \\
    4 \ar@/^/[u] \ar@/^/[r] & 3 \ar@/^/[l] \ar@/^/[u]
  } \quad
  \dgtomg(\gtodg(\cycleg_4)) = \xymatrix {
    1 \aru@/^/[r] \aru@/^/[d] & 2 \aru@/^/[d] \aru@/^/[l] \\
    4 \aru@/^/[u] \aru@/^/[r] & 3 \aru@/^/[l] \aru@/^/[u] \\
  }
  \end{displaymath}
  \begin{displaymath}
  \xymatrix {
      &&&&&&
  } \quad
  \dgtog(\gtodg(\cycleg_4)) = \xymatrix {
      1 \aru[r] & 2 \aru[d] \\
      4 \aru[u] & 3 \aru[l]
    }
  \end{displaymath}
  For a cycle digraph $G$ of length 4, i.e., $G \cong \dcycleg_4$, its underlying graph $\dgtog(G) \cong \cycleg_4$ and the digraph $\gtodg(\dgtog(G))$ are as follows.
  \begin{displaymath}
  G = \xymatrix {
    a \ar[r] & b \ar[d]\\
    c \ar[u] & d \ar[l]
  } \quad
  \dgtog(G) = \xymatrix {
    a \aru[r] & b \aru[d] \\
    c \aru[u] & d \aru[l]
  } \quad
  \gtodg(\dgtog(G)) = \xymatrix {
    a \ar@/^/[r] \ar@/^/[d] & b \ar@/^/[d] \ar@/^/[l] \\
    c \ar@/^/[u] \ar@/^/[r] & d \ar@/^/[l] \ar@/^/[u]
    }
  \end{displaymath}
\end{examp}

\subsection{The automorphism group of a cycle}
\label{sec:prelimautocycle}

The automorphism group of a cycle graph of length $n \ge 3$, that is
$\auto(\cycleg_n) = \auto(\gtodg(\cycleg_n)) = \auto(\dgtomg(\gtodg(\cycleg_n)))$, is a well-known group, called the \emph{Dihedral group} with $2 n$ elements, and denoted by $D_n$.
The group $D_n$ is most commonly represented as the group of symmetries of a regular $n$-gon, consisting of $n$ rotational symmetries (including the identity) and $n$ reflection symmetries.

A comprehensive source is \cite{Conrad2019Dihedral1,Conrad2019Dihedral2}.
We explain here the basic features:
\begin{enumerate}
\item The special case of $\auto(\cycleg_3) = S_3$ is the symmetric group of order $3$, with $3! = 6$ elements, the six permutations of $3$ elements (formally the $6$ bijections from $\tb 13$ to itself).
\item We have the subgroup $\auto(\dcycleg_n)$, given by the automorphisms of the directed cycle of length $n$.
  \begin{enumerate}
  \item This is a cyclic group of order $n$ --- in the geometric picture, these are the $n$ rotations of a regular $n$-gon.
  \item Except for the identity, these ``rotations'' do not have fixed points.
  \item Note we only consider these elements as permutations of the vertices of the cycle graph, or, equivalently, as permutations of the vertices of the $n$-gon, not as geometric maps of the (whole) plane (which would have the centre of the rotation as fixed point).
  \end{enumerate}
\item Additionally to these $n$ rotations, $\auto(\cycleg_n)$ has $n$ elements which in the geometric picture are the $n$ reflections of a regular $n$-gon.
  \begin{enumerate}
  \item For odd $n$, each such ``reflection'' has exactly one fixed point (fixing a vertex of the $n$-gon and the opposite midpoint).
  \item For even $n$, half of the ``reflections'' have exactly two fixed points (fixing a vertex of the $n$-gon and the opposite vertex), while the other half has no fixed point (fixing a midpoint and the opposite midpoint of the $n$-gon), but there exist two neighbouring vertices which are swapped.
  \end{enumerate}
\item Each vertex of $\cycleg_n$ has exactly two automorphism which fix that vertex (the identity and one reflection).
\item The elements of order $2$ of $\auto(\cycleg_n)$ (that is, automorphisms $\sigma$ with $\sigma \ne \id$ and $\sigma \circ \sigma = \id$) are as follows:
  \begin{enumerate}
  \item we always have the $n$ reflections;
  \item for odd $n$ this is all, while for even $n$ we have additionally the rotation about $180$ degrees.
  \end{enumerate}
\end{enumerate}

In Appendix \ref{sec:concdefDn} we give a concrete definition of $D_n$, and show how to derive the above basic facts.

\section{Implication digraphs of 2-CNFs}
\label{sec:idg}

In propositional logic, a clause $C = \set{x,y}$, corresponding to $x \lor y$, is equivalent to the implications $\ol x \ra y$ as well as $\ol y \ra x$.
This is the basis of representing 2-CNFs via implication digraphs.
The general encoding of unit- or binary-clauses via literal pairs $(x,y)$ with $x \ne y$ (due to clauses being clashfree) is made explicit in the following definition, using ``$E(C)$'' for the set of arcs corresponding to clause $C$, and ``$\Cl(e)$'' for the clause corresponding to arc $e$:
\begin{defi}\label{def:relbinclsarcs}
  For the purpose of this definition, let $\Bcl := \set{C \in \Cl : 1 \le \abs{C} \le 2}$ be the set of clauses of length $1$ or $2$, and let $\Blit := \Lit^2 \sm \id_{\Lit}$ be the set of pairs of different literals.

  For $C \in \Bcl$, with $C = \set{x,y}$, let $E(C) := \set{(\ol x, y), (\ol y, x)} \in \Blit$.
  Note that for $\set{x,y}$ the set-term $\set{(\ol x,y),(\ol y,x)}$ is well-defined, since for $\set{y,x}$ we obtain the same set $\set{(\ol y,x),(\ol x,y)}$ (written differently).
  And that for $x = y$ (i.e., $C = \set{x}$ is a unit-clause) we obtain $E(C) = \set{(\ol x, x)}$.

  And for $e = (x,y) \in \Blit$ let $\Cl(e) := \set{\ol x, y} \in \Bcl$.
\end{defi}
Remarks:
\begin{enumerate}
\item $E(C) = \set{e \in \Blit : \Cl(e) = C}$.
\item $\Cl((x,y)) = \Cl((\ol y, \ol x))$.
\item $E(\Cl((x,y))) = \set{(x,y), (\ol y, \ol x)}$.
\item For $C \ne C'$ we have $E(C) \cap E(C') = \es$.
\end{enumerate}

Implication digraphs represent 2-CNFs by interpreting the clauses as implications:
\begin{defi}\label{def:idg}
  For $F \in \Bclss$ the \textbf{implication digraph} \bmm{\idg(F)} is defined by
  \begin{itemize}
  \item $V(\idg(F)) := \lit(F)$
  \item $E(\idg(F)) := \bc_{C \in F} E(C)$.
\end{itemize}
\end{defi}
Some fundamental properties of $\idg(F)$ are as follows:
\begin{enumerate}
\item $\abs{V(\idg(F))} = 2 n(F)$, $\abs{E(\idg(F))} = \sum_{C \in F} \abs C$.
\item For every arc $(x,y) \in E(\idg(F))$ we have $\set{\ol x,y} \in F$.
\item The arc $\ol y \ra x$ is the \textbf{contraposition} of the arc $\ol x \ra y$.
\item A binary clause $\set{x,y}$ contributes exactly two (different) arcs, $\ol x \ra y$ and $\ol y \ra x$, being the contraposition of each other.
\item While a unit-clause $\set{x}$ contributes exactly one arc $\ol x \ra x$, being the contraposition of itself.
\item For every $\set{x_1,x_2} \in F$ (possibly $x_1 = x_2$) we have $(\ol{x_1},x_2) \in E(\idg(F))$.
  This is given by having both arcs associated with $\set{x_1,x_2}$ (while selecting one arc wouldn't be possible, given that clauses are sets).
\item For $F_1, F_2 \in \Bclss$ we have $E(\idg(F_1 \square F_2)) = E(\idg(F_1)) \square E(\idg(F_2))$ for $\square \in \set{\cup, \cap, \sm}$.
\item For a literal $x \in \lit(F)$ its degree $\ldeg_F(x)$ is the in-degree of vertex $x$ in $\idg(F)$, and the out-degree of vertex $\ol x$, while for a variable $v \in \var(F)$ its degree $\vdeg_F(v)$ is the degree of vertex $v$ as well as the degree of vertex $\ol v$ in $\idg(F)$.
\item If there is a walk from a literal $x$ to a literal $y$ in $\idg(F)$, then by contraposition there is a walk from $\ol y$ to $\ol x$ in $\idg(F)$.
\item If there is a walk from a literal $x$ to a literal $y$ in $\idg(F)$, then $F$ implies the implication $x \ra y$, that is, for every satisfying assignment $\vp$ of $F$ holds that if $\vp$ sets $x$ to true, then $\vp$ also sets $y$ to true.
  Especially if there is a walk from a literal $\ol x$ to the literal $x$ (in the easiest case given by a unit-clause $\set{x} \in F$), then the literal $x$ is ``forced'', that is, every satisfying assignment of $F$ must set $x$ to true.
\end{enumerate}

In an implication digraph, a closed walk resp.\ cycle with two clashing literals (i.e., a literal and its complement) is called \textbf{contradictory}.
For completeness, we sketch a proof of the following fundamental fact:
\begin{lem}[\cite{AspvallPlassTarjan1979Binary}]\label{lem:Aspvall}
  A clause-set $F \in \Bclss$ is unsatisfiable iff there exists a contradictory walk in $\idg(F)$ (equivalently, iff there is a strongly connected component containing a variable and its complement).
\end{lem}
\begin{prf}
If there is a walk from some literal $x$ to $\ol x$ and back, then both literals $x, \ol x$ had to be true in a satisfying assignment, and thus $F$ is unsatisfiable.
Now assume that $\idg(F)$ does not have a contradictory walk, and we have to show that $F$ is satisfiable.
If for any literal $x$ there are walks from $x$ to $y$ and from $x$ to $\ol y$, then by contraposition there is a walk from $y$ to $\ol x$, and thus a walk from $x$ to $\ol x$.
Note that we can not have this property for both $x$ and $\ol x$ (by assumption).
So for any variable $v \in \var(F)$ there is $x \in \set{v, \ol v}$ such that there is no walk from $x$ to $\ol x$.
Set $x$ and all literals reachable from $x$ (which do not contradict each other) to true.
This partial assignment is an ``autarky'' for $F$, that is, every clause ``touched'' by the partial assignment (having assigned at least one of its literals) is indeed satisfied by this partial assignment, since if in a clause $\set{x,y}$ (possibly $x = y$) we set $x$ to false, that is, $\ol x$ to true, then due to the arc $\ol x \ra y$ we also set $y$ to true in the assignment.
So this partial assignment just removes clauses from $F$, and repeating this process (to the new clause-set obtained from $F$, which again does not have contradictory walks) we can satisfy all of $F$.
\Qed
\end{prf}

The refinement of Lemma \ref{lem:Aspvall} to contradictory \emph{cycles} (not repeating vertices) is proven in Theorem \ref{thm:Aspvallref}.
We conclude this introduction into $\idg(F)$ by a discussion of the implication digraph as an isomorphism invariant.
By forgetting complementation and translating clauses into arcs we have:
\begin{lem}\label{lem:2CNFiso}
  For $F_1,F_2 \in \Bclss$ holds: if $F_1 \cong F_2$ then $\idg(F_1) \cong \idg(F_2)$.
  More precisely, $\isos(F_1, F_2) \sse \isos(\idg(F_1), \idg(F_2))$.
\end{lem}
The reverse direction of Lemma \ref{lem:2CNFiso} does not hold in general, and so the isomorphism type of implication digraphs is not a ``complete isomorphism invariant'' for 2-CNFs, as the following example shows:
\begin{examp}\label{examp:idg2CNF}
  We consider any digraph $G$ which is the disjoint union of two (directed) cycles, and which can be obtained as an implication digraph.
  If the cycles have \emph{different} lengths, then they cannot be the contraposition of each other, and so each cycle must be a contradictory cycle (since for every literal its complement must be in the same cycle).
  That is, every $F \in \Bclss$ with $\idg(F) \cong G$ is unsatisfiable in this case.
  We assume now that the cycles have \emph{equal} length.
  So we have two possibilities, namely that the cycles are the contraposition of each other, or they both are contradictory.
  The first case corresponds to a satisfiable 2-CNF, while the second case yields an unsatisfiable 2-CNF as before.
  For example consider $F_1 = \set{\set{-1,2},\set{-2,3},\set{-3,4},\set{1,-4}}$, $F_2 = \set{\set{-1},\set{1,2},\set{-2},\set{-3},\set{3,4},\set{-4}}$.
  The implication digraphs are
  \begin{displaymath}
  \idg(F_1) = \xymatrix @C=1.7em @R=1.2em {
    1 \ar[r] & 2 \ar[d] & -1 \ar[r] & -4 \ar[d] \\
    4 \ar[u] & 3 \ar[l] & -2 \ar[u] & -3 \ar[l]
  } \quad
  \idg(F_2) = \xymatrix @C=1.7em @R=1.2em {
    1 \ar[r] & -1 \ar[d] & 3 \ar[r] & -3 \ar[d] \\
        -2 \ar[u] & 2 \ar[l] & -4 \ar[u] & 4 \ar[l]
  }
  \end{displaymath}
  $\idg(F_2)$ has two contradictory cycles, and so $F_2$ is unsatisfiable, while $F_1$ is satisfiable.
  Therefore $F_1 \not\cong F_2$, while $\idg(F_1) \cong \idg(F_2)$.

  Now an example with no unit-clause is as follows.
  Consider any digraph with precisely two components (as undirected graph), each isomorphic to $\gtodg(\cycleg_4)$ (recall Example \ref{exam:dgraph}).
  One possibility is that the two components are contrapositions of each other (corresponding to a satisfiable 2-CNF), while another possibility is that each of the components has a contradictory cycle (corresponding to an unsatisfiable 2-CNF).
  For example
   \begin{displaymath}
     F'_1 = \set{\set{-1,2}, \set{1,-2}, \set{-2,3}, \set{2,-3}, \set{-3,4}, \set{3,-4}, \set{-1,4}, \set{1,-4}},
   \end{displaymath}
  \begin{displaymath}
  F'_2 = \set{\set{-1,2}, \set{1,-2}, \set{1,2}, \set{-1,-2},\set{-3,4}, \set{3,-4}, \set{3,4}, \set{-3,-4}}
  \end{displaymath}
  have the following implication digraphs:
  \begin{displaymath}
  \idg(F'_1) = \xymatrix @C=1.8em @R=1.8em {
    1 \ar@/^/[r] \ar@/^/[d] & 2 \ar@/^/[d] \ar@/^/[l] & -1 \ar@/^/[r] \ar@/^/[d] & -2 \ar@/^/[d] \ar@/^/[l] \\
    4 \ar@/^/[u] \ar@/^/[r] & 3 \ar@/^/[l] \ar@/^/[u] & -4 \ar@/^/[u] \ar@/^/[r] & -3 \ar@/^/[l] \ar@/^/[u]
  } \
  \idg(F'_2) = \xymatrix @C=1.8em @R=1.8em {
    1 \ar@/^/[r] \ar@/^/[d] & 2 \ar@/^/[d] \ar@/^/[l] & 3 \ar@/^/[r] \ar@/^/[d] & 4 \ar@/^/[d] \ar@/^/[l] \\
    -2 \ar@/^/[u] \ar@/^/[r] & -1 \ar@/^/[l] \ar@/^/[u] & -4 \ar@/^/[u] \ar@/^/[r] & -3 \ar@/^/[l] \ar@/^/[u]
  }
  \end{displaymath}
  $\idg(F'_1)$ has no contradictory closed walk, and so $F'_1$ is satisfiable, while $\idg(F'_2)$ has $2 + 2 = 4$ contradictory cycles (recall that cycles don't repeat vertices except of begin and end), and thus $F'_2$ is unsatisfiable.
  Therefore $\idg(F'_1) \cong \idg(F'_2)$, but $F'_1 \not\cong F'_2$.

  We remark that $F_1, F'_1$ are in a class of 2-CNFs with a simple structure, namely 2-CNFs where all clauses are mixed binary. The absence of positive and negative clauses in a 2-CNF $F$ yields completely disjoint positive and negative parts of the implication digraph (recall that a binary clause corresponds to two arcs). Therefore $\idg(F)$ has no contradictory cycle and so $F$ is satisfiable (see \cite{AbbasizanjaniKullmann2016MUconference} for more details).

  Finally we restrict ourselves to a digraph $G$ with one component. If $G$ is strongly connected then $G$ corresponds to an unsatisfiable 2-CNF.
  Examples are $F''_1 = \set{\set{1},\set{-1},\set{2},\set{-2},\set{1,2}}$, $F''_2 = \set{\set{1},\set{-2},\set{-1,2},\set{1,-2}}$, with the implication digraphs shown below.
  We see that $\idg(F''_1) \cong \idg(F''_2)$, while $F''_1 \not\cong F''_2$ (as they have different number of unit-clauses).
  \begin{displaymath}
  \idg(F''_1) = \xymatrix @C=3em @R=1.8em {
    1 \ar@/^/[r] & -1 \ar[d] \ar@/^/[l] \\
    2 \ar[u] \ar@/^/[r] & -2 \ar@/^/[l]
  } \quad
  \idg(F''_2) = \xymatrix @C=3em @R=1.8em {
    1 \ar@/^/[r] & 2 \ar[d] \ar@/^/[l] \\
    -1 \ar[u] \ar@/^/[r] & -2 \ar@/^/[l]
  }
  \end{displaymath}
\end{examp}

\subsection{The implication graph}
\label{sec:impgraph}

\begin{defi}\label{def:ig}
  For $F \in \Bclss$ the \textbf{implication graph} is\\\hspace*{\fill}$\bmm{\ig(F)} := \dgtog(\idg(F))$.\hspace*{\fill}
\end{defi}
So as with $\idg(F)$ we have $V(\ig(F)) = \lit(F)$, while for a binary clause $\set{x,y} \in F$ we have the two (different) edges $\set{\ol x, y}, \set{x, \ol y} \in E(\ig(F))$ (which is the same as obtained for $\set{\ol x, \ol y}$), and for a unary clause $\set{x} \in F$ we have the single edge $\set{\ol x, x} \in E(\ig(F))$ (which is the same as obtained for $\set{\ol x}$).

A contraction of two arcs into one edge, when transitioning from the implication digraph $\idg(F)$ to the implication graph $\ig(F)$, happens thus exactly for antiparallel arcs:
\begin{lem}\label{lem:antiparallel}
  For $F \in \Bclss$ antiparallel arcs in $\idg(F)$ correspond exactly to clauses $C \in F$ with $\ol C \in F$, which means the two following cases:
  \begin{enumerate}
  \item For $C = \set{x}$, that is, complementary unit-clauses $\set{x},\set{\ol x} \in F$, we have that $\idg(F)$ contains the cycle of length $2$ between $x, \ol x$, while $\ig(F)$ contains an edge between them.
  \item For $C = \set{x,y}$, that is, equivalence-clauses $\set{x,y}, \set{\ol x, \ol y} \in F$ (corresponding to $x \lra \ol y$), we obtain cycles of length $2$ between $\ol x, y$ and $x, \ol y$ in $\idg(F)$, and single edges in $\ig(F)$.
  \end{enumerate}
\end{lem}
\begin{corol}\label{cor:antiparallel}
  If $F \in \Bclss$ does not contain clauses $C \in F$ with also $\ol C \in F$, then we have $\abs{E(\ig(F))} = \abs{E(\idg(F))}$.
\end{corol}

\subsection{Digraphs with given skew-symmetry}
\label{sec:digraphsgivenss}

When adding a notion of complementation to digraphs, then we obtain basically the same as 2-CNFs, as we now make precise:
\begin{defi}[\cite{GoldbergKarzanov1996SkewSymmetry}]\label{def:SkewSymG}
  A \textbf{skew-symmetry} of a digraph $G$ is a bijection $\sigma: V(G) \ra V(G)$ with the following properties:
  \begin{enumerate}
  \item $\sigma$ is its own inverse (involution), i.e., $\fa\, v \in V(G) : \sigma(\sigma(v)) = v$;
  \item for every vertex $v \in V(G)$ we have $\sigma(v) \ne v$ (i.e., $\sigma$ has no fixed-point);
  \item for every arc $(a,b) \in E(G)$ holds $\sigma((a,b)) := (\sigma(b),\sigma(a)) \in E(G)$.
  \end{enumerate}
  We refer to $\sigma((a,b))$ as the ``contraposition'' of $(a,b)$, while we call $\sigma(v)$ the ``complement'' of $v$.

  A digraph $G$ is called \textbf{skew-symmetric}, if there exists a skew-symmetry $\sigma$ for $G$, while a \textbf{digraph with skew-symmetry} is a pair $(G,\sigma)$.
  An \textbf{isomorphism} $f: (G,\sigma) \ra (G',\sigma')$ of digraphs with skew-symmetries is a digraph-isomorphism $f: G \ra G'$ such that for all $v \in V(G)$ holds $f(\sigma(v)) = \sigma'(f(v))$.
\end{defi}
Equivalently, a skew-symmetry for $G$ is exactly an isomorphism $f: G \ra \trans G$ (thus $G$ is self-converse), where $f$ as a map (from $V(G)$ to itself) is an involution and fixed-point free (note that the induced arc-map is injective, and every injective map of a finite set to itself is bijective).

A digraph may have no skew-symmetry (e.g., digraphs with an odd number of vertices), exactly one (e.g., every path-digraph of odd length), or many (for a complete digraph $G$, having all possible arcs, every fixed-point-free involution of $V(G)$ is a skew-symmetry).

\begin{defi}\label{def:regularpath}
  A path resp.\ cycle in a digraph with skew-symmetry is called \textbf{arc-regular}, if it does not contain an arc and its contraposition at the same time, while it is called \textbf{vertex-regular}, if it does not contain a vertex and its complement at the same time.
  (So vertex-regularity implies arc-regularity, but not the other way around.)
\end{defi}

For $F \in \Bclss$ the implication digraph $\idg(F)$ has a natural skew-sym\-met\-ry, namely the complementation of literals:
\begin{defi}\label{def:sidg}
  For $F \in \Bclss$ let $\bmm{\sidg(F)} := (\idg(F), (\ol x)_{x \in \lit(F)})$ (``skew-symmetric implication digraph'') be the associated digraph with skew-symmetry.
\end{defi}

2-CNFs are basically the same as digraphs with given skew-symmetry, and the following lemma is our foundation for that:
\begin{lem}\label{lem:sidg}
  For all $F, F' \in \Bclss$ holds $\isos(F,F') = \isos(\sidg(F), \sidg(F'))$.
\end{lem}
\begin{prf}
Let $\FF_1 := \isos(F,F')$ and $\FF_2 := \isos(\sidg(F), \sidg(F'))$.
For both $\FF_1, \FF_2$ the defining condition, that we have a complement-preserving bijection from $\lit(F)$ to $\lit(F')$, is the same: For $\FF_1$ this is part of the clause-set-structure, while for $\FF_2$ this is exactly achieved by the condition of preserving the skew-symmetry.
So it remains to show that the additional conditions of preserving the clauses resp.\ the arcs are equivalent.

\noindent First consider $f \in \FF_1$.
We show $f \in \FF_2$:
\begin{enumerate}
\item Consider $(x_1,x_2) \in E(\idg(F))$.
  So $\set{\ol{x_1}, x_2} \in F \Ra \set{f(\ol{x_1}), f(x_2)} \in F' \Ra (\ol{f(\ol{x_1})}, f(x_2)) = (f(x_1), f(x_2)) \in E(\idg(F'))$.
\item Consider $(y_1,y_2) \in E(\idg(F'))$.
  So $\set{\ol{y_1}, y_2} \in E(\idg(F'))$.
  Thus there is $\set{x_1,x_2} \in F$ with $f(x_1) = \ol{y_1}$ and $f(x_2) = y_2$.
  We get $(\ol{x_1}, x_2) \in E(\idg(F))$ with $f(\ol{x_1}) = y_1$, and thus $f((\ol{x_1},x_2)) = (y_1,y_2)$.
\end{enumerate}
Second consider $f \in \FF_2$.
We show $f \in \FF_1$:
\begin{enumerate}
\item Consider $\set{x_1,x_2} \in F$.
  So $(\ol{x_1},x_2) \in E(\idg(F)) \Ra (f(\ol{x_1}),f(x_2)) \in E(\idg(F')) \Ra \set{\ol{f(\ol{x_1})}, f(x_2)} = \set{f(x_1), f(x_2)} \in F'$.
\item Consider $\set{y_1,y_2} \in F'$.
  So $(\ol{y_1},y_2) \in E(\idg(F'))$.
  Thus there is $(x_1,x_2) \in E(\idg(F))$ with $f(x_1) = \ol{y_1}$ and $f(x_2) = y_2$.
  We get $\set{\ol{x_1},x_2} \in F$ with $f(\ol{x_1}) = y_1$, and thus $f(\set{\ol{x_1},x_2}) = \set{y_1,y_2}$.
  \Qed
\end{enumerate}
\end{prf}

\subsection{Digraphs with complementation}
\label{sec:digraphscompl}

Given now a digraph with skew-symmetry $(G,\sigma)$, in order to obtain $F(G,\sigma) \in \Bclss$, we have to choose the signs of the literals (since our clause-sets are based on variables, that is, positive literals), and we have to handle that our clause-sets don't allow for ``formal variables'' (which do not actually occur):
\begin{defi}\label{def:digraphwithcompl}
  A \textbf{digraph on literals} is a digraph $G$ with $V(G) \sse \Lit$ (vertices are literals), such that $V(G)$ is closed under complementation (i.e., $\ol{V(G)} = V(G)$), and $G$ does not have isolated vertices (i.e., vertices of degree zero).
  A \textbf{digraph with complementation} is a digraph with skew-symmetry $(G,\sigma)$, such that $G$ is a digraph on literals, and $\sigma(x) = \ol x$ for all $x \in V(G)$.
  Only the digraph $G$ needs to be mentioned here ($\sigma$ is implicitly given).
  The set of all digraphs with complementation is denoted by $\DiCo$.
\end{defi}

So a digraph with complementation is a digraph $G$, where vertices are literals, and where complementation is a skew-symmetry of the graph, that is, for all $(a,b) \in E(G)$ also $(\ol b, \ol a) \in E(G)$ holds.
Digraphs with complementation are (special) digraphs with (given) skew-symmetry, so their isomorphisms need to respect complementation.

By re-using the literals of clause-sets as vertices, and standardising skew-symmetries as complementation, we have thus enabled also the digraphs to distinguish between positive and negative literals.
For the mathematics of this article this is of no relevance, but it enables us to use seamlessly the standard conventions on CNFs.
\begin{lem}\label{lem:everydissdico}
  For every digraph with skew-symmetry $(G,\sigma)$, such that $G$ has no isolated vertices, there is $G' \in \DiCo$ with $(G,\sigma) \cong G'$.
\end{lem}
\begin{prf}
$V(G)$ is partitioned into 2-element-subsets $\set{x,y}$ with $y = \sigma(x)$ (and $x = \sigma(y)$).
Choose a subset $P \sse V(G)$ which intersects each such 2-element-subset at exactly one element (thus $\abs P = \frac 12 \abs{V(G)}$).
Thus $P \cup \sigma(P) = V(G)$ with $P \cap \sigma(P) = \es$.
Choose any injection $f_0: P \ra \Va$, and extend $f_0$ to a bijection $f: V(G) \ra f_0(P) \cup \ol{f_0(P)}$ by $f(\sigma(v)) := \ol{f_0(v)}$ for $v \in P$.
Now let the digraph $G'$ have vertex-set $V(G') := f(V(G))$, and obtain the arcs via transport:
$E(G') := \set{(f(a), f(b)) : (a,b) \in E(G)}$.
By definition $f$ is a digraph-isomorphism from $G$ to $G'$.
$G'$ is a digraph with complementation, since for $(a,b) \in E(G')$ we have $(f^{-1}(a), f^{-1}(b)) \in E(G)$, thus $(\sigma(f^{-1}(b)), \sigma(f^{-1}(a))) \in E(G)$, whence $(\ol b, \ol a) = (f(\sigma(f^{-1}(b))), f(\sigma(f^{-1}(a)))) \in E(G')$.
By definition $f$ translates application of $\sigma$ into application of complementation, and thus is also an isomorphisms of digraphs with skew-symmetries from $(G,\sigma)$ to $G'$.
\Qed
\end{prf}

Between 2-CNFs and digraphs with complementation we have now a direct bijective relation:
\begin{defi}\label{def:F(G)}
  For $F \in \DiCo$ let $F(G) := \set{\set{\ol a, b} : (a,b) \in E(G)} \in \Bclss$ be the associated 2-CNF.
\end{defi}
The basic properties have all easy proofs:
\begin{lem}\label{lem:fromdigraphtocls}
  Concerning $\sidg(F)$ and $F(G)$ we have:
  \begin{enumerate}
  \item For $G \in \DiCo$ holds $\sidg(F(G)) = G$.
  \item For $F \in \Bclss$ holds $\sidg(F) \in \DiCo$ and $F(\sidg(F)) = F$.
  \item Thus the maps $F \in \Bclss \mapsto \sidg(F) \in \DiCo$ and $G \in \DiCo \mapsto F(G) \in \Bclss$ are inverse bijections.
  \end{enumerate}
\end{lem}

Given a digraph with skew-symmetry $(G,\sigma)$, by Lemma \ref{lem:everydissdico} we can assume w.l.o.g., that we have indeed a digraph with complementation (just given by $G$), and by Lemma \ref{lem:fromdigraphtocls} we can always consider such $G$ together with its equivalent representation $F(G)$, expressing graph-theoretic properties via properties of 2-clause-sets (or vice versa).
For example, an arc $(x,y)$ is mapped by complementation to itself, i.e., $\ol{(x,y)} = (\ol y, \ol x) = (x,y)$, iff $x = \ol y$, iff the arc corresponds to the unit-clause $\set{y}$.

\begin{corol}
  For $G, G' \in \DiCo$ holds $\isos(G,G') = \isos(F(G), F(G'))$.
\end{corol}
\begin{prf}
We have $\isos(G,G') = \isos(\sidg(F(G)), \sidg(F(G'))) = \isos(F(G), F(G'))$ by Lemma \ref{lem:sidg}.
\Qed
\end{prf}

\begin{examp}\label{examp:idg2CNFmult}
  For an example of a digraph (over literals) with multiple complementations, we continue Example \ref{examp:idg2CNF}, by considering the digraph $G$ consisting of two disjoint cycles $v_1 \ra \dots \ra v_4 \ra v_1$ and $w_1 \ra \dots \ra w_4 \ra w_1$ of length four.
  We have seen two complementations, given by $G \cong \idg(F)$ and $G \cong \idg(F')$.
  Now what are \textbf{all} complementations for $G$?

  These are exactly those isomorphisms from $G$ to $\trans G$, which as vertex-maps (ignoring the arcs) are fixed-point-free involutions.
  Obviously an isomorphism $f: G \ra \trans G$ is determined by specifying $f(v_1)$ and $f(w_1)$, which can either be both in the opposite cycle (Type I) or both in the same cycle (Type II) --- then $f$ has just to follow the cycles for the other assignments.
  Thus we have $4 \cdot 4 = 16$ possibilities for Type I, and $4 \cdot 4 = 16$ possibilities for Type II (always just considering digraph-isomorphisms).
  Now let's add the conditions of being an involution and fixed-point free.

  For Type I, when assigning $f(v_1) = w_i$, then we have to assign $f(w_i) = v_1$ to get an involution, and so we get $4$ self-inverse isomorphisms of Type I.
  These are automatically fixed-point free, so we get $4$ complementations.

  For Type II, we get an involution concerning $f(v_1)$ for all four choices $f(v_1) = f(v_i)$, where $f(v_1) \in \set{v_1, v_3}$ is excluded due to having a fixed-point.
  This makes two choices $f(v_1) \in \set{v_2,v_4}$, and further two choices $f(w_1) \in \set{w_2,w_4}$, which makes together $2 \cdot 2 = 4$ complementations of Type II.

  In other words, there are $4$ complementations yielding a digraph with skew-symmetry isomorphic to $\sidg(F)$ (i.e., Type I), namely one can choose $\ol{v_1} = w_i$ for any $i$.
  And there are $2 \cdot 2 = 4$ complementations yielding $\sidg(F')$ (Type II), namely one can choose $\ol{v_1} = v_2$ or $\ol{v_2} = v_4$ for the first cycle, and similarly for the second cycle.
  Altogether $G$ has exactly $4+4=8$ complementations, and they yield exactly two isomorphism-types of digraphs with skew-symmetry.
\end{examp}

\section{Nearly unique skew-symmetries}
\label{sec:nearlyunique}

A fundamental observation for our article is that for $F \in \Bclss$ such that $\idg(F)$ has exactly one skew-symmetry, $F$ can be reconstructed (up to isomorphism) from the unlabelled $\idg(F)$, and thus here $\idg(F)$ is a complete isomorphism-invariant for $F$.
Indeed we also have that for such $F$ the isomorphisms are \emph{exactly} the same as for the associated implication digraphs:
\begin{defi}\label{def:transporttrans}
  Consider a set $X$, a transformation $\sigma: X \ra X$, and a bijection $f: X \ra Y$.
  Then the \textbf{transport} $\sigma_f: Y \ra Y$ is defined by $\sigma_f(y) := f(\sigma(f^{-1}(y))) \in Y$ for $y \in Y$,
\end{defi}
If we have an isomorphism $f: G \ra G'$ between digraphs, then $\sigma$ is a skew-symmetry for $G$ iff $\sigma_f$ is a skew-symmetry for $G'$; so if $(G,\sigma)$ is a digraph with skew-symmetry, then so is $(G', \sigma_f)$, and $f$ is an isomorphisms between these digraphs with skew-symmetry.
\begin{lem}\label{lem:isoexactoness}
  Consider $F, F' \in \Bclss$ such that both $\idg(F)$ and $\idg(F')$ have exactly one skew-symmetry.
  Then $\isos(F, F') = \isos(\idg(F), \idg(F'))$.
\end{lem}
\begin{prf}
In general every isomorphism from $F$ to $F'$ is an isomorphism from $\idg(F)$ to $\idg(F')$ (Lemma \ref{lem:2CNFiso}); so assume that $f$ is an isomorphism from $\idg(F)$ to $\idg(F')$, and we have to show that $f$ is an isomorphism from $F$ to $F'$.
Let $\sigma$ be the unique skew-symmetry of $\idg(F)$; thus $\sidg(F) = (\idg(F), \sigma)$.
And due to the uniqueness of the skew-symmetry for $F'$, we also have $\sidg(F') = (\idg(F'), \sigma_f)$.
Since $f$ is an isomorphism from $(\idg(F),\sigma)$ to $(\idg(F'),\sigma_f)$, by Lemma \ref{lem:sidg} $f$ is also an isomorphism from $F$ to $F'$.
\Qed
\end{prf}

The assumption of Lemma \ref{lem:isoexactoness} of having exactly one skew-symmetry \emph{at all} is too strong for us --- in our applications, we have uniqueness if we \emph{ignore} those skew-symmetries which correspond to unit-clauses (clauses of length one):
\begin{defi}\label{def:unitfree}
  An arc $(a,b) \in E(G)$ of a digraph $G$ is called a \textbf{unit} w.r.t.\ a skew-symmetry $\sigma$ of $G$ if $b = \sigma(a)$.
  A skew-symmetry $\sigma$ of a digraph $G$ is called \textbf{unit-free} if there are no units w.r.t.\ $\sigma$.
\end{defi}
Note that $b = \sigma(a)$ is equivalent to $a = \sigma(b)$.
Accordingly, we call $F \in \Bclss$ \textbf{unit-free} (or ``unit-clause-free'') if for all $C \in F$ holds $\abs C = 2$.
Obviously $F$ is unit-free iff $\sidg(F)$ is unit-free, and more generally the numbers of arcs of $\sidg(F)$ which are units equals the number of unit-clauses of $F$.
And for a digraph $G$ with complementation holds that $G$ is unit-free if $F(G)$ is unit-free (and the number of unit-clauses of $F(G)$ equals the number of units of $G$).
If $(G,\sigma)$ is unit-free, and $f$ is an isomorphism from $(G,\sigma)$ to $(G',\sigma')$, then also $(G',\sigma')$ is unit-free (generally, an isomorphism maintains the number of units).

\begin{lem}\label{lem:isoexactoneunitfree}
  Consider $F, F' \in \Bclss$ such that
  \begin{enumerate}
  \item[(i)] both $\idg(F)$ and $\idg(F')$ have exactly one unit-free skew-symmetry;
  \item[(ii)] $F, F'$ are unit-free.
  \end{enumerate}
  Then $\isos(F, F') = \isos(\idg(F), \idg(F'))$.
\end{lem}
\begin{prf}
Again assume that $f$ is an isomorphism from $\idg(F)$ to $\idg(F')$, and we have to show that $f$ is an isomorphism from $F$ to $F'$.
Let $\sigma$ be the unique unit-free skew-symmetry of $\idg(F)$; since $F$ has no units, thus $\sidg(F) = (\idg(F), \sigma)$.
And due to the uniqueness of the unit-free skew-symmetry for $F'$, we also have $\sidg(F') = (\idg(F'), \sigma_f)$.
Since $f$ is an isomorphism from $(\idg(F),\sigma)$ to $(\idg(F'),\sigma_f)$, by Lemma \ref{lem:sidg} $f$ is also an isomorphism from $F$ to $F'$.
\Qed
\end{prf}

\section{Skew-symmetries of basic digraphs}
\label{sec:skewsymbasic}

In preparation for showing later that our main class of digraphs has exactly one unit-free skew-symmetry, we determine here the skew-symmetries of some elementary digraph-classes.
Recall, the skew-symmetries of a digraph $G$ are those digraph-isomorphisms $f: G \ra \trans{G}$, which as permutations of $V(G)$ are involutions and don't have fixed-points.
Path digraphs with skew-symmetry have exactly one unit:
\begin{lem}\label{lem:pathskeysym}
  Consider a path digraph $G$ with $n \ge 1$ vertices.
  $G$ has exactly one isomorphism $G \ra \trans G$ (exactly one anti-automorphism), and thus $G$ is self-converse.
  Exactly for even $n$ this is a (the unique) skew-symmetry, and it has exactly one unit (while for odd $n$ there is no skew-symmetry).
\end{lem}
\begin{prf}
Consider even $n$.
The unique isomorphism from $G$ to $\trans G$ inverts the order of the vertices (first to last etc.; proven in Corollary \ref{cor:isosdirpath}).
Obviously it is an involution.
Furthermore it has no fixed-point, and exactly one unit (proven in Lemma \ref{lem:fixedpointsautoCn}, Part 5b; note that for a vertex $v$ and a skew-symmetry $\sigma$ we have a unit $(v,\sigma(v))$ iff $\sigma(v) - v = 1$ and $v \ne -1$ (this ``back-arc'' is not there for the path-graph), so one of the two near-fixed points from Part 5b, using $x = -1$, is excluded).
\Qed
\end{prf}

If we have a non-disjoint union of two directed cycles, where the intersection is a path, then a skew-symmetry of the whole digraph induces a skew-symmetry of the intersection, and thus has a unit:
\begin{lem}\label{lem:twocyclesunit}
  Consider a digraph $G$ which is the union of two directed cycle graphs $G_1, G_2$, i.e., $V(G) = V(G_1) \cup V(G_2)$ and $E(G) = E(G_1) \cup E(G_2)$, such that the overlap $V(G_1) \cap V(G_2)$ is not empty, and the induced subdigraph on it is a path graph of length $\abs{V(G_1) \cap V(G_2)} - 1$.
  Then every skew-symmetry of $G$ has a unit.
\end{lem}
\begin{prf}
The transposition $\trans G$ is the union of the two directed cycle graphs $\trans{G_1}, \trans{G_2}$, where $V_0 := V(\trans{G_1}) \cap V(\trans{G_2}) = V(G_1) \cap V(G_2)$.
Let $G'$ resp.\ $G''$ be the induced subdigraph of $V_0$ in $G$ resp.\ $\trans G$.

Now consider a skew-symmetry $\sigma$ of $G$, that is, an isomorphism $\sigma$ from $G$ to $\trans G$.
Since the vertices of $V_0$ are exactly the vertices in $G$ as well as in $\trans G$ which lie on exactly two cycles, we have $\sigma(V_0) = V_0$, and the restriction $\sigma' := \sigma \rstr{V_0}$ is an isomorphism from $G'$ to $G'' = \trans{G'}$.
Furthermore, $\sigma'$ is an involution and fixed-point free, thus is a skew-symmetry of $G'$, and thus by Lemma \ref{lem:pathskeysym} has a unit.
And every unit of $\sigma'$ is also a unit of $\sigma$.
\Qed
\end{prf}

We conclude by determining the skew-symmetries of a cycle digraph:
\begin{lem}\label{lem:CycleSkewsym}
  Consider a cycle digraph $G$ with $n \ge 2$ vertices.
  If $n$ is odd then there is no skew-symmetry.
  For even $n$ there are exactly $n/2$ skew-symmetries, and each has exactly two units.
\end{lem}
\begin{prf}
Assume $G = \dcycleg_n$ and $n \ge 3$ (for $n = 2$ the assertions are obvious).
The isomorphisms from $G$ to $\trans G$ are given by the $n$ rotations composed with one fixed isomorphism from $G$ to $\trans G$, where one can use the rotation ``anticlockwise'', i.e., $1 \mapsto 1, 2 \mapsto n, \dots, n \mapsto 2$.
This yields that precisely the $n$ reflections of the (undirected) cycle $\cycleg_n$ are the isomorphisms from $G$ to $\trans G$ (proven in Corollary \ref{cor:isoZn'}).
They all are involutions, and exactly half of them are fixed-point free; this is proven in Lemma \ref{lem:fixedpointsautoCn}, Part 5, where also the two units are determined (note that for a vertex $v$ and a skew-symmetry $\sigma$ we have a unit $(v,\sigma(v))$ iff $\sigma(v) - v = 1$).
\Qed
\end{prf}

In Theorem \ref{thm:2mudigraph} we will show that all skew-symmetries of a cycle digraph indeed yield isomorphic minimally unsatisfiable clause-sets (and thus, if a cycle digraph $G$ can be provided with a skew-symmetry $\sigma$, then $(G,\sigma)$ is unique up to isomorphism).

\section{2-MUs of deficiency one}
\label{sec:2-MU(1)}

We now come to he characterisation of the isomorphism types of $F \in \Bmusati{\delta=1}$.
By \cite[Corollary 13]{DDK98} and \cite[Lemma C.3]{Ku99dK} \textbf{1-singular DP-reduction}, i.e., DP-reduction for variables occurring exactly once positively and once negatively, applied to any MU $F$, results in $\set{\bot}$ iff $\delta(F) = 1$.
So we can generate (exactly) all of $\Musati{\delta=1}$ by inverse 1-singular DP-reduction, that is, starting from $\set{\bot}$, and repeatedly replacing clauses $C \in F$ for already constructed $F$ by clauses $C', C''$, which fulfil
\begin{itemize}
\item $v \in C'$, $\ol v \in C''$ for some $v \in \Va \sm \var(F)$;
\item $(C' \sm \set{v}) \cup (C'' \sm \set{\ol v}) = C$.
\end{itemize}
From this follows easily (by induction over the construction) the well-known fact that clauses $C, D \in F \in \Musati{\delta=1}$ clash in at most one literal, that is $\abs{C \cap \ol D} \le 1$.
We obtain by Corollary \ref{cor:antiparallel} that antiparallel arcs in the implication digraph of 2-MUs occur exactly in the one-variable case:
\begin{lem}\label{lem:antiparallelarcsd12MU}
  For $F \in \Bmusati{\delta=1}$ with $n(F) \ge 2$ there are no antiparallel arcs in $\idg(F)$ (and thus $\abs{E(\ig(F))} = \abs{E(\idg(F))}$).
\end{lem}

We generate (exactly) all of $\Bmusati{\delta=1}$ by inverse 1-singular DP-reduction, starting with $\set{\bot}$ as above, when we always make sure that only clauses of length at most $2$ are obtained.
The fundamental observation here is, that once we have created some $F$ containing a clause of length at least three, that is, $F \notin \Pcls 2$, then every further $F$ will also have this property, since one of $C', C''$ must contain at least two literals from $C$ (if $C$ has at length at least three), and this together with $v$ resp.\ $\ol v$ makes at least three literals.

That is, we start with the empty clause, and repeatedly replace a single clause $C$ already generated by two clauses $v \in C'$, $\ol v \in C''$ for
\begin{displaymath}
  (C' \sm \set{v}) \cup (C'' \sm \set{\ol v}) = C \text{ and } \abs{C'}, \abs{C''} \le 2,
\end{displaymath}
and a new variable $v$.
The clause-sets generated this way, starting with $\set{\set{\bot}}$, together exactly yield $\Bmusati{\delta=1}$.

Here we consider generating the elements of $\Bmusatds$ (without $\set{\bot}$), and so the starting point are the 2-MUs with precisely one variable, namely $\set{\set{v}, \set{\ol v}}$.
We need indeed not to create all of $\Bmusatds$, but only up to isomorphism.
We have w.l.o.g.\ the following cases, for $\abs{C} = 1,2$ (replacements of a unit-clause or of a binary clause), obtaining three rules T (``\textbf{transfer} unit-clause''), E (``\textbf{eliminate} unit-clause''), S (``\textbf{stretch} binary clause''):
\begin{enumerate}
\item[(i)] If $C = \set{x}$, then:
  \begin{enumerate}
  \item[\textbf{Rule T:}] $C' = \set{x,v}$, $C'' = \set{\ol v}$
  \item[\textbf{Rule E:}] $C' = \set{x,v}$, $C'' = \set{x, \ol v}$.
  \end{enumerate}
\item[(ii)] \textbf{Rule S:} If $C = \set{x,y}$, $x \ne y$, then: $C' = \set{x,v}$, $C'' = \set{y,\ol v}$.
\end{enumerate}
These rules can be applied to an arbitrary clause-set $F$ and a chosen $C \in F$, replacing $C$ by the clauses $C', C''$, using $v \in \Va \sm \var(F)$.
The above ``w.l.o.g.'' here just means that all created unit-clauses (by Rule T) are negative.
Since the rules can not increase the number of initial unit-clauses, we get:
\begin{lem}\label{lem:atmosttwounits}
  $F \in \Bmusats$ with $\delta(F) = 1$ has at most two unit-clauses.
\end{lem}

\begin{corol}\label{cor:atmosttwo units}
  We have the following restrictions on the applications of Rules T, E, S in the generation process of $\Bmusatds$ up to isomorphism:
  \begin{itemize}
  \item Rule E can be applied at most twice.
  \item After applying Rule E twice, no application of Rule T is possible.
  \item After applying Rule E at most once, Rule T is applicable arbitrarily often.
  \item Rule S can only be applied after at least one application of Rules T or E, and then can be applied arbitrarily often.
  \end{itemize}
\end{corol}

For later use we note the following facts on unit-clauses in 2-MUs:
\begin{lem}[{{\cite[Lemma 3.1]{KBWS20162CNFJ}}}]\label{lem:nounit2MU}
  $F \in \Bmusat$ with a unit-clause has $\delta(F) = 1$.
\end{lem}
\begin{prf}
If $F$ has a unit-clause, then unit-clause propagation creates the empty clause, since otherwise $F$ had a non-trivial autarky, as in the proof of Lemma \ref{lem:Aspvall}, contradicting MU.
Thus $F$ is renamable Horn, and so $\delta(F) = 1$.
\Qed
\end{prf}

We obtain an alternative proof of the generalisation of Lemma \ref{lem:atmosttwounits} (this statement appeared first in \cite[Proposition 3, Page 48]{KBZ2002b}):
\begin{corol}[{{\cite[Lemma 8]{Liberatore2008Redundanz}}}]\label{cor:nounit2MU}
  $F \in \Bmusat$ has at most two unit-clauses.
\end{corol}
\begin{prf}
If $F$ has a unit-clause, then $\delta(F) = 1$, and thus, as observed in Lemma \ref{lem:atmosttwounits}, has at most two unit-clauses.
\Qed
\end{prf}

\subsection{Standardising the generation process}
\label{sec:standgen}

We now (further) standardise the process, to minimise the number of case distinctions needed.
It would be possible to start only with variable $v=1$, that is, with $\set{\set{1},\set{-1}}$, and for each new variable to choose the next natural number.
In our examples we will sometimes proceed in this way, but in general it is more convenient to have a free choice of variables (restricting this doesn't save anything from a proof perspective).
To really simplify the generation, the application of rules need to be restricted.

For each rule, the clause we choose ($C$ above) is the \textbf{main clause}, while the \textbf{side clauses} are the replacement clauses ($C', C''$ above).

First we note that the generation process can be restricted w.l.o.g.\ to have three consecutive phases for the three rules:
\begin{itemize}
\item If Rule S is followed by Rule T or Rule E, then we can swap the rule-applications, as the side clauses for Rule S are binary and thus disjoint with the main clause for Rule T or Rule E.
\item If Rule E is followed by Rule T, then also here we can swap the rule-applications, because of the disjointness of the side clauses of Rule E and the main clause of Rule T.
\end{itemize}
So we can assume that a generation process has first applications of Rule T, then at most two applications of Rule E, and then applications of Rule S.

Furthermore, two consecutive applications of Rule T can be replaced by one application of Rule T followed by one application of Rule S, since, assuming that Rule T is first applied to $\set{1}$, then to $\set{-2}$:
\begin{eqnarray*}
  & \set{1} \leadsto^\mr{T} \set{1,2}, \set{-2} \leadsto^\mr{T} & \set{1,2}, \set{-2,3}, \set{-3} \quad \text{is simulated by}\\
  & \set{1} \leadsto^\mr{T} \set{1,3}, \set{-3} \leadsto^\mr{S} & \set{1,2}, \set{-2,3}, \set{-3}.
\end{eqnarray*}

The last point in this standardisation process is to consider exactly one (initial) application of Rule T (so we have still exactly two unit-clauses), followed by at least one application of Rule E.
Here it doesn't matter, whether the first application of Rule E uses as main clause the original unit-clause or the new unit-clause produced by Rule T, while we note that the unit-clause for a second application of Rule E would be unique (since it would eliminate the last unit-clause).
The reason is that both clause-sets are isomorphic: the first case yields
\begin{displaymath}
  \set{\set{1}, \set{-1}} \leadsto^\mr{A} \set{\set{1,2}, \set{-2}, \ul{\set{-1}}} \leadsto^\mr{B} \set{\set{1,2}, \set{-2}, \ul{\set{-1,3}, \set{-1,-3}}},
\end{displaymath}
the second case yields
\begin{displaymath}
  \set{\set{1,2}, \ul{\set{-2}}, \set{-1}} \leadsto^\mr{B} \set{\set{1,2}, \ul{\set{-2,3},\set{-2,-3}}, \set{-1}},
\end{displaymath}
and the isomorphism swaps variables $1$ and $2$.
We summarise:
\begin{lem}\label{lem:genpr2MU1}
  We can generate up to isomorphism the elements of $\Bmusatds$ by a sequence of applications of Rules T, E, S, with the following restrictions:
  \begin{enumerate}
  \item First at most one application of Rule T,
  \item then at most two applications of Rule E,
  \item and finally arbitrarily many application of Rule S (if at least one application of Rules T or E took place).
  \end{enumerate}
  If we have applications of Rule T and then at least one Rule E, then as main clause of the first Rule E the new unit-clause is used.
\end{lem}

We obtain five basic clause-sets of $F \in \Bmusatds$, according to the number of applications of Rules T, E (while Rule S is applied arbitrarily often):
\begin{corol}\label{cor:fiveclss}
  The five starting points for the applications of Rule S are as follows, showing the sequence of applications of Rules T, E, and after the colon the number of unit-clauses:
\begin{description}
\item[(T)] $\set{\set{1,2},\set{-2},\set{-1}}$ : 2.
\item[(E)] $\set{\set{1,2},\set{1,-2},\set{-1}}$ : 1.
\item[(TE)] $\set{\set{1,2},\set{-2,3},\set{-2,-3},\set{-1}}$ : 1.
\item[(EE)] $\set{\set{1,2}.\set{1,-2},\set{-1,3},\set{-1,-3}}$ : 0.
\item[(TEE)] $\set{\set{1,2},\set{-2,3},\set{-2,-3},\set{-1,4},\set{-1,-4}}$ : 0.
\end{description}
Every element of $\Bmusatds$ is either isomorphic to $\set{\set{1},\set{-1}}$, or to one clause-set obtained from the above five clause-sets (T) -- (TEE) by applications of Rule S, applied $n \ge 0$ times.
\end{corol}

Concerning Rule S, which replaces $\set{x,y}$, which can be considered as the implication $\neg x \ra y$, by the two implications $\neg x \ra v \land v \ra y$ (with $v$ a new variable), it is easy to see that it produces just a chain as follows:
\begin{lem}\label{lem:2cnfchain}
  Applying Rule S $n \ge 1$ times to the clause-set $\set{\set{x,y}}$, in any order, yields a clause-set with $n+1$ clauses, isomorphic to
  \begin{displaymath}
    S_{\set{x,y}}^n := \setb{\set{x,1},\set{-1,2}, \dots, \set{-(n-1),n},\set{-n,y}}
\end{displaymath}
(assuming here that the auxiliary variables $1, \dots, n$ are new).

  And since applications of Rule S don't interfere, if we have an arbitrary clause-set $F$, then $n \ge 0$ applications of Rule S to $F$, in any order, yield a clause-set isomorphic to the union of $F_1 \cup \dots F_m \cup F^*$, where this union is disjoint, $n = n_1 + \dots + n_m$, $n_i \in \NN$, $C_i \in F$ are chosen with $\abs{C_i} = 2$, $F_i := S_{\set{C_i}}^{n_i}$ with appropriately renamed (new) auxiliary variables, and $F^* := F \sm \set{C_1, \dots, C_m}$.
\end{lem}
As a (sub-)formula-transformation, Lemma \ref{lem:2cnfchain} says that the implication $\neg x \ra y$ is expanded to $\neg x \ra 1 \ra \dots \ra n \ra y$.

\subsection{Four families}
\label{sec:fourfamilies}

To apply Lemma \ref{lem:2cnfchain} to the cases of Corollary \ref{cor:fiveclss} is the task of the following three subsections, comprising the three main cases of 2, 1, 0 unit-clauses.
Instead of obtaining five families, we consider four families, by handling Cases (E), (TE) together:
\begin{defi}\label{def:2MU(1)}
  Define the uniform 2-CNF $M_n$ with $n-1$ clauses for $n \in \NN$ (using integers as literals) as $M_n := \set{\set{-1,2}, \dots,\set{-(n-1),n}}$, that is, $M$ is the implication chain $1 \ra \dots \ra n$.
  Now the four families are:
  \begin{enumerate}
  \item[I] $\bmm{\Hp_n} := M_n \cup \set{\set{1}, \set{-n}}$ for $n \ge 1$ (Case (T) plus the trivial case; Subsection \ref{sec:twounits}).
  \item[II] $\bmm{\Hl_{n,i}} := M_n \cup \set{\set{1}, \set{-n,-i}}$ for $n \ge 2$, $1 \le i \le n-1$ (Cases (E), (TE); Subsection \ref{sec:oneunit}).
  \item[III] $\bmm{\Shl_{n,i}} := M_n \cup \set{\set{1,i}, \set{-n,-i}}$ for $n \ge 3$, $2 \le i \le \frac{n+1}2$ (Case (EE); Subsection \ref{sec:nounits}).
  \item[IV] $\bmm{\Ub_{n,x,y}} := M_n \cup \set{\set{1,x}, \set{-n,-y}}$ for $n \ge 4$, $2 \le x < y \le n-1$, and $x + y \le n+1$ (Case (TEE); Subsection \ref{sec:nounits}).
  \end{enumerate}
\end{defi}
Note that in Definition \ref{def:2MU(1)} the upper index is the number of unit-clauses, while the first lower index is the number of variables.
The four families I - IV help simplifying the argumentation and distinguishing isomorphism types (and they exactly correspond to the four homeomorphism types, as later shown in Theorem \ref{thm:homeodef1}).
For general understanding it is helpful to see how one can comprise all four families into the final family $\Ub_{n,x,y}$, namely we obtain $\Hp_n$ for $x = 1, y = n$, $\Hl_{n,i}$ for $x = 1, y = i$, and $\Shl_{n,i}$ for $x = i, y = i$.
The proof of unsatisfiability (which also clearly shows minimal unsatisfiability) for this general form of $\Ub_{n,x,y}$ is as follows:
\begin{enumerate}
\item From $\set{1,x}$ and the initial part of $M_n$, until $x$, we get $\set{x}$.
\item From the middle part of $M_n$, from $x$ to $y$, we then get $\set{y}$.
\item From $\set{-n,-y}$ and the final part of $M_n$, starting with $y$, we get $\set{-y}$.
\end{enumerate}

A different account of the four families emphasises the implication chains as follows.
Here different names mean different underlying variables, ``$\ra \dots \ra$'' means chains of binary clauses (as implications), with (completely) new variables (different from the mentioned ones, and disjoint for different chains.
We state the degree of variables different from $2$:
\begin{enumerate}
\item[I] $x;\, \ol x$\\
  $x;\, x \ra \cdots \ra y;\, \ol y$ .
\item[II] $x;\, x \ra \cdots \ra \ol x$ : $\var(x)$ of degree $3$\\
  $x;\, x \ra \cdots \ra y \ra \cdots \ra \ol y$ : $\var(y)$ of degree $3$.
\item[III] $x \ra \cdots \ra \ol x \ra \cdots \ra x$ : $\var(x)$ of degree $4$.
\item[IV] $x \ra \cdots \ra \ol x \ra \cdots \ra y \ra \cdots \ra \ol y$ : $\var(x), \var(y)$ of degree $3$.
\end{enumerate}

\subsection{Two unit-clauses (Family I)}
\label{sec:twounits}

First we investigate Case (T), leading to family $\Hp_n$ for $n \ge 2$, while the trivial case is $n = 1$; if one wants to discriminate here, then $n=1$ yields \textbf{Family Ia}, while otherwise we have \textbf{Family Ib}.

\begin{lem}\label{lem:firstfamily}
  Applying Rule S arbitrarily often to Case (T), plus the single clause-set $\set{\set{1},\set{-1}}$, yields up to isomorphism exactly the family $\Hp_n$ (Family I in Definition \ref{def:2MU(1)}), that is, the formula $(1) \land (1 \ra \dots \ra n) \land (-n)$.
\end{lem}
\begin{prf}
We start with $\set{\set{1,2},\set{-2},\set{-1}}$ (Corollary \ref{cor:fiveclss}, Case (T)).
Consider $n \ge 2$.
We want to apply Lemma \ref{lem:2cnfchain} for $n-2$ application of Rule (S).
Rename the starting point to $\set{\set{-1,n},\set{-n},\set{1}}$ (swapping variables $n$ and $2$, and flipping variable $1$).
Replacing $\set{-1,n}$, using new variables $2, \dots, n-1$, we obtain the $n-1$ clauses \mathlist{\set{-1,2},\set{-2,3},\dots,\set{-(n-1),n}} (Lemma \ref{lem:2cnfchain}).
Together with the two old clauses $\set{1},\set{-n}$ this is $\Hp_{n}$.
While $\Hp_1 = \set{\set{1},\set{-1}}$.
\Qed
\end{prf}

Since the other four starting points of Corollary \ref{cor:fiveclss} have at most one unit-clause, we have shown (first shown in \cite[Lemma 5.1, Part 1]{KBZ2002b}):
\begin{lem}\label{lem:2unit}
  For $F \in \Bmusati{\delta=1}$ holds $F \cong \Hp_{n(F)}$ iff $F$ has two unit-clauses, i.e., $u(F) = 2$.
  For these $F$, all variables have degree $2$ (i.e., $n_2(F) = n(F)$).
\end{lem}
The implication digraph of $\Hp_n$ is a cycle digraph with $2n$ vertices and $2n$ edges, shown as follows.
Here arcs from unit-clauses are drawn as double-arcs\footnote{if multi-digraphs would be used, then unit-clauses indeed would yield two parallel arcs}:
\begin{displaymath}
  \idg(\Hp_n) = \xymatrix @C=1.5em @R=0.9em {
    1 \ar[r] & 2 \ar[r] & \cdots  \ar[r] & n-1 \ar[r] & n \ar@{=>}[d]\\
    -1 \ar@{=>}[u] & -2 \ar[l] & \cdots \ar[l] & -(n-1) \ar[l] & -n \ar[l]
  }
\end{displaymath}

\begin{corol}\label{cor:firstfamily}
  The isomorphism type of the implication graph of $\Hp_{n}$ ($n \ge 1$), that is, of 2-MUs with two unit-clauses, is that of a cycle graph of length $2 n$ for $n \ge 2$ (Family Ib), while for $n=1$, that is, Family Ia, it is the complete graph with $2$ vertices.
  The implication digraphs of these 2-MUs have exactly one contradictory cycle (which is equal to its own contraposition).
\end{corol}

As an interesting application we can characterise exactly when implication digraphs are cycles:
\begin{thm}\label{thm:2mudigraph}
  For $F \in \Bclss$ the implication digraph is a cycle digraph iff $F$ is isomorphic to $\Hp_{n(F)}$.
\end{thm}
\begin{prf}
It remains to consider the case that $\idg(F)$ is a cycle digraph, and we have to show $F \cong \Hp_{n(F)}$.
$F$ has a contradictory cycle, and thus is unsatisfiable.
Removal of any clause from $F$ would remove at least one arc, resulting in a digraph without a cycle (at all), which would be satisfiable --- whence $F$ is indeed minimally unsatisfiable.
By Lemma \ref{lem:CycleSkewsym} we know that $F$ has exactly two unit-clauses, and thus we get $2 (c(F) - 2) + 2 = 2 c(F) - 2$ many arcs in $\idg(F)$.
Since $\idg(F)$ is a cycle digraph, it has $2 n(F)$ many vertices and $2 n(F)$ many arcs.
Thus $2 c(F) - 2 = 2 n(F)$, that is $n(F) = c(F) - 1$, and we get $\delta(F) = 1$.
So by Lemma \ref{lem:2unit} indeed we can conclude that $F \cong \Hp_{n(F)}$ holds.
\Qed
\end{prf}

\begin{examp}\label{exp:2mudigraph}
  If $\ig(F)$ is a cycle graph, then there are many possibilities for $F$.
  For example $\Hp_2 = \set{\set{1},\set{-1,2},\set{-2}}$ has as implication graph the cycle
  \begin{displaymath}
    \ig(\Hp_2) = \xymatrix @C=1.5em @R=0.9em {
      1 \aru[r] & 2 \aru[d]\\
      -1 \aru[u] & -2 \aru[l]
    },
  \end{displaymath}
  while the satisfiable $\set{\set{1},\set{1,-2},\set{-2}}$ has the same implication graph.
\end{examp}

\subsection{One unit-clause (Family II)}
\label{sec:oneunit}

We now come to Cases (E), (TE), i.e., exactly one unit-clause, which together yield Family II from Definition \ref{def:2MU(1)}; if one wants to be more specific, then Case (E) yields \textbf{Family IIa} (Subsection \ref{sec:FamilyIIa}), while Case (TE) yields \textbf{Family IIb} (Subsection \ref{sec:FamilyIIb}).

First we note that if we have one chain based on $\set{x,z}$ and another chain based on $\set{y,\ol z}$, then we can merge these chains:
\begin{lem}\label{lem:2cnfchainmerge}
  Applying Rule S $n \ge 0$ times, in any order, to the two binary clauses $\set{\set{x,z}, \set{y,\ol z}}$ (with $x, y$ arbitrary literals), yields  the $n+2$ clauses $\set{\set{x,1},\set{-1,2}, \dots, \set{-n,n+1},\set{-(n+1),y}}$ up to isomorphism (assuming new variables $1, \dots, n+1$).
\end{lem}
As a (sub-)formula-transformation, Lemma \ref{lem:2cnfchainmerge} says that the implication-chain $\neg x \ra z \ra y$ is expanded (and renamed) to $\neg x \ra 1 \ra \dots \ra n+1 \ra y$.

\subsubsection{Family IIa}
\label{sec:FamilyIIa}

\begin{lem}\label{lem:secondfamily1}
  Applying Rule S $(n-2)$-times for $n \ge 2$ to Case (E) yields up to isomorphism exactly the family $\Hl_{n,1}$ (Family II in Definition \ref{def:2MU(1)} with index $i=1$), that is, the formula $(1) \land (1 \ra \dots \ra n) \land (n \ra -1)$.
  This family is called IIa.
\end{lem}
\begin{prf}
We start with $\set{\set{1,2},\set{1,-2},\set{-1}}$ (Corollary \ref{cor:fiveclss}, Case (E)).
Swap variables $2$ and $z$, and flip variable $1$, obtaining
\begin{displaymath}
  \set{\set{-1,z},\set{-1,-z},\set{1}}.
\end{displaymath}
By Lemma \ref{lem:2cnfchainmerge}, applying Rule (S) $n-2$ times to $\set{-1,z},\set{-1,-z}$, using new variables $2, \dots, n$, we obtain the $n$ clauses $\set{-1,2},\set{-2,3},\dots,\set{-(n-1),n},\set{-n,-1}$, which together with the old clause $\set{1}$ is $\Hl_{n,1}$.
\Qed
\end{prf}

The implication digraph of $\Hl_{n,1}$ has $2n$ vertices and $2n+1$ edges, and consists of two cycle digraphs of length $n+1$, which overlap in a path of length $1$; two vertices have degree $3$ (namely $1, -1$), all other vertices have degree $2$:
\begin{gather*}
  \idg(\Hl_{n,1}) = \xymatrix @C=1.5em @R=0.8em {
    1 \ar[r] \ar[drr] & \cdots \ar[r] & n \ar[dll]\\
    -1 \ar@{=>}[u] & \cdots \ar[l] & -n \ar[l]
  } =
  \xymatrix @C=1em @R=0.4em {
    n \ar[rd] & \ar[l] & \ar@{..}[l] & 2 \ar[l] \\
     & -1 \ar@{=>}[r] & 1 \ar[ur] \ar[dr] \\
    -2 \ar[ru] & \ar[l] & \ar@{..}[l] & -n \ar[l]
  }
\end{gather*}

\subsubsection{Family IIb}
\label{sec:FamilyIIb}

\begin{lem}\label{lem:secondfamily2}
  Applying Rule S $(n-3)$-times for $n \ge 3$ to Case (TE) yields up to isomorphism exactly the family $\Hl_{n,i}$, $2 \le i \le n-1$ (Family II in Definition \ref{def:2MU(1)} with index $i \ge 2$), that is, the formula $(1) \land (1 \ra \dots \ra n) \land (n \ra -i)$.
  This family is called IIb.
\end{lem}
\begin{prf}
We start with $\set{\set{1,2},\set{-2,3},\set{-2,-3},\set{-1}}$ (Corollary \ref{cor:fiveclss}, Case (TE)).
Swap variables $2$ and $i$, and $3$ and $z$, and flip variable $1$, obtaining
\begin{displaymath}
  \set{\set{-1,i},\set{-i,z},\set{-i,-z},\set{1}}.
\end{displaymath}
Let $p := i - 2$ (so $0 \le p \le n-3$) and $q := n-3-p$ (so $0 \le q \le n-3$ and $p+q=n-3$).
By Lemma \ref{lem:2cnfchain}, applying Rule (S) $p$-times and using new variables $2,\dots,p+1$ (note $p+1=i-1$), from $\set{-1,i}$ we obtain (now as formula) $1 \ra \dots \ra i$.
And by Lemma \ref{lem:2cnfchainmerge}, applying Rule (S) $q$-times and using new variables $i+1, \dots, i+1+q$ (note $i+1+q=(p+3)+(n-3-p)=n$), from ${\set{-i,z},\set{-i,-z}}$ we obtain the formula $i \ra i+1 \ra \dots \ra n \ra -i$, which is equivalent to $(1 \ra \dots \ra n) \land (i \ra -n)$ plus the old clause $(1)$.
\Qed
\end{prf}

We note that with $i=n$ we would get $\Hp_n = \Hl_{n,n}$, but we avoid this degeneration, so that we can distinguish the implication digraphs.
The implication digraph of $\Hl_{n,i}$, $i \ge 2$, has $2n$ vertices and $2n+1$ edges, and consists of two cycle digraphs of length $n+i$, which overlap in a path of length $2i - 1 \ge 3$; two vertices have degree $3$ (namely $i, -i$), all other vertices have degree $2$:
\begin{gather*}
  \idg(\Hl_{n,i}) = \xymatrix @C=1.5em @R=0.8em {
    1 \ar[r] & \cdots \ar[r] & i \ar[r] \ar[drr] & \cdots \ar[r] & n \ar[dll]\\
    -1 \ar@{=>}[u] & \cdots \ar[l] & -i \ar[l] & \cdots \ar[l] & -n \ar[l]
  } = \\
  \xymatrix @C=1em @R=0.4em {
    & n \ar[ld] & n-1 \ar[l] && \cdots \ar[ll] & i+2 \ar[l] & i+1 \ar[l] & \\
    -i  \ar[r] & -(i-1) \ar[r] & \cdots \ar[r] & -1 \ar@{=>}[r] & 1 \ar[r] &\cdots \ar[r] & i-1 \ar[r] & i \ar[ld] \ar[ul] \\
    & -(i+1) \ar[lu] & -(i+2) \ar[l] && \cdots \ar[ll] & -(n-1) \ar[l] & -n \ar[l]
  }
\end{gather*}
The lengths of the three paths between the two vertices of degree $3$ are $n-i+1, 2 i - 1, n-i+1$ (which sums to $2n+1$).

\subsubsection{Both subfamilies together}
\label{sec:IIaIIb}

Since we covered exactly the cases of Corollary \ref{cor:fiveclss} with one unit-clause in Lemmas \ref{lem:secondfamily1}, \ref{lem:secondfamily2}, we have shown (first shown in \cite[Lemma 5.1, Part 2]{KBZ2002b}):
\begin{lem}\label{lem:1unit}
  For $F \in \Bmusati{\delta=1}$ holds $F \cong \Hl_{n(F),i}$, for some $1 \le i < n(F)$ (Family II), iff $u(F) = 2$.
  These $F$ have $n_3(F) = 1$, $n(F) = n_2(F) + n_3(F)$.
\end{lem}

\begin{corol}\label{cor:secondfamily}
  The isomorphism type of the implication graph of $\Hl_{n,i}$ ($n \ge 2$, $1 \le i \le n-1$), that is, of 2-MUs with exactly one unit-clause, is that of a graph with $2 n$ vertices and $2 n + 1$ edges, with exactly two vertices of degree $3$ and all other vertices of degree $2$, and with three vertex-disjoint paths between them, two of length $n-i+1 \ge 2$, one of length $2i-1 \ge 1$.
  Thus $n, i$ can be uniquely computed from the graph.
  The implication digraphs of these 2-MUs have exactly two contradictory cycles (being contrapositions of each other).
\end{corol}

\subsection{Zero unit-clauses (Families III, IV)}
\label{sec:nounits}

Finally we come to Cases (EE), (TEE) (without unit-clauses, and thus all clause-sets considered are 2-uniform).

\subsubsection{Family III}
\label{sec:familyIII}

\begin{lem}\label{lem:thirdfamily}
  Applying Rule S $(n-3)$-times for $n \ge 3$ to Case (EE) yields up to isomorphism exactly the family $\Shl_{n,i}$ (Family III in Definition \ref{def:2MU(1)}), that is, the formula $(-i \ra 1) \land (1 \ra \dots \ra n) \land (n \ra -i)$ (with $2 \le i \le \frac{n+1}2$).
  This formula is exactly (up to isomorphism) the implication chain $x \ra \dots \ra -x \ra \dots \ra x$, where both dotted parts contain only new variables, and where $i-1 \ge 1$ is the number of elements of the first dotted part and $n-i \ge 1$ of the second.
\end{lem}
\begin{prf}
We start with $\set{\set{1,2},\set{1,-2},\set{-1,3},\set{-1,-3}}$ (Corollary \ref{cor:fiveclss}, Case (EE)).
Swap variables $2, z_1$ and $3, z_2$, and flip variable $1$, obtaining
\begin{displaymath}
  \set{\set{-1,z_1},\set{-1,-z_1},\set{1,z_2},\set{1,-z_2}}.
\end{displaymath}
Consider first $2 \le i \le n-1$ (not just $i \le \frac{n+1}2$).
Apply Lemma \ref{lem:2cnfchainmerge} $(i-2)$-times to the first two clauses (replacing $z_1$), obtaining the implication chain $1 \ra \dots \ra -1$, with $i-1$ variables in the dotted part.
And apply Lemma \ref{lem:2cnfchainmerge} $(n-i-1)$-times to the last two clauses (replacing $z_2$), obtaining the implication chain $-1 \ra \dots \ra 1$, with $n-i$ variables in the dotted part.
So we have shown that we obtain exactly all the $\Shl_{n,i}$ (with the extended range for $i$), and it remains to show that the cases with $i > \frac{n+1}2$ (equivalently $n-i<i-1$) are isomorphic to cases with $2 \le i' \le \frac{n+1}2$.
And this is easily done by flipping all literals --- now the two dotted parts swapped place, and we are done.
\Qed
\end{prf}

The implication digraph of $\Shl_{n,i}$ has $2 n$ vertices and $2 n + 2$ edges, and two vertices have degree $4$, while all other vertices have degree $2$:
\begin{gather*}
  \idg(\Shl_{n,i})) = \xymatrix @C=1.5em @R=0.8em {
    1 \ar[r] & \cdots \ar[r] & i \ar[r] \ar[drr] & \cdots \ar[r] & n \ar[dll]\\
    -1 \ar[urr] & \cdots \ar[l] & -i \ar[l] \ar[ull] & \cdots \ar[l] & -n \ar[l]
  } = \\
  \xymatrix @C=1.5em @R=0.3em {
    & 1 \ar[r] & 2 \ar[r] & \cdots \ar[r] & i-2 \ar[r] & i-1 \ar[rdd] & \\
    & -(i-1) \ar[r] & -(i-2) \ar[r] & \cdots \ar[r] & -2 \ar[r] & -1 \ar[rd] & \\
    -i \ar[ru] \ar[ruu] &&&&&& i \ar[ld] \ar[ldd]\\
    & -(i+1) \ar[lu] & -(i+2) \ar[l] & \cdots \ar[l] & -(n-1) \ar[l] & -n \ar[l] & \\
    & n \ar[luu] & n-1 \ar[l] & \cdots \ar[l] & i+2 \ar[l] & i+1 \ar[l]
  }
\end{gather*}

\begin{corol}\label{cor:thirdfamily}
  The isomorphism type of the implication graph of $\Shl_{n,i}$ ($n \ge 3$, $2 \le i \le \frac{n+1}2$; Family III), that is, of 2-MUs of deficiency one without unit-clauses and with two variables of degree four, is that of a graph with $2 n$ vertices and $2 n + 2$ edges, with exactly two vertices of degree $4$ and all other vertices of degree $2$, and with four vertex-disjoint paths between them, two of length $i \ge 2$, two of length $n-i+1 \ge 2$, where we always have $i \le n-i+1$.
  Thus $n, i$ can be uniquely computed from the graph.
  The implication digraphs of these 2-MUs have exactly four contradictory cycles (grouped into two pairs which are contrapositions of each other).
\end{corol}

\subsubsection{Family IV}
\label{sec:familyIV}

\begin{lem}\label{lem:fourthfamily}
  Applying Rule S $(n-4)$-times for $n \ge 4$ to Case (TEE) yields up to isomorphism exactly the family $\Ub_{n,x,y}$ (Family IV in Definition \ref{def:2MU(1)}), that is, the formula $(-x \ra 1) \land (1 \ra \dots \ra n) \land (n \ra -y)$ (with $2 \le x < y \le n-1$ and $x + y \le n+1$).
  This formula is exactly (up to isomorphism) the implication chain
  \begin{displaymath}
    a \ra \overbrace{\dots}^p \ra -a \ra \overbrace{\dots}^r \ra b \ra \overbrace{\dots}^q \ra -b
  \end{displaymath}
  (with $\var(a) \ne \var(b)$), where all dotted parts contain only new variables, namely $p = n-y \ge 1$ in the front, $r = y-x-1 \ge 0$ in the middle, and $q = x-1 \ge 1$ in the back (note $n = p+r+q+2$).
  The additional condition $x + y \le n+1$ just means $q \le p$, that is, the front part contains at least as many variables as the back part.
\end{lem}
\begin{prf}
Writing the two special implications of the formula for $\Ub_{n,x,y}$ as $-1 \ra x$ and $n \ra -y$, and reorganising the implication chain $M_n = 1 \ra \dots \ra n$ as $y \ra \dots \ra n$ together with $-y \ra \dots -1$ (which includes $-x$), we obtain $y \ra \dots \ra n \ra -y \ra \dots \ra -x \ra \dots \ra -1 \ra x$, which is the alternative form $a \ra \dots \ra -a \ra \dots \ra b \ra \dots \ra -b$, using $a := y$ and $b := -x$.
Let $p \ge 1$, $r \ge 0$, $q \ge 1$ be the number of (new) variables in respectively the first, middle, back dotted parts of the alternative form.
By the above translation we obtain
\begin{itemize}
\item  $p = \abs{\set{y+1,\dots,n}} = n-y$ ($\Lra y = n-p$)
\item $r = \abs{\set{y-1,\dots,x+1}} = y-x-1$
\item $q = \abs{\set{x-1,\dots,1}} = x-1$ ($\Lra x =q+1$).
\end{itemize}
The condition $1 \le p \le n-3$ is thus equivalent to $3 \le y \le n-1$, while $1 \le q \le n-3$ is equivalent to $2 \le x \le n-1$.
And $x < y$ is equivalent to $p+q+1 < n$ (recall $p+r+q+2=n$).

The \emph{additional} condition $x+y \le n+1$ is equivalent to $x \le n-y+1$, which is equivalent to $q \le p$:
Indeed, in case of $q > p$, by contraposition and flipping all signs, we see that we can read the implication chain for the alternative form backwards, and thus obtain an isomorphic chain now with $q \le p$.

It remains to show that exactly the alternative forms are produced.
Start with $\set{\set{1,2},\set{-2,3},\set{-2,-3},\set{-1,4},\set{-1,-4}}$ (Corollary \ref{cor:fiveclss}, Case (TEE)), renamed to $\set{\set{a,b},\set{-b,z_1},\set{-b,-z_1},\set{-a,z_2},\set{-a,-z_2}}$.
Applying Lemma \ref{lem:2cnfchain} to the first clause we obtain $-a \ra \dots \ra b$ (dots possibly empty).
And applying Lemma \ref{lem:2cnfchainmerge} to the two other blocks of two clauses each we obtain $b \ra \dots \ra -b$ and $a \ra \dots \ra -a$ (dots nonempty).
\Qed
\end{prf}

Allowing degenerations, we have $\Hp_n = \Ub_{n,1,n}$, $\Hl_{n,i} = \Ub_{n,1,i}$ and $\Shl_{n,i} = \Ub_{n,i,i}$.
The implication digraph of $\Ub_{n,x,y}$ has $2 n$ vertices and $2 n + 2$ edges, and four vertices have degree 3, while all other vertices have degree $2$:
\begin{gather*}
  \idg(\Ub_{n,x,y}) = \xymatrix @C=1.5em @R=0.8em {
    1 \ar[r] & \cdots \ar[r] & x \ar[r] & \cdots \ar[r] & y \ar[drr] \ar[r] & \cdots \ar[r] &  n \ar[dll]\\
    -1 \ar[urr] & \cdots \ar[l] &  -x \ar[l] \ar[ull] & \cdots \ar[l] & -y \ar[l] & \cdots \ar[l] & -n \ar[l]
  } = \\
  \hspace{-3em}
  \xymatrix @C=1.1em @R=0.6em {
    & 1 \ar[r] & 2 \ar[r] & \cdots \ar[r] & x-2 \ar[r] & x-1 \ar[rd] & \\
    -x \ar[ru] \ar[r] & -(x-1) \ar[r] & -(x-2) \ar[r] & \cdots \ar[r] & -2 \ar[r] & -1 \ar[r] & x \ar[d]\\
    -(x+1) \ar[u] &&&&&& x+1 \ar[d]^{\dots}_{\dots} \\
    -(y-1) \ar[u]^{\dots}_{\dots} &&&&&& y-1 \ar[d] \\
    -y \ar[u] & -(y+1) \ar[l] & -(y+2) \ar[l] & \cdots \ar[l] & -(n-1) \ar[l] & -n \ar[l] & y \ar[l] \ar[ld] \\
    & n \ar[lu] & n-1 \ar[l] & \cdots \ar[l] & y+2 \ar[l] & y+1 \ar[l]
  }
\end{gather*}

\begin{corol}\label{cor:fourthfamily}
  The isomorphism type of the implication graph of $\Ub_{n,x,y}$ ($n \ge 4$, $2 \le x < y \le n-1$, $x+y \le n+1$), that is, of 2-MUs of deficiency one without unit-clauses and with four variables of degree three, is that of a graph with $2 n$ vertices and $2 n + 2$ edges, with exactly four vertices of degree $3$ and all other vertices of degree $2$, and with the following path structure between the four degrees-$3$-vertices (yielding, as in the other cases, a partitioning of the edge-set):
  \begin{enumerate}
  \item There are exactly $6$ such paths, two of length $y-x \ge 1$, two of length $x \ge 2$, two of length $n - y + 1 \ge 2$.
    (Adding up to $2n+2$.)
  \item These paths are vertex-disjoint, except for begin and end.
  \item Calling such paths ``parallel'', if they have the same begin and end, this equivalence relation yields four equivalence classes:
    The two paths $A, B$ of length $y - x$ are two singleton-classes, while the two paths $C_1, C_2$ of length $x$ are parallel, and the two paths $D_1, D_2$ of length $n - y + 1$ are parallel.
  \item Paths $A, B$ share no vertex.
  \end{enumerate}
  We have $x \le n-y+1$.
  Thus $n, x, y$ can be uniquely computed from the graph.
  The implication digraphs of these 2-MUs have exactly four contradictory cycles (grouped into two pairs which are contrapositions of each other).
\end{corol}

\subsubsection{Both families together}
\label{sec:familyIIIFamilyIV}

Since we covered exactly the cases of Corollary \ref{cor:fiveclss} with zero unit-clause in Lemmas \ref{lem:thirdfamily}, \ref{lem:fourthfamily}, we have shown the classification of 2-uniform elements of $\Bmusati{\delta=1}$, summarised in the following lemma:
\begin{lem}\label{lem:2uniform}
  For 2-uniform $F \in \Bmusati{\delta=1}$ with $n := n(F)$ holds:
  \begin{itemize}
  \item If $n_4(F) \ge 1$, then $n \ge 3$ and $F \cong \Shl_{n,i}$ for some $i \in \set{2, \dots, \frac{n+1}2}$ (Family III; and indeed $n_4(F) = 1$ and $n(F) = n_2(F) + n_4(F)$).
  \item Otherwise $n \ge 4$ and $F \cong \Ub_{n,x,y}$ for some $x,y \in \set{2, \dots, n-1}$ with $x < y$ and $x+y \le n+1$ (Family IV; and $n_3(F) = 2$ and $n(F) = n_2(F) + n_3(F)$).
  \end{itemize}
\end{lem}

\subsection{Classification}
\label{sec:d1class}

Altogether we have achieved the classification of $\Bmusatds$ by the four Families I - IV:

\begin{thm}\label{thn:classd1}
  Consider input $F \in \Bmusatds$.
  Let $L(F)$ denote a parameter-list of length $0,1,2,$ according to the applicable case for $\GUb{u(F)}_{n(F),L(F)}$ (Definition \ref{def:2MU(1)}, the four Families I - IV).
  In linear time the unique parameter-list $P(F) = (u(F), n(F), L(F))$ can be computed, such that
  \begin{displaymath}
    F \cong \GUb{u(F)}_{n(F),L(F)}
  \end{displaymath}
  holds.
  This parameter-list is a complete isomorphism invariant for $F$, that is, for $F, F' \in \Bmusatds$ holds $F \cong F'$ iff $P(F) = P(F')$, and also all possible parameter values occur.

  The map $\canonCLS: \Bmusatds \ra \Bmusatds$ given by $\canonCLS(F) := \GUb{u(F)}_{n(F),L(F)}$, is a linear-time computable clause-set-canonisation, that is, for $F, F' \in \Bmusatds$ holds $F \cong F'$ iff $\canonCLS(F) = \canonCLS(F')$.

  The map $F \in \Bmusatds \mapsto \canonCLS'(F) := \ig(\canonCLS(F))$ to the class of graphs is a linear-time computable graph-canonisation, that is, for $F, F' \in \Bmusatds$ holds $F \cong F'$ iff $\canonCLS'(F) = \canonCLS'(F')$.
  From $\canonCLS'(F)$ in linear time $F$ can be reconstructed up to isomorphism.
\end{thm}
\begin{prf}
Exactly one of the four families from Definition \ref{def:2MU(1)} applies to $F$ (up to isomorphism), with the overall case distinction given in Lemmas \ref{lem:2unit}, \ref{lem:1unit}, and \ref{lem:2uniform}.
Indeed, the four families $\Hp_n, \Hl_{n,i}, \Shl_{n,i}, \Ub_{n,x,y}$ are separated by vertex-degrees in the implication graph, since their degree-spectra as triples in $(\nni^3)$ for the numbers of degree-2/3/4-vertices, with ``$\infty$'' meaning ``unbounded'', are resp.\ $(\infty,0,0)$, $(\infty,2,0)$, $(\infty,0,2)$ and $(\infty,4,0)$.
Now the parameters can be recovered from the implication graphs by Corollaries \ref{cor:firstfamily}, \ref{cor:secondfamily}, \ref{cor:thirdfamily}, and \ref{cor:fourthfamily}.
That these computations can be done in linear time, using the adjacency-list representation and the standard random-access model, is easy to see (just repeated graph traversals are sufficient).
 \Qed
\end{prf}

\begin{examp}\label{exam:canon}
  Consider $F := \set{\set{x,y}, \set{\ol x,y}, \set{\ol y,z}, \set{\ol z,\ol y}} \in \Bmusatds$.
  We have $n(F) = 3$, $u(F) = 0$ (number of unit-clauses) and $L(F) = 2$ (length of the paths between the two vertices of degree 4 in $\idg(F)$).
  So $\canonCLS(F) := \GUb{0}_{3,2}$.
\end{examp}

Since the characteristic degree-spectra of the four Families I - IV are useful, we explicitly state them here:
\begin{corol}\label{cor:def1degreespec}
  Every $F \in \Bmusatds$ belongs exactly to one of the four Families I - IV according to the following conditions on the degrees:
  \begin{enumerate}
  \item[I] $n(F) = n_2(F)$.
  \item[II] $n_3(F) = 1$, $n(F) = n_2(F) + 1$.
  \item[III] $n_4(F) = 1$, $n(F) = n_2(F) + 1$.
  \item[IV] $n_3(F) = 2$, $n(F) = n_2(F) + 2$.
  \end{enumerate}
\end{corol}
We will see later in Corollary \ref{cor:deffromdegr}, that indeed these degree-spectra characterise deficiency $1$.

By adding up the contributions we obtain the exact number of isomorphism types of 2-MUs of deficiency one as follows:
\begin{corol}\label{cor:numdef1}
  The exact number of isomorphism types of $F \in \Bmusati{\delta=1}$ with given $n(F) = n \in \NNZ$ is
  \begin{displaymath}
    \begin{cases}
      1 & \text{if } n = 0\\
      \frac 14 n (n+2) & \text{if $n$ is even}\\
      \frac 14 (n+1)^2 & \text{if $n$ is odd.}
    \end{cases}
  \end{displaymath}
  This is \href{https://oeis.org/A076921}{A076921} (where that sequence starts with index $1$).
  More precisely, the number of isomorphism types for the four families in Definition \ref{def:2MU(1)} are given by $\mr{ci}_j(n)$, for $1 \le j \le 4$, as follows:
  \begin{enumerate}
  \item[I] $\mr{ci}_1(n) = 1$;
  \item[II] $\mr{ci}_2(n) = n-1$;
  \item[III] $\mr{ci}_3(n) = \frac{n-1}2$ for odd $n \ge 3$, and $\mr{ci}_3(n) = \frac{n-2}2$ for even $n \ge 4$;
  \item[IV] $\mr{ci}_4(n) = \frac 14 (n-2)^2$ for even $n \ge 4$, and $\mr{ci}_4(n) = \frac 14 (n-1)(n-3)$ for odd $n \ge 5$.
  \end{enumerate}
\end{corol}

\section{Singular DP-reduction and smoothing}
\label{sec:sDPred}

The fundamental tool for the analysis of MUs is ``singular DP-reduction'', i.e., the reduction $F \in \Musat \leadsto \dpi v(F)$ for ``singular variables'' $v$:
in general $\dpi v(F) \notin \Musat$, however if $v$ is a \textbf{singular variable}, i.e., $\ldeg_F(v) = 1$ or $\ldeg_F(\ol v) = 1$ holds, then $\dpi v(F) \in \Musat$ is guaranteed.

The application of DP-reduction for singular variables in $F \in \Musat$ is called \textbf{singular DP-reduction}.
These singular reductions for MUs do not yield tautological resolvents, and neither between the resolvents nor between resolvents and old clauses a contraction happens; recall that we are using clause-\emph{sets}, where as a result of DP-reduction, \emph{in general} two previously different clauses can become equal, and so more clauses might disappear than expected.
Thus by one singular DP-reduction, applied to an MU, exactly one variable and one clause disappears, leaving the deficiency invariant.
So the class of MUs with fixed deficiency $k \ge 1$ is stable under singular DP-reduction (\cite[Lemma 9]{KullmannZhao2012ConfluenceJ}).

An $F \in \Musat$ is called \textbf{nonsingular}, if $F$ does not contain a singular variable; the set of all nonsingular MUs is denoted by $\bmm{\Musatns} \subset \Musat$.
For $F \in \Musat$, the set of all nonsingular MUs reachable from $F$ by singular DP-reduction is denoted by $\bmm{\sdp(F)} \subset \Musatns$.
So for any $F' \in \sdp(F)$ we have $\delta(F') = \delta(F)$.
A fundamental lemma in \cite{KullmannZhao2012ConfluenceJ} is that the elements of $\sdp(F)$ all have the same number of variables (but in general they are non-isomorphic).

For $F \in \Bmusatds$, by definition a variable $v$ is singular iff vertex $v$ in $\idg(F)$ has in- or out-degree $1$.
Since $\Pcls 2$ is stable under resolution, also the classes $\Bmusati{\delta=k}$ are stable under singular DP-reduction.

The basic result, established in \cite{KBZ2002b,AbbasizanjaniKullmann2016MUconference}, for this work is that for $F \in \Bmusat$ and $F' \in \sdp(F)$ holds:
\begin{itemize}
\item If $\delta(F) = 1$ then $F' = \set{\bot}$;
\item if $\delta(F) \ge 2$ then $F' \cong \Bpt{\delta(F)}$.
\end{itemize}
In Section \ref{sec:2-MU(1)} we characterised all 2-MUs with deficiency one, based on reversal of 1-singular DP-reduction.
To generate all 2-MUs for higher deficiencies, general singular DP-reduction (for 2-CNFs) has to be reversed.

\subsection{1-singular DP-reduction}
\label{sec:1sDPred}

For an MU $F$, the nicest case of singular DP-reduction is the confluent case, that is, $\abs{\sdp(F)} = 1$.
By \cite[Section 5]{KullmannZhao2012ConfluenceJ} we have confluence, when performing only ``1-singular DP-reduction'', that is, applying singular DP-reduction only in case of 1-singular variables (i.e., of degree $2$).
It is indeed easy to see that 1-singular DP-reduction for any clause-set never strictly increases any literal-degree, and thus it it confluent.
We denote by $\bmm{\Musatnos} \subset \Musat$ the set of \textbf{non-1-singular} $F \in \Musat$, i.e., where every variable of $F$ has degree at least $3$ (while for nonsingular $F \in \Musatns$, where we note $\Musatns \subset \Musatnos$, every variable has degree at least $4$).
For $\mc C \sse \Musat$ we use $\bmm{\mc{C}^+} := \mc C \cap \Musatnos$.
We use $\bmm{\1dp(F)} \in \Musatnos$ for $F \in \Musat$ to denote the (unique) non-1-singular MU obtained by (repeated) 1-singular DP-reduction from $F$.
The basis for Section \ref{sec:2-MU(1)} is that for all $F \in \Musat$ holds $\1dp(F) = \set{\bot}$ iff $\delta(F) = 1$.
\begin{lem}\label{lem:mupstable}
  $\Musatnos$ is stable under singular DP-reduction.
\end{lem}
\begin{prf}
Since $\Musat$ is stable under singular DP-reduction, we have to show that it is not possible that for $F \in \Musatnos$ after one singular DP-reduction a 1-singular variable appears.
Now the only possibility of a singular DP-reduction on $v$ with main clause $v \in C \in F$ and side-clauses $\ol v \in D_1, \dots, D_m \in F$ to (strictly) decrease the degree of a literal $x \in \lit(F)$ is that $x \in C \cap D_1 \cap \dots \cap D_m$ --- but since $F \in \Musatnos$, we have $m \ge 2$, and thus the literal-degree of $x$ in $\dpi{v}(F)$ is at least two.
So $\dpi{v}(F) \in \Musatnos$.
\Qed
\end{prf}

The analysis of singular DP-reduction for a class $\mc{C} \sse \Musat$, where always stability of $\mc C$ under singular DP-reduction is assumed, now can proceed by first considering the simple confluent reduction $F \in \mc C \leadsto \1dp(F) \in \mc{C}^+$ and characterising the elements of $\mc{C}^+$.
The second stage then can start with $\mc{C}^+ \sse \mc C$, and need only to consider ``non-1-singular DP-reductions'' (that is, singular DP-reductions which are not 1-singular DP-reductions) to arrive at $\mc{C}' = \mc{C} \cap \Musatns$.

\subsection{Smoothing of (multi-)graphs}
\label{sec:smooting}

We will now see that a general reduction operation for graphs, strongly related to the concept of ``homeomorphism'' of graphs, covers most cases of 1-singular DP-reduction for 2-CNFs.
Indeed it is essential to consider \emph{multi}graphs here, which allow loops and parallel edges.
Recall the definitions from Subsection \ref{sec:graph}: a multigraph is a pair $(V,E)$ with $V$ the vertex-set (as usual), and $E$ a map which assigns to every possible edge its multiplicity (that is, $E: \binom{V(G)}{1,2} \ra \NNZ$); the underlying graph is $\ug((V,E))$.
\begin{defi}\label{def:bouquetetc}
  We will use the notation $B_n$ for a \textbf{bouquet graph} with one vertex and $n \ge 0$ loops, and $D_n$ for a \textbf{dipole graph} with two vertices and $n$ parallel edges between them; in order to fully specify these multigraphs, we use $B_n(v)$ and $D_n(v,w)$, supplying the vertices.
  Additionally we use $D^2(a,b,c,d)$ for the \textbf{double dipole graph}: a cycle $a,b,c,d,a$, with additional edges between $a,b$ and $c,d$ (so that altogether there are $4$ vertices and $6$ edges).
\end{defi}
Following \cite[Section 7.2.4, D37]{GrossYellen2003Graphhandbuch}, ``smoothing'' removes linear vertices (vertices of degree $2$), which have no loops attached, by connecting the two incident edges:
\begin{defi}\label{def:loopless}
  A \textbf{loopless} vertex in a multigraph $G$ is an element $v \in V(G)$ with $v \notin \enachbarn_G(v)$ (equivalently, with $E(G)(\set{v}) = 0$).
  The set of loopless linear vertices of $G$ is denoted by $\llin(G) := \set{v \in V(G) : \vdeg_G(v) = 2 \land v \notin \enachbarn_G(v)}$.
\end{defi}
\begin{defi}\label{def:smoothing}
  Consider a multigraph $G$ and $v \in \llin(G)$.
  Now  a \textbf{smoothing step on $v$ in $G$} (also ``smoothing out/away $v$''), obtaining the multigraph $\smooth_v(G)$, considers the vertices $u, w$ with $\enachbarn_G(v) = \set{u,w}$ (note that possibly $u = w$, but $v \notin \set{u,w}$), and lets
  \begin{enumerate}
  \item $V(\smooth_v(G)) := V(G) \sm \set{v}$;
  \item for $e \in \binom{V(G)}{1,2}$ with $v \notin e$ (these are the possible edges for $\smooth_v(G)$):
    \begin{displaymath}
      E(\smooth_v(G))(e) :=
      \begin{cases}
        E(G)(e) + 1 & \text{if } e = \set{u,w}\\
        E(G)(e) & \text{otherwise}
      \end{cases}.
    \end{displaymath}
  \end{enumerate}
\end{defi}

\begin{examp}\label{exam:smooth-extream}
  Examples for multigraphs $G$ with $\llin(G) = \es$ (so no smoothing is applicable) are:
  \begin{itemize}
  \item the empty multigraph $G = (\es, \es)$;
  \item every $B_n$, $n \ge 0$ ($B_1$ does have a linear vertex, but this vertex has a loop);
  \item every $D_n$ for $n \ge 0$, $n \ne 2$;
  \item every $D^2$.
  \end{itemize}
  For $G = \xymatrix { v \aru@/^/[r] & u \aru@/^/[l]} = D_2(v,u)$, there are precisely two possible smoothing steps, yielding either $\smooth_u(G) = \xymatrix @C=1em { v \aru@(ur,dr)[]&} = B_1(v)$ or $\smooth_v(G) = \xymatrix @C=1em { u \aru@(ur,dr)[]&} = B_1(u)$ (and no further smoothing steps are applicable).

  For $G = \xymatrix { v \aru@/^/[r] & u \aru@/^/[l] \aru@/^/[r] & w \aru@/^/[l]}$ the two smoothing steps yield $\smooth_{v,w}(G) = \smooth_{w,v}(G) = B_2(u) = \xymatrix @C=1em { & u \aru@(ur,dr)[] \aru@(ul,dl)[] &}$.
\end{examp}

If (in Definition \ref{def:smoothing}) $u \ne w$, then the degree of the remaining vertices in $\smooth_v(G)$ is unchanged and no loop is created or destroyed, while in case of $u = w$ (that is, $v$ has two parallel edges to $u$ and no other incident edges), the degrees of vertices $\ne u$ are unchanged (and loops unchanged), while the degree of $u$ is decremented by one, and the two edges with $v$ are replaced by a single loop.
Thus smoothing never creates new loopless linear vertices (no new vertices to which smoothing can be applied), and it destroys an existing smoothing-possibility exactly in the special situation, where the smoothing vertex $v$ is part of an isolated 2-cycle (where its only neighbour $u$, which was loopless linear before, is no longer eligible after the reduction):
\begin{defi}\label{def:isolatedcycle}
  Consider a multigraph $G$.
  An \textbf{isolated $n$-cycle} for $n \ge 1$ is a submultigraph $S$ of $G$, which is a cycle multigraph of length $n$, where in case of $n=1$ the single vertex of $S$ is of degree $1$ (that is, has exactly one loop and no other incident edge), while for $n \ge 2$ each vertex of $S$ is linear in $G$ (that is, has no other edge incident with it in $G$ than the two edges already in $S$).
  We say that $G$ \textbf{has no isolated cycle} if $G$ does not have any isolated $n$-cycle for $n \ge 2$.
\end{defi}
So $G$ has no isolated cycle iff no connected component of $G$ is a cycle multigraph of length at least two (while any isolated vertices, with or without loops, are irrelevant here).
\begin{lem}\label{lem:looplessinv}
  Consider a multigraph $G$ and $v \in \llin(G)$, where $\enachbarn_G(v) = \set{u,w}$:
  \begin{enumerate}
  \item If $u = w$ and $\enachbarn_G(u) = \set{v}$, that is, $u, v$ form an isolated 2-cycle of $G$, then also $u \in \llin(G)$ and $\llin(\smooth_v(G)) = \llin(G) \sm \set{v,u}$.
  \item Otherwise we have $\llin(\smooth_v(G)) = \llin(G) \sm \set{v}$.
  \end{enumerate}
\end{lem}
\begin{defi}\label{def:irregularsmoothing}
  A \textbf{irregular vertex} of a multigraph $G$ is a vertex which is part of an isolated 2-cycle, while a \textbf{regular vertex} is a loopless linear vertex which is not part of an isolated 2-cycle.
  In the context of a smoothing reduction sequence, these notions refer to the current multigraph at this point in the reduction sequence, and we speak of an ``(ir)regular smoothing step''.
\end{defi}

Performing smoothing steps on a multigraph $G$ with no isolated cycles as long as possible results in a multigraph $G'$ (with $V(G') \sse V(G)$), where $G'$ is \emph{uniquely determined}:
\begin{lem}\label{lem:smoothingConfluent}
  Smoothing of a multigraph $G$ with no isolated cycles has only regular smoothing steps, and is confluent, that is, the resulting multigraph \bmm{\smooth(G)} does not depend on the choices of smoothing steps performed.
  And $V(\smooth(G)) = V(G) \sm \llin(G)$ (precisely the loopless linear vertices already in $G$ are removed).
\end{lem}
\begin{prf}
If the smoothing process would arrive at an irregular step (according to Lemma \ref{lem:looplessinv}), where some smoothing possibility is lost, namely at an isolated 2-cycle, then this 2-cycle must have been originally an isolated $n$-cycle for some $n \ge 2$ in $G$.
\Qed
\end{prf}

\begin{examp}\label{exam:smoothConfluent}
  Smoothing of the graph $G$ yields the multigraph $G'$:
  \begin{displaymath}
   G = \xymatrix @C=1.8em @R=2em {
    5 \aru[r] & 6 \aru[r] & 1 \aru[d] \aru[r] & 7 \aru[d] & 9 \aru[d] \aru[dl]\\
    4 \aru[u] & 3 \aru[l] & 2 \aru[l] & 8 \aru[l] & 10 \aru[l]
  } \qquad \qquad
  G' = \xymatrix @C=1.8em @R=2em {
    1 \aru[d] \aru[rd] &\\
    2 \aru@/^1.1pc/[u] & 8 \aru[l] \aru@(ur,dr)[]
  }
  \end{displaymath}
\end{examp}
To perform smoothing steps until completion for a cycle multigraph $C$ (all vertices of $C$ are linear) of length at least two, exactly one $v \in C$ is chosen, and the whole cycle $C$ is replaced by a loop at $v$.
\begin{examp}\label{exam:smoothisocycle}
  The smoothing of the cycle multigraphs $\cycleg_n$, $n \ge 1$, uses $n-1$ steps altogether, and all except for the final step are regular.
  The result is exactly one of $B_1(i)$, $1 \le i \le n$.
  For example smoothing of $\cycleg_3$ yields one of the following multigraphs:
  \begin{displaymath}
  \xymatrix @C=3em { 1 \aru@(ur,dr)[] & 2 \aru@(ur,dr)[] & 3 \aru@(ur,dr)[]}
  \end{displaymath}
\end{examp}
The general statement is as follows.
\begin{lem}\label{lem:smoothingCycle}
  Consider a cycle multigraph $G$ and any linear order on vertices, where the last element is $v \in V(G)$.
  Smoothing of $G$, following the linear order, ends up in $B_1(v)$.
  So smoothing of $G$ for any order is confluent up to isomorphism.
\end{lem}

To summarise, we come to the definition of the whole process:
\begin{defi}\label{def:smooth}
  For a (finite) multigraph $G$ by \bmm{\smooth(G)} we denote the unique multigraph obtained from $G$ by performing smoothing steps as long as possible, using any order, but for isolated cycles using some choice of the final vertex (the result of the smoothing of the isolated cycle, according to Lemma \ref{lem:smoothingCycle}), so that $\smooth(G)$ becomes a function.
\end{defi}
So by Lemmas \ref{lem:smoothingConfluent}, \ref{lem:smoothingCycle}, the result of $\smooth(G)$ does not depend an choices iff $G$ has no isolated cycles.
While for each isolated cycle a choice is needed to determine the resulting surviving vertex (all these choices result in isomorphic final results, for each isolated cycle we get one isolated $B_1$).

Obviously, smoothing of a multigraph with more than one connected components is equal to the disjoint union of smoothing for each component.
\begin{lem}\label{lem:smoothingComponent}
  For a multigraph $G$ with connected components $G_1,\dots, G_m$, $m \in \NN$ we have $\smooth(G) = \addbcup_{1 \le i \le m} \smooth(G_i)$.
\end{lem}
Smoothing is an isomorphism-invariant:
\begin{lem}\label{lem:sm(G)iso}
  For two multigraphs $G, G'$, if $G \cong G'$ then $\smooth(G) \cong \smooth(G')$.
\end{lem}
Following \cite[Section 7.2.4, D38]{GrossYellen2003Graphhandbuch}, two multigraphs $G, G'$ are called \textbf{homeomorphic}, if $\smooth(G) \cong \smooth(G')$.
So two isomorphic multigraphs are homeomorphic, but not vice versa.

\subsection{The implication multigraph}
\label{sec:impmulti}

\begin{examp}\label{exam:smoothing}
  Consider our running example $F$ and its non-1-singular normalform $\1dp(F)$ (Subsection \ref{sec:introruningexample}).
  The multigraph $\gtomg(\idg(F))$, shown below, has four linear vertices $5,-5,6,-6$, which are removed by smoothing as follows.
  \begin{displaymath}
    \xymatrix @C=1.4em @R=0.9em {
       &&& 4 \aru[d] &&& \\
       & 1 \aru@/^1.1pc/[rru] \aru@/_1.7pc/[dddd] && 6 \aru[d] && 3 \aru@/_1.1pc/[llu] \aru@/^1.7pc/[dddd] & \\
       &&& 2 \aru@/^1.1pc/[llu] \aru@/_1.1pc/[rru] &&& \\
       &5 \aru@/_/[uu] &&&& -5 \aru@/^/[uu]&\\
       &&& -2 \aru[d] &&& \\
       & -3 \aru@/^1.1pc/[rru] \aru@/_/[uu] && -6 \aru[d] && -1 \aru@/_1.1pc/[llu] \aru@/^/[uu] & \\
       &&& -4 \aru@/^1.1pc/[llu] \aru@/_1.1pc/[rru] &&&
    }
    \xymatrix @C=1.5em @R=0.6em {
       &&& 4 \aru[dd] &&& \\
       & 1 \aru@/^1.1pc/[rru] \aru@/^1.1pc/[dddd] &&&& 3 \aru@/_1.1pc/[llu] \aru@/_1.1pc/[dddd] & \\
       &&& 2 \aru@/^1.15pc/[llu] \aru@/_1.1pc/[rru] &&& \\
       &&&&&&\\
       &&& -2 \aru[dd] &&& \\
       & -3 \aru@/^1.1pc/[rru] \aru@/^1.1pc/[uuuu] &&&& -1 \aru@/_1.1pc/[llu] \aru@/_1.1pc/[uuuu] & \\
       &&& -4 \aru@/^1.1pc/[llu] \aru@/_1.1pc/[rru] &&&&
  }
  \end{displaymath}
  The result $\smooth(\gtomg(\idg(F)))$ is isomorphic to the multigraph $\gtomg(\idg(\1dp(F)))$ which has no linear vertices (see Section \ref{sec:introruningexample} for the implication digraph of $\1dp(F)$).
  That is, the results of smoothing for the implication digraph of $F$ and $\1dp(F)$ are isomorphic, and so their multigraphs are homeomorphic.
\end{examp}

In order to present the connection between 1-singular DP-reduction for 2-MUs $F$ and smoothing of the implication graphs, we need to handle antiparallel arcs in the implication digraphs, which become parallel edges in the implication multigraphs.
Recall their (general, for all of 2-CNF) characterisation in Lemma \ref{lem:antiparallel} (they come exactly from clauses $C \in F$, where also $\ol C \in F$ holds).
For deficiency $1$, antiparallel arcs in 2-MUs are very rare: by Lemma \ref{lem:antiparallelarcsd12MU} we have them exactly for the one-variable cases (they are broken by any move introducing any additional variable).
But for deficiency $k \ge 2$ we start with $k$ directed 2-cycles (i.e., $k$ pairs of antiparallel arcs) and their $k$ transpositions, and so they are an essential feature here; in Example \ref{exam:smoothing} we started originally with three directed 2-cycles and their contrapositions (described in Subsection \ref{sec:introruningexample} as $\idg(\Bpt 3)$), then broke all of them, arriving at the digraph of Example \ref{exam:smoothing}, but one 2-cycle re-emerged after smoothing.
\begin{defi}\label{def:img(F)}
  For $F \in \Bclss$ the \textbf{implication multigraph} \bmm{\img(F)} is
  \begin{displaymath}
    \img(F) := \dgtomg(\idg(F)).
  \end{displaymath}
\end{defi}

For a variable $v$ its variable-degree $\vdeg_F(v)$ equals the degree of vertex $v$ as well as the degree of vertex $\ol v$ in $\img(F)$:
\begin{displaymath}
  \vdeg_F(v) = \deg_{\img(F)}(v) = \deg_{\img(F)}(\ol v).
\end{displaymath}
Thus a vertex $x$ is linear in $\img(F)$ iff the vertex $\ol x$ is linear in $\img(F)$ (this comes from the skew-symmetry), while a variable $v$ is 1-singular in $F$ iff both vertices $v, \ol v$ are linear in $\img(F)$.
If $f: F \ra F'$ is an isomorphism between $F, F' \in \Pcls 2$, then $f: \img(F) \ra \img(F')$ is also an isomorphism.

\begin{defi}\label{def:smoothF}
  For $F \in \Bclss$ we define
  \begin{displaymath}
    \bmm{\smooth(F)} := \smooth(\img(F))
  \end{displaymath}
  (the \textbf{smoothed implication multigraph}; recall Definition \ref{def:smooth}), and similarly we define $\smooth_v(F) := \smooth_v(\img(F))$.
  We call $\smooth(F)$ the \textbf{homeomorphism type} of $F$ (up to isomorphism).
\end{defi}

\subsection{Smoothing as 1-singular DP-reduction}
\label{sec:smoothas1singDP}

\begin{defi}\label{def:varosing}
  For $F \in \Cls$ let $\bmm{\varosing(F)} := \set{v \in \var(F) : \ldeg_F(v) = \ldeg_{\ol v}(F) = 1}$ be the set of 1-singular variables.
\end{defi}
For general clause-sets $F \in \Pcls 2$, 1-singular DP-reduction for $v \in \varosing(F)$ (the reduction $F \leadsto \dpi{v}(F)$ for $v \in \varosing(F)$) allows the following degenerations:
\begin{description}
\item[Blocking] $\set{v,x}, \set{\ol v,\ol x} \in F$ for $\var(v) \ne \var(x)$: $\dpi{v}(F) = F \sm \set{\set{v,x}, \set{\ol v,\ol x}}$, not producing a resolvent (which would be tautological).
\item[Contraction] For $\set{v,x}, \set{\ol v, y} \in F$ with $\ol x \not= y$ the resolvent is already in $F$, that is, $\set{x, y} \in F$: also here $\dpi{v}(F) = F \sm \set{\set{v,x}, \set{\ol v, y}}$ holds.
\end{description}
For $F \in \Bmusat$ these situations can not occur (since otherwise the two occurrences of $v$ could be removed from $F$ without affecting unsatisfiability):
\begin{defi}\label{def:nongegenrate}
  For $F \in \Pcls 2$, we call $v \in \varosing(F)$ \textbf{nondegenerate}, if for the two occurrences $\set{v,x}, \set{\ol v,y} \in F$ (note that only $x \ne \ol v$ and $y \ne v$ is guaranteed here in general) holds $\ol x \ne y$ and $\set{x,y} \notin F$.
  So in this case always the new clause $\set{x,y}$ is produced by 1-singular DP-reduction on $v$.

  A nondegenerate $v$ is called \textbf{subsumptive} if $x = v$ or $y = \ol v$ or $x = y$ holds, otherwise \textbf{nonsubsumptive}.
\end{defi}
Every 1-singular variable in 2-MUs is nondegenerate.
1-Singular DP-reduction for general MUs does not remove 1-singular variables other than $v$, but in general it can create new 1-singular variables.
For 2-CNFs however this can only happen in one situation (with obvious proof):
\begin{lem}\label{lem:2cnfvarosing}
  Consider $F \in \Pcls 2$ and a nondegenerate $v \in \varosing(F)$, and let $F' := \dpi{v}(F)$.
  Call a literal $x$ ``full for $v$'', if $\set{v,x}, \set{\ol v, x} \in F$ (so we have the special case of a subsumptive $v$, where both parent clauses are non-unit).
  \begin{enumerate}
  \item\label{lem:2cnfvarosing1} The new clause produced by $\dpi{v}(F)$ is a unit-clause or the empty clause iff $v$ is subsumptive (so it is a binary clause iff $v$ is nonsubsumptive).
  \item\label{lem:2cnfvarosing2} If $x$ is full for $v$, then $\ldeg_{F'}(x) = \ldeg_F(x) - 1$, while for all other literals $y \notin \set{v,\ol v, x}$ holds $\ldeg_{F'}(y) = \ldeg_F(y)$.
  \item\label{lem:2cnfvarosing3} If no literal is full for $v$, then for all literals $x$ with $x \notin \set{v, \ol v}$ holds $\ldeg_{F'}(x) = \ldeg_F(x)$.
  \end{enumerate}
\end{lem}

\begin{corol}\label{cor:2cnfvarosing}
  In the situation of Lemma \ref{lem:2cnfvarosing}:
  \begin{enumerate}
  \item \label{cor:2cnfvarosing1}If $x$ is full for $v$, and $\ldeg_F(x) = 2$ and $\ldeg_F(\ol x) = 1$ holds, then $\varosing(F') = (\varosing(F) \sm \set{v}) \cup \set{\var(x)}$ (and thus $\abs{\varosing(F')} = \abs{\varosing(F)}$).
  \item\label{cor:2cnfvarosing2} If no literal is full for $v$, then $\varosing(F') = \varosing(F) \sm \set{v}$ (and thus $\abs{\varosing(F')} = \abs{\varosing(F)}-1$).
  \end{enumerate}
\end{corol}

One step of 1-singular DP-reduction on $v$ can be simulated on the implication multigraph by two smoothing steps on $v, \ol v$ (in any order), except for two special cases, as explained now (recall the overloading of $\smooth(G)$ by $\smooth(F)$ according to Definition \ref{def:smoothF}):
\begin{lem}\label{lem:corresp1sdpsmooth}
  Consider $F \in \Bclss$ and a nondegenerate $v \in \varosing(F)$.
  \begin{enumerate}
  \item\label{lem:corresp1sdpsmooth1} Assume the two occurrences of $v$ are unit-clauses.
    Then we have an isolated 2-cycle in $\img(F)$ formed by vertices $v$ and $\ol v$, and thus exactly one of $v$ or $\ol v$ can be smoothed out (recall Lemma \ref{lem:smoothingCycle}), and this by an irregular step.
    So in $\dpi{v}(F)$ the two clauses $\set{v}, \set{\ol v}$ are replaced by the empty clause, while in $\smooth_v(F)$ the 2-cycle $v, \ol v$ is replaced by vertex $\ol v$ with loop (and no other incident edges; accordingly the 2-cycle is replaced in $\smooth_{\ol v}(F)$ by vertex $v$ with loop).
  \item\label{lem:corresp1sdpsmooth2} Assume $x$ is full for $v$.
    So $\idg(F)$ has at least two paths from $\ol x$ to $x$, namely $\ol x \ra v \ra x$ and $\ol x \ra \ol v \ra x$.
    Then the only difference between $\img(\dpi{v}(F))$ and $\smooth_{v,\ol v}(F) = \smooth_{\ol v,v}(F)$, where only regular smoothing steps are used, is that the former has exactly one edge between $\ol x$ and $x$, and the latter has exactly two edges.
  \item\label{lem:corresp1sdpsmoot3} Otherwise we have (using only regular smoothing steps)
    \begin{displaymath}
      \img(\dpi{v}(F)) = \smooth_{v,\ol v}(F) = \smooth_{\ol v,v}(F).
    \end{displaymath}
  \end{enumerate}
\end{lem}
\begin{prf}
Part \ref{lem:corresp1sdpsmooth1} is fully explicit.
For Part \ref{lem:corresp1sdpsmooth2} we need to only note that $\img(F)$ has no edge from $\ol x$ to $x$, since that would mean $\set{x} \in F$, which is excluded by $v$ being nondegenerate.
Part \ref{lem:corresp1sdpsmoot3} follows directly from the definitions.
\Qed
\end{prf}

For 2-MUs with deficiency at least two, smoothing exactly corresponds to (complete) 1-singular DP-reduction:
\begin{thm}\label{thm:smooting2MUsk2}
  Consider $F \in \Bmusat$ with $\delta(F) \ge 2$.
  Then:
  \begin{enumerate}
  \item All iterated 1-singular DP-reductions for $F$ are nonsubsumptive.
  \item $\var(\1dp(F)) = \var(F) \sm \varosing(F)$.
  \item $\smooth(F) = \img(\1dp(F))$; here $\smooth(F)$ is independent of any choices, due to $\img(F)$ not having isolated cycles.
  \end{enumerate}
\end{thm}
\begin{prf}
Since singular DP-reduction for MUs maintains the deficiency, by Lemma \ref{lem:nounit2MU} never a unit-clause can be produced by iterated 1-singular DP-reductions, and thus by Lemma \ref{lem:2cnfvarosing}, Part \ref{lem:2cnfvarosing1}, all these reductions are nonsubsumptive.
Thus by Corollary \ref{cor:2cnfvarosing}, Part \ref{cor:2cnfvarosing2}, no new 1-singular variables are created in the reduction-process, and the reduced variables are exactly those in $\varosing(F)$.
Turning to the smoothing process, transforming the 1-singular DP-reduction via Lemma \ref{lem:corresp1sdpsmooth}, Part \ref{lem:corresp1sdpsmoot3}, into a smoothing sequence, this process uses only regular smoothing steps, and arrives at a result without linear vertices.
Thus the original $\img(F)$ had no isolated cycles.
\Qed
\end{prf}

The classification of homeomorphism types for 2-MUs of deficiency at least $2$ will be investigated, at the level of their implications graphs (in a generalised setting), in Subsection \ref{sec:homeotypes}, with the conclusion in Corollary \ref{cor:binbracelet}.

We now turn to consider $\smooth(F)$ for $F \in \Bmusats_{\delta=1}$.
We know $\1dp(F) = \set{\bot}$, but smoothing shows a more differentiated picture, namely we obtain exactly four homeomorphism types, corresponding to the four Families I - IV from Section \ref{sec:2-MU(1)}.
\begin{thm}\label{thm:homeodef1}
  $\smooth(F)$ for $F \in \Bmusatds$ is equal to exactly one of the following homeomorphism types, which are the homeomorphism types of the four Families $\Hp_n, \Hl_{n,i}, \Shl_{n,i}, \Ub_{n,x,y}$:\vspace{1ex}
  \begin{displaymath}
    \xymatrix { \bullet \aru@(ur,dr)[] } \quad\quad\quad\;\;
    \xymatrix { \bullet \aru@/^/[r] \aru@/_/[r] \aru[r] & \bullet } \quad\quad
    \xymatrix { \bullet \aru@/^4pt/[r] \aru@/^8pt/[r] \aru@/_4pt/[r] \aru@/_8pt/[r] & \bullet } \quad\quad
    \xymatrix @R=0.5em { \bullet \aru[d] \aru@/_4pt/[r] \aru@/^4pt/[r] & \bullet \aru[d]\\
      \bullet \aru@/_4pt/[r] \aru@/^4pt/[r] & \bullet }
  \end{displaymath}
  (that is, $B_1$, $D_3$, $D_4$, $D^2$; recall Definition \ref{def:bouquetetc}).
  More precisely, $\smooth(F)$ is determined as follows for the four Families (recall their degree-characterisations in Corollary \ref{cor:def1degreespec}):
  \begin{enumerate}
  \item[I] $F$ has two unit-clauses:
    This is the only case where $\img(F)$ has an isolated cycle (and indeed $\img(F)$ is a cycle).
    Exactly for all the literals $x \in \lit(F)$ we can obtain, by appropriate choices, $\smooth(F) = B_1(x)$.

    For all other Families, $\smooth(F)$ is independent of the choices made for smoothing.
  \item[II] $F$ has exactly one variable $v$ of degree $3$:
    $\smooth(F) = D_3(v, \ol v)$.
  \item[III] $F$ has exactly one variable $v$ of degree $4$:
    $\smooth(F) = D_4(v, \ol v)$.
  \item[IV] $F$ has exactly two variables $v, w$ of degree $3$:
    In $\idg(F)$, from vertex $v$ there is exactly one path to $w$ (of length $L$), and one path to $\ol w$ (of length $L'$).
    We have that one of $L, L'$ is $y-x$, while the other is at least $y - x + 2$, using the characterisation of $\idg(F)$ in Corollary \ref{cor:fourthfamily}: let literal $u \in \set{w, \ol w}$ be the literal with the shorter path.
    Now we have $\smooth(F) = D^2(\ol v, v, u, \ol u)$.
  \end{enumerate}
\end{thm}
\begin{prf}
By Lemma \ref{lem:corresp1sdpsmooth} on $F$ we can safely perform two types of 1-singular DP-reductions:
\begin{enumerate}
\item[(i)] nonsubsumptive reductions;
\item[(ii)] subsumptive reductions in case exactly one unit-clause is involved;
\end{enumerate}
while this being simulated by regular smoothing steps.

Considering Corollary \ref{cor:fiveclss}, by Reduction (i), reversing Rule S, we arrive at the five Cases (T), (E), (TE), (EE), (TEE).
Reduction (ii) reverses Rule T, and thus (T) is transformed to () (the starting point), while (TE), which is equivalent to (ET), gets reduced to (E), which we already have.
So we have the following four cases for the reduction results (applying (i), (ii) in any order as long as possible) of $F$:
\begin{enumerate}
\item[I] $() = \set{\set{v},\set{\ol v}}$:
  $F \cong \Hp_{n(F)}$ (equivalently, $F$ has two unit-clauses), exactly the case where $F$ has an isolated cycle (and indeed $\img(F)$ is the isolated cycle).
  This case is also characterised by $\varosing(F) = \var(F)$; $v \in \var(F)$ is any of the variables.
  Irregular smoothing yields one of $B_1(v), B_1(\ol v)$.
\item[II] $(E) = \set{\set{x}, \set{\ol x,y}, \set{\ol x, \ol y}}$:
  $F \cong \Hl_{n(F),i}$ (equivalently, $F$ has exactly one unit-clause; equivalently, $F$ has exactly one variable of degree $3$); here $\set{x} \in F$.
  Regular smoothing (of $y, \ol y$) yields $D_3(x, \ol x)$.
\item[III] $(EE) = \set{\set{v,x},\set{v,\ol x},\set{\ol v,y},\set{\ol v,\ol y}}$:
  $F \cong \Shl_{n(F),i}$ (equivalently, $F$ has exactly one variable of degree $4$, which is $v$).
  Regular smoothing (of $x, \ol x, y, \ol y$) yields $D_4(v, \ol v)$.
\item [IV]$(TEE) = \set{\set{x,y},\set{\ol x,a},\set{\ol x,\ol a},\set{\ol y,b},\set{\ol y,\ol b}}$:
  $F \cong \Ub_{n(F),i,j}$ (equivalently, $F$ has exactly two variables of degree $3$, which are $\var(x), \var(y)$).
  Regular smoothing (of $a, \ol a, b, \ol b$) yields $D_2(x, \ol x, y, \ol y)$.
  \Qed
\end{enumerate}
\end{prf}

\section{Weak double cycles}
\label{sec:2MU-WDC}

In this section we present elementary properties of ``weak double cycles'' (WDCs), a class of digraphs which captures naturally the implication digraphs of 2-MUs of deficiency at least two.
Starting from an undirected cycle of length $m$, we first make it a digraph (each edge becomes a pair of antiparallel arcs), and then split vertices and arcs, and arrive at $m$-WDCs.
This is explained in Subsection \ref{sec:wdcbasicdefs}.
So WDCs are made of ``small cycles'' within ``big cycles'', which is investigated in Subsection \ref{sec:cyclestrucWDC}.
This cycle-structure enables us to get a handle on the isomorphisms between WDCs in Subsection \ref{sec:isoswdcs}.
For an interesting class of WDCs, the direction of the arcs doesn't matter, as shown in Subsection \ref{sec:forgetinfo}.
This leads to the classification of this type of WDCs via ``bracelets'' in Subsection \ref{sec:homeotypes}.

\subsection{The basic definition}
\label{sec:wdcbasicdefs}

The operation of \textbf{splitting a vertex} $x$ in a digraph $G$ consists of replacing $x$ by two new vertices $u,v$ and an arc $(u,v)$, such that all arcs coming into $x$ come into $u$, and all arcs going out of $x$ go out of $v$.
\textbf{Splitting an arc} in $G$ replaces an arc $(x,y) \in E(G)$ by $(x,v),(v,y)$ for a new (linear) vertex $v$.
These two operations look locally as follows:
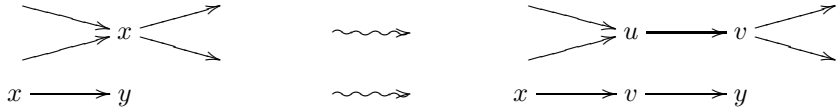
\begin{displaymath}
  \xymatrix @C=3em @R=0.3em {
  \ar[rd] &&&&& \ar[rd] &&&& \\
  & x \ar[ru] \ar[rd] && \ar@{~>}[r] &&& u \ar[r] & v \ar[ru] \ar[rd] & \\
  \ar[ru] &&&&& \ar[ru] &&&& \\
  x \ar[r] & y && \ar@{~>}[r] && x \ar[r] & v \ar[r] & y
  }
\end{displaymath}
``New vertices'' always means ``completely new'', for the whole process.

\begin{lem}\label{lem:splittingmaintain}
  The operations of splitting of vertices or arcs, applied to arbitrary digraphs, maintain the following properties:
  \begin{enumerate}
  \item whether the digraph is strongly connected or not;
  \item the number of cycles;
  \item whether the digraph has sinks resp.\ sources or not.
  \end{enumerate}
\end{lem}

The main class of digraphs studied here is obtained from ``double cycles'' by the above two operations, where a \textbf{double $m$-cycle} for $m \ge 3$ is a digraph isomorphic to $\gtodg(\cycleg_m)$; so we have $m$ vertices and $2 m $ arcs.
Double cycles are strongly connected, every vertex has degree $4$, and for every arc also the reverse arc exists (this characterises the class of double cycles).
\begin{defi}[\cite{SeymourThomassen1987GeradeGerichteteGraphen}]\label{def:wdc}
  An \textbf{$m$-weak-double-cycle} (WDC) is a digraph obtained from some double $m$-cycle ($m \ge 3$) by splitting vertices or arcs (possibly none).
\end{defi}
WDCs are strongly connected, and if $G$ is a WDC so is $\trans G$.
For every digraph, splitting of vertices of degree at most $3$, which are neither sources nor sinks (i.e., have ingoing and outgoing arcs), can be simulated by splitting of arcs.
Thus we only apply splitting of vertices to vertices of degree at least $4$ (our digraphs do not contain sources or sinks):
\begin{lem}\label{lem:standWDC}
  The class of $m$-WDCs ($m \ge 3$) is exactly created by starting with some double $m$-cycle $G_0$, and applying two rounds of operations, creating digraphs $G$, which are exactly all $m$-WDCs:
  \begin{enumerate}
  \item The first round consists of splitting vertices of degree $4$.
    (Note that the vertex degrees in $G$ are at most $4$, and the vertices of degree $4$ are the elements of $V(G) \cap V(G_0)$.)
    The obtained digraphs are nonlinear.
  \item The second round consists of splitting arcs (an arbitrary number of times; introducing linear vertices).
  \end{enumerate}
\end{lem}

\begin{lem}\label{lem:degreesWDC}
  Consider $m \ge 3$ and a WDC $G$, created according to Lemma \ref{lem:standWDC} by $0 \le \alpha \le m$ many splitting of vertices of degree $4$, followed by $\beta \ge 0$ many splitting of arcs.
  \begin{enumerate}
  \item $\abs{V(G)} = m + \alpha + \beta$.
  \item $\abs{E(G)} = 2 m + \alpha + \beta$.
  \item All vertices have degrees $2,3,4$ (i.e., $n_2(G) + n_3(G) + n_4(G) = \abs{V(G)}$), where
    \begin{enumerate}
    \item $n_4(G) = m - \alpha$;
    \item $n_3(G) = 2 \alpha$;
    \item $n_2(G) = \beta$ (the number of linear vertices).
    \end{enumerate}
  \item So $m = m(G)$, $\alpha = \alpha(G)$, $\beta = \beta(G)$ are isomorphism-invariants of $G$.
  \item $\smooth(G)$ is the nonlinear WDC obtained by only applying the same $\alpha$-many vertex-splittings (smoothing exactly reverses the $\beta$-many arc-splittings).
  \end{enumerate}
\end{lem}

\begin{corol}\label{cor:nonlinearWDCs}
  A digraph $G$ is a WDC iff $\smooth(G)$ is a nonlinear WDC.
\end{corol}

\subsection{The cycle structure of WDCs}
\label{sec:cyclestrucWDC}

We now explain the basic cycle structure of $m$-WDCs, namely the ``small'' and the ``big'' cycles.
Recalls that WDCs are digraphs, and thus the cycles in them are also directed.
Initially, a double $m$-cycle has
\begin{itemize}
\item $m$ \emph{small cycles} of length $2$;
\item two \emph{big cycles} of length $m$ (each covering all vertices and half of the arcs).
\end{itemize}
These $m + 2$ cycles are precisely all the cycles in the double $m$-cycle.
By Lemma \ref{lem:splittingmaintain}, thus any $m$-WDC has precisely $m + 2$ cycles (and thus $m$ is indeed an isomorphism-invariant).
The degree-$4$-vertices are possibly replaced by two degree-$3$-vertices, via the splitting of vertices (note that if not said otherwise, degrees are always measured in the whole graph).
\begin{itemize}
\item So the small cycles are characterised by having exactly $2+k$ nonlinear vertices for some $0 \le k \le 2$ (how many of its vertices have been split), with $2-k$ vertices of degree $4$ and $2k$ vertices of degree $3$.
\item And the big cycles are characterised by having exactly $m+A$ nonlinear vertices , with $m-A$ vertices of degree $4$ and $2 A$ vertices of degree $3$.
  Note that both big cycles use all nonlinear vertices
\end{itemize}

\begin{lem}\label{lem:smallbigspectra}
  Let $d_G(C) \in \NNZ^2$ for a cycle $C$ of a digraph $G$ be the number of degree-3 and degree-4 vertices of $C$ (with degrees taken in $G$).
  Consider $m \ge 3$.
  \begin{enumerate}
  \item The set of $d_G(C)$ for $G$ a $m$-WDC and $C$ a small cycle of $G$ is $\set{(2k, 2-k) : k \in \set{0,1,2}}$.
  \item The set of $d_G(C)$ for $G$ a $m$-WDC and $C$ a big cycle of $G$ is $\set{(2k', m-k') : k' \in \set{0,\dots,m}}$.
  \item These sets of degree-pairs are disjoint (and thus we can distinguish between small and big cycles in an $m$-WDC by just looking at their vertex-degrees).
  \end{enumerate}
\end{lem}
\begin{prf}
Assume $(2k, 2-k) = (2k', m-k')$. Thus $k = k'$, and then $2-k = 2-k' < m-k'$.
\Qed
\end{prf}

For a naming convention, we choose the number of degree-$4$-vertices:
\begin{defi}\label{def:smallcycle}
  For $k \in \set{0,1,2}$ we call a cycle $C$ in a WDC $G$ a \textbf{$k$-small cycle} if $C$ has exactly $k$ degree-$4$-vertices and $4-2k$ degree-$3$-vertices (note that $k$ is uniquely determined here).
  A \textbf{small cycle} in $G$ is a $k$-small cycle for some $0 \le k \le 2$.
  The set of all small cycles is denoted by $\scs(G)$, and the subset of $k$-small-cycles by $\scs_k(G)$.

  A \textbf{big cycle} in $G$ is a cycle which is not a small cycle.
  The set of all big cycles is denoted by $\bcs(G)$.
\end{defi}
So the $2$-small cycles are those where no vertex-splitting occurred, the $1$-small cycles had exactly one vertex-splitting, and the $0$-small cycles had two vertex-splittings.
Note that arc-splittings can have occurred with all three types of small cycles (and recall that vertex splittings are done only to degree-$4$-vertices).

By induction over the construction in Lemma \ref{lem:standWDC} one can easily show that every small cycle is naturally partitioned into four path-subgraphs as follows, based on the observation that 
\begin{defi}\label{def:partsmallcycles}
  Consider a small cycle $C$ in an $m$-WDC $G$ ($m \ge 3$; recall that $C$ is a subdigraph of $G$):
  \begin{enumerate}
  \item There are exactly two other cycles $C', C''$ with $V(G) \cap V(C') \ne \es$ and $V(G) \cap V(C'') \ne \es$ (we note $V(C') \cap V(C'') = \es$ iff $m \ge 4$); here the intersections are pathdigraphs $P', P''$.
    The set of \textbf{overlaps} of $C$ is $\set{P', P''}$, denoted by $\overlaps_G(C) := \set{P', P''}$.
  \item An overlap is \textbf{trivial} if it contains exactly one vertex.
    The length $\length(P) \in \NNZ$ is the number of arcs, so $P$ is trivial iff $\length(P) = 0$.
    If $C$ is $k$-small, then exactly $k$ of the overlaps of $C$ are trivial.
    \begin{displaymath}
      \xymatrix @C=3em @R=0.9em {
        \bullet \ar@/^2pc/[rr] && \bullet \ar@{-->}_{P'}[dd] && \bullet \ar@/_2pc/^{A_2}[ll] \ar@/^2pc/[rr] && \bullet \ar@{-->}[dd]\\
        & C' && C && C'' &\\
        \bullet \ar@{-->}[uu] && \bullet \ar@/^2pc/[ll] \ar@/_2pc/^{A_1}[rr] && \bullet \ar@{-->}_{P''}[uu] && \bullet \ar@/^2pc/[ll]
      }
    \end{displaymath}
  \item There is a (unique) path $A_1$ in $C$ from $\plast(P')$ to $\pfirst(P'')$, and a (unique) path $A_2$ from $\plast(P'')$ to $\pfirst(P')$; both paths have length at least one (contain at least one arc).
    The set of \textbf{extended interiors} is $\set{A_1, A_2}$.
    (We note $A_1 \ne A_2$, $E(A_1) \cap E(A_2) = \es$, and $V(A_1) \cap V(A_2) = \es$ iff both overlaps are nontrivial.)
  \item Obtain the subdigraphs $A', A''$ from $A_1, A_2$ by removal of first and last vertices; thus both these subdigraphs of $C$ might have the empty vertex-set (then they are called \textbf{trivial}), or otherwise they are paths.
    The set of \textbf{interiors} of $C$ is $\set{A', A''}$, denoted by $\interiors_G(C) := \set{A', A''}$.

    Note that $A' = A''$ iff $V(A') = V(A'') = \es$ (i.e., both $A', A''$ are trivial) iff the paths $A_1, A_2$ both have length one.
  \end{enumerate}
  So $V(P'), V(P''), V(A'), V(A'')$ are pairwise disjoint, and their union is $V(C)$ (while $\abs{E(P')} + \abs{E(P'')} + \abs{E(A')} + \abs{E(A'')} = \abs{E(C)} - 2 - \ve' - \ve''$, where $\ve', \ve'' \in \set{0,1}$ indicate whether $A'$ resp.\ $A''$ are nontrivial).
\end{defi}

Via overlaps and interiors we obtain a refinement of the parameters $m(G)$, $\alpha(G)$, $\beta(G)$ from Lemma \ref{lem:degreesWDC}:
\begin{lem}\label{lem:degreesWDCrefine}
  Consider a WDC $G$.
  \begin{enumerate}
  \item $m(G) = \abs{\scs(G)}$.
  \item $\alpha(G) = \abs{\set{P \in \bc_{C \in \scs(G)} \overlaps(C) : \pfirst(P) \ne \plast(P)}}$ (the number of nontrivial overlaps).
  \item Let $\beta'(G) := \abs{\bc_{C \in \scs(G)} \bc_{P \in \overlaps(C)} V(P) \sm \set{\pfirst(P),\plast(P)}} \in \NNZ$ (number of vertices in overlaps not being endpoints).
  \item Let $\beta''(G) := \abs{\bc_{C \in \scs(G)} \bc_{A \in \interiors(C)} V(A)} \in \NNZ$ (number of vertices in interiors).
  \item Now we have $\beta'(G) + \beta''(G) = \beta(G)$.
  \item Also $\beta'(G), \beta''(G)$ are isomorphism invariants of $G$.
  \end{enumerate}
\end{lem}

From the decomposition of small cycles into overlaps and interiors we obtain:
\begin{lem}\label{lem:classvertswdc}
  Consider a WDC $G$ and a vertex $v \in V(G)$.
  \begin{enumerate}
  \item Either $v \in V(P)$ for some overlap $P$ of $G$.
    Then $v$ is contained in exactly four cycles, the two small cycles sharing the overlap $P$, and both big cycles.
  \item Otherwise $v$ is element of some interior, and is contained in exactly one small and one big cycle.
  \end{enumerate}
\end{lem}

\begin{corol}\label{cor:classarcswdc}
  Consider a WDC $G$ and an arc $e \in E(G)$.
  \begin{enumerate}
  \item\label{cor:classarcswdc1} Either $e \in E(P)$ for some (nontrivial) overlap $P$ of $G$.
    Then $e$ is contained in exactly four cycles, namely the two small cycles sharing the overlap $P$, and both big cycles.
  \item\label{cor:classarcswdc2} Otherwise $e$ is contained in exactly one small and one big cycle.
  \end{enumerate}
\end{corol}

The interaction between small and big cycles can be described as follows:
\begin{lem}\label{lem:interplaybigsmall}
  Consider a WDC $G$.
  \begin{enumerate}
  \item\label{lem:interplaybigsmall1} The intersection of the two big cycles (as a subdigraph) is exactly the (vertex-disjoint) union of the overlaps.
  \item\label{lem:interplaybigsmall2} The intersection of a big cycle and a small cycle is the union the two overlaps of the small cycle and one of two extended interiors.
  \end{enumerate}
\end{lem}

The big cycles can be visualised as having ``clockwise'' and ``anticlockwise'' direction, using the natural planar drawing of WDCs.
Both contain all the overlapping vertices between small cycles, and alternately choose the ``outer section'' and the ``inner section'' of a small cycle.
The two big cycles intersect exactly in the overlaps.

\subsection{Isomorphisms between WDCs}
\label{sec:isoswdcs}

Isomorphisms map cycles to cycles, with the details as follows:
\begin{lem}\label{lem:mapsmallcycle}
  Consider an isomorphism $f: G \ra G'$ between WDCs.
  \begin{enumerate}
  \item\label{lem:mapsmallcycle1} The image under $f$ of a big cycle of $G$ is a big cycle of $G'$.
  \item Fixing the big cycles of $G$ as $B_0, B_1$ and of $G'$ as $B_0', B_1'$, we call an arbitrary isomorphism $f: G \ra G'$ \emph{positive} if $f(B_\ve) = B_\ve'$ for both $\ve$, and \emph{negative} otherwise (then $f(B_\ve) = B_{\ol \ve}'$).
  \item\label{lem:mapsmallcycle2} Consider a $k$-small cycle $C$ of $G$, together with the restriction $f_C$ of $f$ to $C$.
    Then $f_C$ is an isomorphism from $C$ to a $k$-small cycle $C'$ of $G'$.
  \item\label{lem:mapsmallcycle3} $f_C$ maps the overlaps of $C$ to the overlaps of $C'$, and also the interiors of $C$ to the interiors of $C'$.
    The restrictions of $f_C$ to these four subdigraphs are isomorphisms.
  \item\label{lem:mapsmallcycle4} Let $\set{P_1',P_1''}$ be the overlaps of $C$, and let $\set{P_2',P_2''}$ be the overlaps of $C'$.
    Then $f(\pfirst(P_1')) \in \set{f(\pfirst(P_2')),f(\pfirst(P_2''))}$.
  \item\label{lem:mapsmallcycle5} If for an isomorphism $f': G \ra G'$ mapping $C$ to $C'$ holds $f'(\pfirst(P_1')) \neq f(\pfirst(P_1'))$, then $f, f'$ have different (opposite) orientations.
  \end{enumerate}
\end{lem}
\begin{prf}
Given the above invariants of small and big cycles, and the invariants of overlaps and interiors, only Part \ref{lem:mapsmallcycle5} remains.
Consider the extended interior following directly the overlap containing $f(\pfirst(P_1'))$ resp.\ $f'(\pfirst(P_1'))$.
By assumption these extended interiors are different, thus arc-disjoint.
By Lemma \ref{lem:interplaybigsmall}, Part \ref{lem:interplaybigsmall2}, they are thus subdigraphs of different big cycles.
\Qed
\end{prf}

We recall that in isomorphism between cycle digraphs is uniquely determined by knowing for any single vertex in the domain its image.
\begin{lem}\label{lem:detisoonecycle}
  Consider WDCs $G, G'$, together with a small cycle $C$ in $G$ and a small cycle $C'$ in $G'$, and an isomorphism $f_0: C \ra C'$.
  \begin{enumerate}
  \item\label{lem:detisoonecycle1} There is at most one continuation of $f_0$ to an isomorphism $f: G \ra G'$.
  \item\label{lem:detisoonecycle2} The existence of $f$ can be decided in linear time, and in the positive case $f$ can also be computed in linear time.
  \item\label{lem:detisoonecycle3} An isomorphism $f: G \ra G'$ is uniquely represented by any element $(v,w) \in V(C) \times V(C') \sse V(G) \times V(G')$, where $f(v) = w$.
    If $v$ (and thus also $w$) is not an element of the interior of $C$ (resp.\ $C'$), then the small cycles $C, C'$ must also be indicated.
  \end{enumerate}
\end{lem}
\begin{prf}
Every small cycle in a WDC intersects both big cycles, and the big cycles together cover all vertices.
This shows Part \ref{lem:detisoonecycle1}, while Part \ref{lem:detisoonecycle2} is obvious (``just follow the arcs''), and Part \ref{lem:detisoonecycle3} follows from Part \ref{lem:detisoonecycle1}.
\Qed
\end{prf}

We obtain a simple procedure for deciding isomorphism of WDCs:
\begin{corol}\label{cor:WDCpolyIso}
  The class of WDCs has isomorphism decision in quadratic time.
  If $G, G'$ are $m$-WDCs ($m \ge 3$), then $\abs{\isos(G,G')} \le 2 m$, and also the elements of $\isos(G,G')$ can be enumerated in quadratic time.
\end{corol}
\begin{prf}
If $G, G'$ don't have the same $m$, then $\isos(G,G') = \es$.
Otherwise choose any small cycle $C_0$ in $G$, and run through the small cycles $C_1, \dots, C_m$ in $G'$, checking for the two possible isomorphisms mapping $C_0$ to $C_1$ according to Lemma \ref{lem:mapsmallcycle}, Part \ref{lem:mapsmallcycle4}, which can be done in linear time according to Lemma \ref{lem:detisoonecycle}.
The isomorphisms could be represented here very simply by their full table (the quadratic time allows for that).
\Qed
\end{prf}

From the fact that the maximum degree $d$ of WDCs is $d=4$, we obtain by \cite{Luks1982IsoBounded} polytime decision of the isomorphism problem, but no concrete bound for the given $d$, since the Big-Ohs in \cite{Luks1982IsoBounded} are not specified; the same holds for the recent improvement \cite{GroheNeuenSchweitzer2018GraphIso}.
The only concrete bound is given for $d=3$ in \cite{GalilHoffmannLuksSchnorrWeber1987GraphIso}, namely a runtime of $O(n^3 \log n)$.

Via Corollary \ref{cor:WDCpolyIso} we calculate in fact also the composition table of the automorphism group $\auto(G)$.
We now show how $\auto(G)$ for an $m$-WDC has a natural embedding into $\auto(\scg(G))$, where the ``small cycle graph'' $\scg(G)$ is a cycle graph of length $m$:
\begin{defi}\label{def:scg}
  For a WDC $G$ let $\scg(G)$ be the graph with vertex-set the small cycles of $G$ (as subdigraphs of $G$), that is, $V(\scg(G)) := \scs(G)$, and with an edge between vertices $C, C'$ if $V(C) \cap V(C') \ne \es$.
  And for an isomorphism $f: G \ra G'$ between WDCs let $\scg(f): \scg(G) \ra \scg(G')$ map a small cycle $C \in V(\scg(G))$ to the small cycle in $V(\scg(G'))$ with vertex-set $f(V(C))$ (recall Lemma \ref{lem:mapsmallcycle}, Part \ref{lem:mapsmallcycle2}).
\end{defi}

The mapping $G \mapsto \scg(G)$ has natural functorial properties:
\begin{lem}\label{lem:scgfunctor}
  Consider WDCs $G, G', G''$.
  \begin{enumerate}
  \item Identities are mapped to identities: $\scg(\id_{V(G)}) = \id_{V(\scg(G))}$.
  \item For $f \in \isos(G,G')$ and $g \in \isos(G',G'')$ holds $\scg(g \circ f) = \scg(g) \circ \scg(f)$.
  \end{enumerate}
\end{lem}

\begin{lem}\label{lem:embedauto}
  For a WDC $G$, the mapping $\Phi: f \in \auto(G) \mapsto \scg(f) \in \auto(\scg(G))$ is a group embedding (an injective group homomorphism).
\end{lem}
\begin{prf}
By Lemma \ref{lem:scgfunctor} we obtain that $\Phi$ is a group homomorphism; it remains to show that the kernel of $\Phi$ is trivial.
So consider an isomorphism $f: G \ra G$ with $\scg(f) = \id_{V(\scg(G))}$; we have to show $f = \id_{V(G)}$.
Consider any small cycle $C$ of $G$.
Let $\set{P_1,P_2}$ be the overlaps of $C$.
If $f(\pfirst(P_1)) = \pfirst(P_1)$, then $f = \id_{V(G)}$ by Lemma \ref{lem:detisoonecycle}, while otherwise by Lemma \ref{lem:mapsmallcycle}, Part \ref{lem:mapsmallcycle5}, $f$ would have the opposite orientation of the identity, and thus would swap the two neighbouring small cycles of $C$, which contradicts that $f$ acts identical on the small cycles.
\Qed
\end{prf}

Thus knowing what an isomorphism is doing on the small cycles is enough to identify the isomorphism:
\begin{corol}\label{cor:injisos}
  For WDCs $G, G'$, the map $\Phi: f \in \isos(G, G') \mapsto \scg(f) \in \isos(\scg(G), \scg(G'))$ is injective.
\end{corol}
\begin{prf}
Consider $f, g \in \isos(G, G')$ with $\Phi(f) = \Phi(g)$.
Thus $g^{-1} \circ f \in \auto(G)$ with $\scg(g^{-1} \circ f) = \Phi(g)^{-1} \circ \Phi(f) = \id$, whence $g^{-1} \circ f = \id$, that is $f = g$.
\Qed
\end{prf}

Since $\auto(\scg(G)) \cong \auto(\cycleg_m)$ is the Dihedral group with $2 m$ elements, we obtain:
\begin{corol}\label{cor:autowdcDih}
  Consider an $m$-WDC $G$ ($m \ge 3$).
  $\auto(G)$ is a subgroup of the Dihedral group $D_m$ with $2 m$ elements (recall Subsection \ref{sec:prelimautocycle}).
  Thus, according to Appendix \ref{sec:subgroupsDn}, $\auto(G)$ is (up to isomorphism) either
  \begin{itemize}
  \item a Dihedral group of order $2 m'$ for $m' \ge 3$, $m' \teilt m$,
  \item or $\ZZ_{m'}$ for $m \ge 1$, $m' \teilt m$,
  \item or the Klein four group $\ZZ_2 \times \ZZ_2$ for $2 \teilt m$,
  \item or $\ZZ_2$.
  \end{itemize}
\end{corol}

\subsection{Forgetting the direction of arcs}
\label{sec:forgetinfo}

\begin{defi}\label{def:emptyint}
  We say that a WDC $G$ has \textbf{empty interior}, if all interiors of all cycles are empty, or, equivalently, if $\beta''(G) = 0$ (Lemma \ref{lem:degreesWDCrefine}).
\end{defi}
So nonlinear WDCs (characterised by $\beta(G) = 0$) have empty interior, but having an empty interior still allows to have linear vertices in the overlaps.
WDCs with empty interiors are characterised by their vertex-sets being disjoint unions of the vertex-sets of their overlaps.

\begin{lem}\label{lem:reconstnonlinWDC}
  Consider a WDC $G$ with empty interior.
  \begin{enumerate}
  \item\label{lem:reconstnonlinWDC1} From the unlabelled $\dgtomg(G)$ (forgetting the direction of the arcs), and even from the unlabelled $\dgtog(G)$ (additionally contracting parallel edges), we can reconstruct $G$ up to isomorphism (in linear time).
  \item\label{lem:reconstnonlinWDC2} For a WDC $G'$ we have $G \cong G' \Lra \dgtomg(G) \cong \dgtomg(G') \Lra \dgtog(G) \cong \dgtog(G')$.
  \end{enumerate}
\end{lem}
\begin{prf}
First consider the case of reconstructing $G$ (up to isomorphism) from the unlabelled $\dgtomg(G)$; ``unlabelled'' means that also $\dgtomg(G)$ is given only up to isomorphism.
The only choices to be made are for the directions of nontrivial overlaps (in $\dgtomg(G)$ we have them as undirected paths).
Consider the equivalence relation $\sim$ on small cycles generated by considering two neighbouring cycles equivalent iff their overlap is nontrivial; so the equivalence class of a small cycle stretches ``to the left and right'' until a one-point connection to a neighbouring cycle is met.
The meaning of $C \sim C'$ is that giving a nontrivial overlap of $C$ a direction determines the direction of $C'$.
So for every equivalence class we have exactly two choices for the direction.
Now it is easy to see that choosing for each equivalence class any direction yields a digraph isomorphic to $G$.

The only change going from $\dgtomg(G)$ to $\dgtog(G)$ is that if a small cycle is just a 2-cycle (both overlaps are trivial) in $G$, then in $\dgtomg(G)$ it becomes a pair of parallel edges, which then in $\dgtog(G)$ degenerates to a single edge --- this special case can be detected, and taken into account in the reconstruction.

This shows Part \ref{lem:reconstnonlinWDC1}.
For Part \ref{lem:reconstnonlinWDC2} we only need to observe that $G'$ (which upfront is just a general WDC) also has empty interior, since this property is invariant under isomorphism.
\Qed
\end{prf}

If a WDC $G$ has empty interior, then so has $\trans G$, and both digraphs have the same underlying multigraph, and thus we obtain:
\begin{corol}\label{cor:WDCeisc}
  WDCs with empty interiors are self-converse.
\end{corol}

\begin{examp}\label{exp:wdcselfconv}
  For a WDC $G$ with empty interior we can give a simple anti-auto\-mor\-phis\-m $\phi$ by using Lemma \ref{lem:pathskeysym}, taking the (disjoint) union of the unique anti-automorphisms for the overlaps (as path digraphs).
  $\phi$ is an involution, and has no fixed points iff all overlaps have odd length (i.e., they have an odd number of arcs).
  So $\phi$ is a skew-symmetry iff all overlaps have odd length, and then we have exactly $m(G)$ many units.

  WDCs with non-empty interiors in general are not self-converse; consider the following WDC $G$ with $m = 3$, $\alpha = 0$ (thus $\beta' = 0$) and $\beta'' = 0 + 1 + 2 = 3$, where the arrows of the extended interiors contain the number of interior vertices (while the interior vertices themselves are not drawn):
  \begin{displaymath}
    G := \xymatrix {
      & \bullet \ar@/_/[dl]|0 \ar@/_/[dr]\\
      \bullet \ar@/_/[ur] \ar@/^/[rr]|1 && \bullet \ar@/^/[ll] \ar@/_/[ul]|2
    }
  \end{displaymath}
  Now the anti-clockwise big cycle has intersection-sizes $0,1,2$ with the interiors of the three small cycles, while the clockwise big cycle has intersection-sizes $0,0,0$.
  For the transpose $\trans G$ these sizes anti-clockwise become $0,0,0$, and clockwise become $0,2,1$.
  Since isomorphisms between WDCs must respect the small cycles, thus $G$ and $\trans G$ are not isomorphic.
\end{examp}

\subsection{Homeomorphism types and (binary) bracelets}
\label{sec:homeotypes}

A ``bracelet'' of length $m \in \NN$ is a tuple of length $m$ of natural numbers, where bracelets are \emph{equivalent}, if one can be obtained from the other by rotation or reflection.
A ``binary bracelet'' has only entries $0, 1$.
The precise details are as follows (mostly according to \cite{Sawada2001Bracelets}):
\begin{defi}\label{def:bracelets}
  Consider $m \in \NN$.
  By $\lexorder$ we denote lexicographical order on $\NNZ^m$ (the elements of $\NNZ^m$ we also consider as words of length $m$).
  Consider the natural action of the Dihedral group $D_m$ on $\NNZ^m$ (by rotations and reflections), that is the map $D_m \times \NNZ^m \ra \NNZ^m$, which we write $g * t \in \NNZ^m$ for $g \in D_m$ and $t \in \NNZ^m$ (see Appendix \ref{sec:operationtuples} for the details).

  Let $\brsim_m$ be the equivalence relation on $\NNZ^m$ induced by this action, that is, two tuples $t, t' \in \NNZ^m$ are equivalent ($t \brsim_m t'$) iff one can be transformed into the other by rotations or reflections, that is, there is $g \in D_m$ with $g * t = t'$.
  \begin{itemize}
  \item A \textbf{bracelet of length $m$} (also ``turnover necklace''; $m$ is also called ``number of beads'') is an element $b$ of $\NNZ^m$, which is the smallest element of its equivalence class, that is, for all $b'$ with $b' \brsim_m b$ holds $b \lexorder b'$.
  \item The set of all bracelets of length $m$ is denote by $\allbr(m)$.
  \item While the set of all bracelets $b$ of length $m$, such that for all $1 \le i \le m$ holds $b_i < k$ for some given $k \in \NN$, is denoted by $\allbr_k(m)$.
    (Thus $\allbr(m) = \bc_{k \in \NN} \allbr_k(m)$.)
    $k$ is called the number of ``colours'' in this context.
  \item $B_k(m) := \abs{\allbr_k(m)}$.
  \end{itemize}
\end{defi}

By \cite{Sawada2001Bracelets} for an $m$-tuple $t$ one can compute the associated bracelet $b(t)$ (that is, the unique element $b(t) \in \allbr(m)$ with $t \brsim_m b(t)$) in linear time.
By \cite[Equation (5.1)]{Sawada2001Bracelets} we have
\begin{displaymath}
  \frac 12 \frac {k^m}m \le B_k(m) \le 2 \frac {k^m}m.
\end{displaymath}

\begin{examp}\label{exp:bracelets}
  The lexicographical order on $\NN_0^2$ is $(0,0) \slexorder (0,1) \slexorder (0,2) \slexorder \cdots \slexorder (1,0) \slexorder (1,1) \slexorder(1,2) \slexorder \cdots$.
  The first binary bracelets are:
  \begin{itemize}
  \item $\allbr_2(1) = \set{0,1}$, $\allbr_2(2) = \set{00,01,11}$, $\allbr_2(3) = \set{000, 001, 011,111}$.
  \item $\allbr_2(4) = \set{0000, 0001, 0011, 0101, 0111, 1111}$.
  \end{itemize}
  Numerical data for $m=1,2,3,\dots, 8$:
  \begin{enumerate}
  \item For $B_2(m)$ (OEIS, \cite[Sequence \href{https://oeis.org/A000029}{A000029}]{Sloane2008OEIS}): $2, 3, 4, 6, 8, 13, 18, 30$.
  \item For $B_3(m)$ (OEIS, \cite[Sequence \href{https://oeis.org/A027671}{A027671}]{Sloane2008OEIS}): $3, 6, 10, 21, 39, 92, 198, 498$.
  \end{enumerate}
  General $k$ is given in OEIS (\cite[Sequence \href{https://oeis.org/A081720}{A081720}]{Sloane2008OEIS}).
\end{examp}

By \cite{GilbertRiordan1961Sequences} the number $N_k(m)$ of necklaces (which don't allow reflections) with $k$ colours and of length $m$ is $N_k(m) = \frac 1m \sum_{d \teilt m} \phi(d) k^{\frac md}$ (OEIS, \cite[Sequence \href{https://oeis.org/A000031}{A000031}]{Sloane2008OEIS}).
For the number of bracelets we get
\begin{equation}
  \label{eq:bracelets}
  B_k(m) = \frac 12 (N_k(m) + \begin{cases} \frac{k+1}2 k^{m/2} & m \text{ even}\\ k^{(m+1)/2} & m \text{ odd} \end{cases} ).
\end{equation}

\begin{examp}\label{exp:compBk}
  Some example computations:
  \begin{itemize}
  \item $N_2(m) = \frac 1m(\sum_{d \teilt m} \phi(d) 2^{\frac md})$.
  \item $N_2(3) = \frac 13 (1 \cdot 2^3 + 2 \cdot 2^1) = 4$, $N_2(4) = \frac 14 (1 \cdot 2^4 + 1 \cdot 2^2 + 2 \cdot 2^1) = 6$, $N_2(5) = \frac 15 (1 \cdot 2^5 + 4 \cdot 2^1) = 8$, $N_2(6) = 14$.
  \item Even $m$: $B_2(m) = \frac 12(N_2(m) + \frac 32 2^{m/2})$.
  \item Odd $m$: $B_2(m) = \frac 12(N_2(m) + 2^{(m+1)/2})$.
  \item $B_2(3) = \frac 12 (4 + 2^2) = 4$, $B_2(4) = \frac 12 (6 + \frac 32 2^2) = 6$.
  \end{itemize}
\end{examp}

\begin{defi}\label{def:bWDC}
  Consider an $m$-WDC $G$ ($m \ge 3$).
  By $\brop(G) \in \allbr(m)$ we denote the \textbf{associated bracelet}, which is obtained by considering one of the two big cycles, choosing some overlap as a starting point, and then putting for all overlaps $P$, starting with the chosen one and following the cycle, the number $\length(P) \in \NNZ$ into a tuple, and from the obtained $m$-tuple determine the associated bracelet.
  This definition is independent of the choice of the cycle (the ``direction'') and the starting point by the definition of $\brsim_m$.

  And by $\brop_2(G) \in \allbr_2(m)$ we denote the \textbf{associated binary bracelet}, which is obtained in the same way, but putting into the tuple $\min(\length(P), 1)$.
\end{defi}

\begin{examp}\label{exp:bWDC}
  Considering the running example Equation \eqref{eq:running} (Subsection \ref{sec:introruningexample}), yielding the WDC $G$ (ignoring the skew-symmetry).
  Starting top left and going clockwise, we obtain $(0,2,0,0,2,0)$ and thus $\brop(G) = (0,0,2,0,0,2)$ and $\brop_2(G) = (0,0,1,0,0,1)$ (see the homeomorphism type in  Equation \eqref{eq:runninghomeo}).
\end{examp}

\begin{thm}\label{thm:binarybrhomeo}
  Two WDCs $G, G'$ are homeomorphic (i.e., $\dgtomg(G), \dgtomg(G')$ are homeomorphic) iff $\brop_2(G) = \brop_2(G')$.
  If $G, G'$ are nonlinear, then they are isomorphic iff $\brop_2(G) = \brop_2(G')$.
  More generally, if $G, G'$ have empty interior, then they are isomorphic iff $\brop(G) = \brop(G')$.
\end{thm}
\begin{prf}
First note that $\brop(G)$ is an isomorphism invariant by Definition \ref{def:bWDC} (the equivalence of bracelets (Definition \ref{def:bracelets}) removes the dependency on the starting point and the direction).
And for $G$ with empty interior, from $\brop(G)$ we can reconstruct $\dgtomg(G)$ up to isomorphism, and thus also $G$ up to isomorphism by Lemma \ref{lem:reconstnonlinWDC}.
Thus for $G, G'$ with empty interior we have $G \cong G' \Lra \brop(G) = \brop(G')$.
If $G$ is nonlinear, then $\brop(G) = \brop_2(G)$.
And since smoothing just removes linear vertices, for a general WDC $G$ holds $\brop_2(\smooth(G)) = \brop_2(G)$.
\Qed
\end{prf}

So via bracelets we have an efficient handle on the isomorphism types of WDCs with empty interior:
\begin{corol}\label{cor:binarybrhomeo1}
  The number of homeomorphism types of $m$-WDCs and the number of isomorphism types of nonlinear WDCs is $B_2(m)$.
  The number of isomorphism types of WDCs with empty interiors and where every overlap has length strictly less than $k \ge 2$ is $B_k(M)$.
\end{corol}

We can also improve the basic runtime for isomorphism testing in Corollary \ref{cor:WDCpolyIso}, by computing the associated bracelets and comparing them for equality:
\begin{corol}\label{cor:binarybrhomeo2}
  Whether two WDCs are homeomorphic, or whether two WDCs with empty interior are isomorphic, can be decided in linear time.
\end{corol}

Via the fast enumeration of bracelets in \cite{Sawada2001Bracelets}, we can also very efficiently the isomorphism types of WDCs with empty interior, but in the article we won't expand on that.

\section{Classifying 2-MUs of higher deficiency}
\label{sec:class2MUk}

We now consider $F \in \Bmusati{\delta=k}$ for $k \ge 2$.
We recall that $F$ is 2-uniform (Lemma \ref{lem:nounit2MU}).
And in $F$ every literal occurs at most twice:
\begin{lem}[{{\cite[Proposition 4, Page 48]{KBZ2002b}}}]
  For $F \in \Bmusat$ and a literal $x$ holds $\ldeg_F(x) \le 2$.
\end{lem}
\begin{prf}
Assume $\ldeg_F(x) \ge 3$.
Set $x$ to false in $F$ (removing the clauses containing $\ol x$, and removing $x$ from the other clauses), and obtain $F_0$.
Now remove some further clauses and obtain $F' \in \Bmusat$ with $F' \sse F_0$.
$F'$ must contain the remains of all the original clauses containing $x$, since if any of these clauses would be missing, then it could be removed from the original $F$ while maintaining the unsatisfiability after setting literal $x$ to false and true, and thus $F$ without that clause would already be unsatisfiable.
Thus $F'$ contains at least three unit-clauses, contradicting Corollary \ref{cor:nounit2MU}.
\Qed
\end{prf}

\subsection{Creation of 2-MUs of deficiency $k \ge 2$}
\label{sec:impldig2muwdcold}

Recall that $\Bmusati{\delta=k}^+$ is the class of 2-MUs, where every variable occurs at least three times.
1-singular DP-reduction (singular DP-reduction for variables occurring exactly two times) reduces $F \in \Bmusati{\delta=k}$ to $F' := \1dp(F) \in \Bmusati{\delta=k}^+$, and be reversing this process we obtain:
\begin{lem}\label{lem:createkfromplus}
  From $\Bmusati{\delta=k}^+$ ($k \ge 2$) we obtain $\Bmusati{\delta=k}$ by repeated applications of 1-singular extension, which means that for $F \in \Bmusati{\delta=k}$ one chooses $\set{x,y} \in F$ ($x \ne y$ holds) and a new variable $v$, and replaces $\set{x,y}$ by $\set{v,x}, \set{\ol v,y}$.
\end{lem}

For the implication digraphs $G := \idg(F)$ and $G' := \idg(F')$ we have $\dgtomg(G') = \smooth(\dgtomg(G))$ by Theorem \ref{thm:smooting2MUsk2}.

Our new starting point is now $F'$, and we perform singular DP-reductions, which by Lemma \ref{lem:mupstable} are necessarily non-1-singular, that is, eliminate variables of degree $3$, also called ``$2$-singular variables'', since there are two side-clauses.
So consider a variable $v$ of degree $3$ in $F$, with occurrences
\begin{displaymath}
  \set{v,x}, \set{\ol v, y}, \set{\ol v, z} \in F.
\end{displaymath}
Again we do not have contraction here, that is $x \ne y$ and $x \ne z$ (otherwise a unit-clause would be created and so $F'$ would have deficiency one, while singular DP-reduction preserves deficiency).
Thus DP-reduction for $v$ increases the literal-degree of $x$ by one, and thus not only literal $v$ occurs only once in $F$, but also literal $x$ (and $\var(x)$ is also 2-singular).
We have shown that a singular DP-reduction for any $F' \in \Bmusati{\delta=k}^+$
\begin{itemize}
\item removes one 2-singular variable (degree-3-variable),
\item and transforms one 2-singular variable (degree-3-variable) into a nonsingular variable (degree-4-variable),
\item while leaving all other literal-degrees unchanged.
\end{itemize}
Since this reduction process for $F'$ ends with a clause-set isomorphic to $\Bpt k$, which has $k$ variables, all of degree $4$ (nonsingular), there can be at most $k$ singular DP-reductions for $F'$.
\begin{lem}\label{lem:genBkmuplus}
  We obtain $\Bmusati{\delta=k}^+$ by starting with any clause-set isomorphic to $\Bpt k$, and then repeatedly applying up to $k$ times the following process for $F \in \Bmusati{\delta=k}^+$: choose some literal $x$ which occurs positively and negatively twice in $F$, and for the occurrences $\set{x,a},\set{x,b},\set{\ol x,c},\set{\ol x, d} \in F$ and a literal $y$ with underlying new variable $\var(y)$, replace the two $x$-clauses by $\set{y, x}, \set{\ol y, a}, \set{\ol y, b}$.
\end{lem}
So altogether we obtain all $F \in \Bmusati{\delta=k}$ by first applying Lemma \ref{lem:genBkmuplus}, obtaining $F' \in \Bmusati{\delta=k}^+$, which is taken as starting point for applying Lemma \ref{lem:createkfromplus}.
Such a generation sequence can be computed in polynomial time for $F$, by first computing $\1dp(F)$, which in turn is reduced by singular DP-reduction, and then reversing the whole reduction sequence.

\begin{corol}\label{cor:deffromdegr}
  For a $2$-uniform $F \in \Bmusat$ holds:
  \begin{enumerate}
  \item $n_3(F)$ is even.
  \item $\delta(F) = n_4(F) + \frac 12 n_3(F)$.
  \end{enumerate}
  While for arbitrary $F \in \Bmusats$ holds $\delta(F) = n_4(F) + \frac 12 (n_3(F) + u(F))$.

  Whether $F \in \Bmusats$ has $\delta(F) = 1$ or not can be decided from $n_3(F), n_4(F)$, namely if $n_3(F) = n_4(F) = 0$, then we have Family I, if $n_3(F) = 1$, then we have Family II, if $n_4(F) = 1$ and $n_3(F) = 0$, then we have Family III, and if we have $n_3(F) = 2$ and $n_4(F) = 0$, then we have Family IV (and in all other cases we have $\delta(F) \ge 2$).
\end{corol}
\begin{prf}
First consider $u(F) = 0$.
If $\delta(F) \ge 2$, then the two assertions follow from $n_4(\Bpt k) = k = n(\Bpt k)$, and the process of Lemma \ref{lem:genBkmuplus} transforming one degree-4-variable into two degree-3-variables.
For $\delta(F) = 1$, the assertion follows from Corollary \ref{cor:def1degreespec} (considering Families III, IV).

For $u(F) \ge 1$ (thus $\delta(F) = 1$) the generalised formula for $\delta(F)$ again follows from Corollary \ref{cor:def1degreespec} (note that Family I has $u(F) = 2$, while Family II has $u(F) = 1$, and both families have $n_4(F) = 0$).
And the separation of deficiency 1 and higher deficiencies also follows from Corollary \ref{cor:def1degreespec}.
\Qed
\end{prf}

\subsection{Implication digraphs of 2-MUs are WDCs}
\label{sec:impldig2muwdcspec}

We now show that the implication digraphs of $F \in \Bmusati{\delta=k}$ are $2k$-WDCs.
To start, we have $\idg(\Bpt k) \cong \gtodg(\cycleg_{2k})$:
\begin{displaymath}
  \idg(\Bpt k) = \xymatrix @C=4em @R=1.7em {
    1 \ar@/^/[r] \ar@/_/[d] & 2 \ar@/_/[r] \ar@/^/[l] & \cdots \ar@/^/[r] \ar@/_/[l] & k-1 \ar@/_/[r] \ar@/^/[l] & k \ar@/^/[d] \ar@/_/[l] \\
    -k \ar@/^/[r] \ar@/_/[u] & {-(k-1)} \ar@/_/[r] \ar@/^/[l] & \cdots \ar@/^/[r] \ar@/_/[l] & -2 \ar@/_/[r] \ar@/^/[l] & -1 \ar@/_/[l] \ar@/^/[u]
  }.
\end{displaymath}
Now consider Lemma \ref{lem:genBkmuplus}.
We replace two clauses $\set{x,a}, \set{x,b}$ by three clauses $\set{y,x}, \set{\ol y, a}, \set{\ol y, b}$.
For the implication digraph this means the transition:
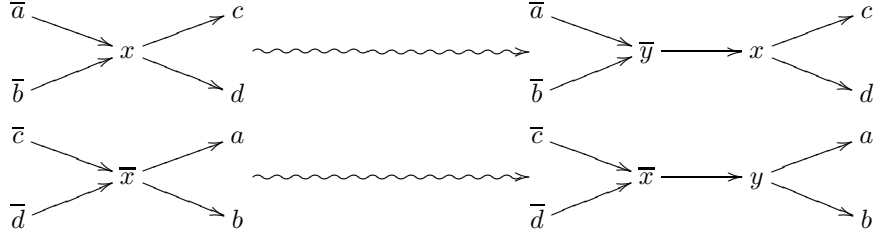
\begin{displaymath}
  \xymatrix @C=3em @R=0.2em {
  \ol a \ar[rd] && c &&& \ol a \ar[rd] &&& c\\
  & x \ar[ru] \ar[rd] & \ar@{~>}[rrr] &&&& \ol y \ar[r] & x \ar[ru] \ar[rd]\\
  \ol b \ar[ru] && d &&& \ol b \ar[ru] &&& d\\
  \ol c \ar[rd] && a &&& \ol c \ar[rd] &&& a\\
  & \ol x \ar[ru] \ar[rd] & \ar@{~>}[rrr] &&&& \ol x \ar[r] & y \ar[ru] \ar[rd]\\
  \ol d \ar[ru] && b &&& \ol d \ar[ru] &&& b
  }
\end{displaymath}
We see that this can be obtained up to isomorphism of digraphs by first, say, splitting vertex $x$, and then splitting vertex $\ol x$ (recall that the vertices in implication digraphs are just placeholders, and also do not know about complementation).
We have shown for $k \ge 2$:
\begin{lem}\label{lem:idgsnWDC}
  The implication digraph of $F \in \Bmusatdp{\delta=k}$ is a nonlinear $2k$-WDC.
\end{lem}
In the same way, obviously one step of 1-singular extension in Lemma \ref{lem:createkfromplus} is captured by two applications of arc-splitting.
\begin{displaymath}
  \xymatrix @C=3em @R=0.6em {
  \ol x \ar[r] & y & \ar@{~>}[r] && \ol x \ar[r] & v \ar[r] & y \\
  \ol y \ar[r] & x & \ar@{~>}[r] && \ol y \ar[r] & \ol v \ar[r] & x
  }
\end{displaymath}
Altogether we have shown
\begin{thm}\label{thm:WDC}
  The implication digraph of $F \in \Bmusati{\delta=k}$ ($k \ge 2$) is a $2k$-WDC.
\end{thm}

\begin{corol}\label{cor:WDCcontcyc}
  The implication digraph of $F \in \Bmusati{\delta=k}$ ($k \ge 2$) has exactly two contradictory cycles (being the contrapositions of each other).
\end{corol}

\begin{examp}\label{exam:bracelet}
  For $\Bpt 3$ the implication digraph $\idg(\Bpt 3)$, shown in Section \ref{sec:introruningexample}, is a $6$-WDC.
  So $\idg(\Bpt 3)$, as a big cycle of $6$ small cycle, corresponds to the binary bracelet $000000$ (the overlap of any two neighbouring cycles is a single vertex).

  Now consider the non-1-singular normalform $\1dp(F)$ of our running example $F$ (Equation \eqref{eq:runninghomeo}).
  The implication digraph $\idg(\1dp(F))$ is a nonlinear $6$-WDC, and has the associated binary bracelet $001001$.
\end{examp}

\subsection{Skew-symmetries of WDCs}
\label{sec:skewsymmwdcs}

We are now ready to prove the main technical result, showing that the implication digraph of $F$ has a unique skew-symmetry (in the precise sense, not just up to isomorphism), which yields the complementation of literals, and thus one can reconstruct $F$ from the (unlabelled) $\idg(F)$.
Since $F$ does not have unit-clauses, we have to exclude skew-symmetries which yield them (otherwise uniqueness wouldn't hold), that is, we consider actually \emph{unit-free} skew-symmetries (Definition \ref{def:unitfree}).
\begin{thm}\label{thm:wdcuniqueskewsym}
  Every WDC has at most one unit-free complementation.
\end{thm}
\begin{prf}
Consider an $m$-WDC $G$ and a unit-free skew-symmetry $\sigma$ for $G$.
We show that $\sigma$ is unique.
$\sigma: G \ra \trans G$ is an isomorphism, where $\trans G$ is also a WDC.
We obtain the induced isomorphism $\sigma' : \scg(G) \ra \scg(\trans G)$ (which by Corollary \ref{cor:injisos} uniquely identifies $\sigma$).
Using the canonical isomorphism $\scg(\trans G) \ra \scg(G)$, mapping each small cycle to the reversed cycle, we obtain the automorphism $\sigma'' \in \auto(\scg(G))$.
We will use the properties of the automorphisms of an $m$-cycle as given in Subsection \ref{sec:prelimautocycle}.
In the remainder of the proof we show that $\sigma''$ must be the point-symmetry, the rotation by 180 degrees.

If $\sigma''$ had a fixed-point (would map one small cycle of $G$ to itself), then $\sigma$ restricted to this cycle would be a skew-symmetry, and so by Lemma \ref{lem:CycleSkewsym}, $\sigma$ would not be unit-free.
If $m$ were odd, then all reflections of an $m$-cycle had a fixed-point, and thus $\sigma''$ were a nontrivial rotation, but for odd $m$ no nontrivial rotation is an involution; so $m$ is even.
It remains to show that $\sigma''$ can't be a reflection, and so assume $\sigma''$ is a reflection.
So $\sigma''$ swaps two neighbouring small cycles. $\sigma$ restricted to these two small cycles would be a skew-symmetry, and thus by Lemma \ref{lem:twocyclesunit}, $\sigma$ would not be unit-free.
\Qed
\end{prf}

We finally have shown the main result of this article:
\begin{thm}\label{thm:idgcompleteisovar}
  For $F, F' \in \Bmusati{\delta=k}$, $k \ge 2$:
  $\isos(F, F') = \isos(\idg(F), \idg(F'))$.
\end{thm}
\begin{prf}
This follows from Theorem \ref{thm:wdcuniqueskewsym} together with Theorem \ref{thm:WDC} and Lemma \ref{lem:isoexactoneunitfree}.
\Qed
\end{prf}

\subsection{Applications}
\label{sec:mainresapps}

\begin{corol}\label{cor:idgcompliso}
  For $F, F' \in \Bmusati{\delta=k}$, $k \ge 1$, holds $F \cong F'$ iff $\idg(F) \cong \idg(F')$, where for $k \ge 2$ the implication digraphs are $2k$-WDCs.
\end{corol}
\begin{prf}
For $k \ge 2$ the assertion follows from Theorems \ref{thm:WDC} and \ref{thm:idgcompleteisovar}, while for $k = 1$ the statement follows from Theorem \ref{thn:classd1}.
\Qed
\end{prf}

\begin{corol}\label{cor:isodec2MU}
  Whether for $F, F' \in \Bmusati{\delta=k}$, $k \ge 1$, holds $F \cong F'$ can be decided in quadratic time.
\end{corol}
\begin{prf}
By Corollary \ref{cor:idgcompliso}, Corollary \ref{cor:WDCpolyIso}, and Theorem \ref{thn:classd1}.
\Qed
\end{prf}

By extending the machinery for bracelets to incorporate the interior points of WDCs, we can indeed achieve isomorphism decision in linear time, but we won't do this in this article.

\begin{corol}\label{cor:automudih}
  For $F, F' \in \Bmusati{\delta=k}$, $k \ge 2$, the number of isomorphisms between $F$ and $F'$ is at most $4 k$.
  The automorphism group of $F$ is a subgroup of the Dihedral group $D_{ 2 k}$ with $4 k$ elements.
\end{corol}
\begin{prf}
By Corollary \ref{cor:autowdcDih}.
\Qed
\end{prf}

\begin{corol}\label{cor:numisotypes}
  The number of isomorphism types of $F \in \Bmusati{\delta=k}$, $k \ge 1$, with (exactly) $n(F) = n \in \NNZ$ variables is $\Theta(n^{3 k - 1})$ (for fixed $k$).
\end{corol}
\begin{prf}
For $k = 1$ this follows from Corollary \ref{cor:numdef1}; so assume $k \ge 2$.
There are $2k$ cycles in $\idg(F)$, with half of them duplicated by skew-symmetry, so that we have $k$ essential cycles.
These cycles are arranged in a big cycle, and so have three non-overlapping parts, say the upper, right, and lower parts, which makes $3k$ numbers.
These number are adding up to $n$, and so the number of isomorphism types is $O(n^{3k-1})$.
By Corollary \ref{cor:automudih} the equivalence-classes are of constant size, yielding the desired $\Theta(n^{3 k - 1})$.
\Qed
\end{prf}

By Theorem \ref{thm:idgcompleteisovar} the isomorphism types of the normalforms of 2-MUs correspond to the isomorphism types of nonlinear skew-symmetric WDCs, and so:
\begin{corol}\label{cor:binbracelet}
  The homeomorphism types of $\Bmusati{\delta=k}$, $k \ge 2$, that is, the isomorphism types of $\Bmusatdp{\delta=k}$, are in one-to-one correspondence with the binary bracelets of length $k$.
\end{corol}
\begin{prf}
The isomorphism types of $\Bmusatdp{\delta=k}$ correspond to the isomorphism types of nonlinear $2k$-WDCs with skew-symmetry.
Isomorphism types of nonlinear $2k$-WDCs correspond to binary bracelets of length $2k$ (recall Theorem \ref{thm:binarybrhomeo}), and due to skew-symmetry, half of them are discarded.
\Qed
\end{prf}

\subsection{The cycle-structure of 2-MUs}
\label{sec:cyclestruct}

Refining Lemma \ref{lem:Aspvall}:
\begin{thm}\label{thm:Aspvallref}
  A clause-set $F \in \Bclss$ is unsatisfiable iff there exists a contradictory cycle in $\idg(F)$.
  More precisely, $F \in \Bmusats$ with $k := \delta(F)$ has
  \begin{itemize}
  \item exactly two contradictory cycles iff either $k \ge 2$, or $k=1$ and $F$ has exactly one unit-clause;
  \item exactly four contradictory cycles if $k=1$ and $F$ has no unit-clause;
  \item exactly one contradictory cycle if $k=1$ and $F$ has two unit-clauses.
  \end{itemize}
  For $k \ge 2$ the contradictory cycles are exactly the big cycles of the $2k$-WDCs.

  Exactly in case of $k=1$ and $F$ having two unit-clauses there is a contradictory cycle which is the contraposition of itself (namely the cycle is just the whole digraph).
  While otherwise the contraposition of contradictory cycle is a different (contradictory) cycle, and yields a partitioning of the contradictory-cycle-set into $2$-element subsets; here a cycle might contain all vertices of the digraph, but never all arcs.

  Either all contradictory cycles are arc-regular or none are.
  The contradictory cycles are arc-regular iff $F$ has no variable of degree $3$ (that is, all variables have degree $2$ or $4$), and in case of $k=1$ we do not have the case of two unit-clauses.
  (Thus for $k=1$ the case of arc-regular contradictory cycles is exactly that of Family III, while for $k \ge 2$ the case of arc-regular contradictory cycles is exactly that of having all overlaps trivial.)

  For $k \ge 2$ the complement-structure of a cycle $C$ and its contraposition $\ol C$ is as follows:
  \begin{enumerate}
  \item The vertices (literals) $x$ on $C$ such that also $\ol x$ is on $C$ are exactly the vertices in the overlaps of the small cycles.
    The path from $x$ to $\ol x$ in $C$ is regular minus the last vertex iff $x$ is the last vertex of the overlap.
  \item The vertices $x$ of $C$ with $\ol x$ not in $C$ are exactly the linear vertices of the implication digraph not in any overlap (that is, the interior vertices).
  \item So for $F$ with empty interior, $C$ and $\ol C$ use all literals of $F$ (both use always both signs), while for vertices in the interior, $C$ selects one sign and $\ol C$ the other.
  \item $C$ uses all clauses.
    A clause is used exactly twice (``in both implication-directions'') iff it is in an overlap (otherwise a clause is used exactly once).
  \end{enumerate}
\end{thm}
\begin{prf}
Concerning the counting, Corollary \ref{cor:WDCcontcyc} handles the case of $k \ge 2$, while the cases with $k = 1$ are handled in Corollaries \ref{cor:firstfamily}, \ref{cor:secondfamily}, \ref{cor:thirdfamily}, \ref{cor:fourthfamily}.
The other details follow by the results of Subsection \ref{sec:cyclestrucWDC}.
\Qed
\end{prf}

\section{Conclusion and outlook}
\label{sec:conclusion}

After having established a solid foundation for the study of implication digraphs of 2-CNFs in Sections \ref{sec:idg}, \ref{sec:nearlyunique}, \ref{sec:skewsymbasic}, we considered in detail the classification of 2-MUs of deficiency $1$ (and their implication digraphs) in Section \ref{sec:2-MU(1)}.
We established the four Families I - IV, the four main types of $F \in \Bmusats$ with $\delta(F) = 1$, with $u(F) = 2$ for I, $u(F) = 1$ for II, and $u(F) = 0$ for III, IV (recall $u(F)$ is the number of unit-clauses).

In Section \ref{sec:sDPred} we considered more generally the close connection between 1-singular DP-reduction for 2-CNFs and smoothing (removal of linear vertices), leading to the homeomorphism type of 2-CNFs.
Concentrating then on 2-MUs, we established the four homeomorphism types of 2-MUs of deficiency $1$ (corresponding to the four Families); here the final steps of smoothing are less powerful than 1-singular DP-reduction, due to the special handling of unit-clauses.
For deficiency $2$ we show that no special cases are needed, and we have a precise correspondence between smoothing and 1-singular DP-reduction.

The focus of the remaining sections was now on deficiency $k \ge 2$.
In Section \ref{sec:sDPred} we established basic facts on weak double cycles (WDCs), the graph-theoretic framework for 2-MUs, with emphasise on their classification.
A complete isomorphism invariant of the class of WDCs with empty interiors, which in their application to 2-MUs strictly include the class of non-1-singular 2-MUs, is given by the associated bracelets, for which strong combinatorial and algorithmic results are available.

In the final Section \ref{sec:class2MUk}  (other than this conclusion) the results on WDCs are applied to 2-MUs (of deficiency at least $2$).
We showed that every WDC has at most one skew-symmetry, when not allowing (resulting) unit-clauses.
We obtain that the isomorphisms between 2-MUs $F,F'$ are exactly the isomorphisms between $\idg(F),\idg(F')$.
We note that for deficiency $1$ we don't have such an exact result, but the fundamental result, that 2-MUs $F, F'$ are isomorphic iff their implication digraphs are isomorphic, holds indeed for all 2-MUs.
As first applications we characterised the automorphism groups of 2-MUs (of deficiency at least $2$), and obtained the asymptotic number of isomorphism types of 2-MUs.
Furthermore a complete invariant for the homeomorphism types of 2-MUs of deficiency $k \ge 2$ are the binary bracelets of length $k$.
We concluded by a detailed picture of the contradictory cycles of 2-MUs (where indeed we not just have contradictory closed walks, but cycles).

\subsection{Next step: generalise bracelets}
\label{sec:outlookgenbrace}

By a naive algorithm we established isomorphism decision for 2-MUs in quadratic time.
But for deficiency $1$ and for the case of empty interiors we actually can do the decision in linear time.
In future work the notion of bracelets has to be generalised, to include the interior vertices, which will then lead to isomorphism decision in linear time (and fast, direct access to the isomorphism types of 2-MUs in general), and general counting formulas.

\subsection{Finding and enumerating MUSs}
\label{sec:outlookgMUS}

The most fundamental open question is the complexity of determining all MUSs (MU-sub-clause-sets) for input $F \in \Pcls 2$.
By the (reasonably) fast isomorphism decision, we can then group the MUSs by their isomorphism type; and once generalised bracelets are in place, a more efficient handle on the isomorphism types of all 2-MUs will be available.
This would yield a list of all isomorphism types of MUSs and their counts, which seems very valuable and is a complete representation of the structure of MUSs of $F$.

A first step at exploiting the detailed information of 2-MUs for finding and enumerating 2-MUSs has been made by \cite{KullmannClewer2026Simple2MUS}, concentrating on the four Families I-IV.
The underlying research program is the quest for finding ``simple MUSs'', with simplicity measured in structural terms (not by length).
It is shown that MUSs which are in the renamable Horn part of 2-MU, Families I+II, can be efficiently found, while as soon as Family III or IV is included, while excluding deficiencies at least $2$, we get NP-completeness (of finding one such MUS).

\paragraph{Acknowledgements}

This work was supported by EPSRC grant EP/S015523/1.

\begin{appendix}

\section{Concrete definition of $D_n$}
\label{sec:concdefDn}

In Section \ref{sec:prelimautocycle} we outlined basic (well-known) facts on $D_n$.
Here we give a concrete representation of $D_n$, such that the basic facts can be directly verified.

Let $\ZZ_n$ be the ring of integers modulo $n$ for $n \ge 2$; if unambiguous by the context, we use integers and their ordinary operations for the elements of $\ZZ_n$, but if needed we use say $2+3 \cong 1 \mod 4$.
If we need to convert $a \in \ZZ_n$ to an integer (for example when computing $\gcd: \NNZ \times \NN \ra \NN$), then we always take the (unique) representative in $\set{0,\dots,n-1}$ modulo $n$.
If we want to emphasise that we only consider the additive group, we use $(\ZZ_n,+)$.

\begin{lem}[{{\cite[Proposition 3.3.1]{IrelandRosen1990NumberTheory}}}]\label{lem:lineqZn}
  The linear equation $a x = b$ in $\ZZ_n$ ($n \in \NN$, $a, b \in \ZZ_n$) has a solution for $x$ (that is, there is $x \in \ZZ_n$ with $a x = b$ in $\ZZ_n$) iff $\gcd(a,n)$ divides $b$, in which case there are exactly $\gcd(a,n)$ many solutions (that is, $\abs{\set{x \in \ZZ_n \mb a x = b}} = \gcd(a,n)$).
  If $x_0$ is any solution, then all solutions are given by $x = x_0 + i \frac{n}{\gcd(a,n)}$ for $i \in \tb{0}{\gcd(a,n)-1}$.
\end{lem}

We now define a concrete representation for the \textbf{Dihedral group} $D_n$, which has $2 n$ elements.
We consider only $n \ge 3$, since we need the degenerations $D_1 \cong (\ZZ_2,+)$ and $D_2 \cong (\ZZ_2,+)^2$ (the ``Klein four group'', the group of four elements, where three elements of order two) only at very few places, and there we make these cases explicit.

\begin{defi}\label{def:Dn}
  Consider $n \in \NN$, $n \ge 3$.
  Let $D_n := \ZZ_n \times \set{-1,+1}$ (just as a set), with $\set{-1,+1} \subset \ZZ_n$ (so $\abs{D_n} = 2 n$):
  \begin{enumerate}
  \item the elements $(x,+1) \in D_n$ are called \textbf{rotations};
  \item the elements $(x,-1) \in D_n$ are called \textbf{reflections}.
  \end{enumerate}
  For $(x,\ve), (x',\ve') \in D_n$ let
  \begin{displaymath}
    (x,\ve) \circ (x',\ve') := (x + \ve \cdot x', \ve \cdot \ve') \in D_n.
  \end{displaymath}
  We use multiplicative notations for the operations of $D_n$.
\end{defi}
\begin{enumerate}
\item $D_n$ is a group:
  \begin{enumerate}
  \item The neutral element is $(0,1)$.
  \item Associativity holds due to $((x,\ve) \circ (x',\ve')) \circ (x'',\ve'') = (x + \ve x' + \ve \ve' x'', \ve \ve' \ve'') = (x,\ve) \circ ((x',\ve') \circ (x'',\ve''))$.
  \item The inverse of $(x,1)$ is $(-x,1)$, while all $(x,-1)$ are self-inverse (thus of order $2$).

    Using multiplicative notation, we write thus $(x,1)^{-1} = (-x,1)$ and $(x,-1)^{-1} = (x,-1)$.
  \end{enumerate}
\item Concerning commutativity we have:
  \begin{enumerate}
  \item $(x,1) \circ (y,1) = (x+y,1) = (y,1) \circ (x,1)$.
  \item $(x,1) \circ (y,-1) = (x+y,-1) = (y,-1) \circ (x,1)^{-1}$.
  \item $(x,-1) \circ (y,-1) = (x-y,1) = ((y,-1) \circ (x,-1))^{-1}$.
  \end{enumerate}
\item $(x,1)^e = (e x, 1)$ for $e \in \ZZ$.
  Thus $(1,1)^n= (0,1)$ (so $(1,1)$ is of order $n$).
\item Using $a := (1,1)$ and $b := (0,-1)$ we thus have
  \begin{enumerate}
  \item $(x,1) = a^x$ and $(x,-1) = a^x \circ b$.
  \item $a^n = 1 = b^2$.
  \item $a \circ b = b \circ a^{-1}$.
  \end{enumerate}
  Thus, as is well-known, $D_n$ is a (``the'') Dihedral group with $2 n$ elements.
\item The elements of order $2$ of $D_n$ (the $a \in D_n \sm \set{1}$ with $a^2 = 1$) are as follows:
  \begin{enumerate}
  \item Every reflection is of order $2$.
  \item For a rotation $(x,1)$ to be self-inverse means $x + x = 0$, that is $2 x = 0$; by Lemma \ref{lem:lineqZn} we have two cases:
    \begin{enumerate}
    \item If $n$ is odd, then the linear equation $2 x = 0$ has exactly the solution $x = 0$, that is, there is no rotation of order $2$,
    \item If $n$ is even, then there is exactly one non-trivial solution $x = \frac n2$, and thus we have exactly one rotation of order $2$.
    \end{enumerate}
  \end{enumerate}
\end{enumerate}

\section{Isomorphisms between basic graphs}
\label{sec:isosbasic}

In Subsection \ref{sec:defcycles} we defined the cycle graph $\cycleg_n$ as having vertex-set $V(\cycleg_n) = \tb 1n$ (so that the vertices could also be used as variables).
Regarding $D_n$, we define the isomorphic variation $C_n \cong \cycleg_n$ with $V(C_n) := \ZZ_n$ and
\begin{displaymath}
  E(C_n) := \set{\set{u,v} \in \pot(\ZZ_n) : v - u = \pm 1} = \set{\set{u,v} \in \pot(\ZZ_n) : v - u = 1}
\end{displaymath}
for $n \ge 3$.
Now we can define the natural action of $D_n$ on the vertex-set of $C_n$, which is an action by automorphisms:
\begin{defi}\label{def:actDnCn}
  For $(x,\ve) \in D_n$ and $v \in V(C_n)$ let
  \begin{displaymath}
    (x,\ve) * v := x + \ve v \in V(C_n).
  \end{displaymath}
  And by $A_{(x,\ve)}^n : V(C_n) \ra V(C_n)$ we denote the map $v \in V(C_n) \mapsto A_{(x,\ve)}(v) := (x,\ve) * v \in V(C_n)$.
\end{defi}
\begin{enumerate}
\item The maps $A = A_{(x,\ve)}^n$ with $v \in V(C_n) \mapsto A(v) = x + \ve v$ are determined by
  \begin{enumerate}
  \item $x = A(0)$;
  \item $\ve$ is the ``movement direction'': $A(v+1) = A(v) + \ve$ for all $v \in V(C_n)$.
  \end{enumerate}
\item The action $*: D_n \times V(C_n) \ra V(C_n)$ is an action of a group:
  \begin{enumerate}
  \item the neutral element of $D_n$ acts as the identity: $(0,1) * y = y$;
  \item $((x,\ve) \circ (x',\ve')) * y = (x,\ve) * ((x',\ve') * y) = x + \ve x' + \ve \ve' y$.
  \end{enumerate}
\item Thus all $A_{(x,\ve)}^n$ are permutations of $V(C_n)$, i.e., $A_{(x,\ve)}^n \in S(V(C_n))$, where $S(X)$ for a set $X$ is defined as the set of all bijections from $X$ to $X$ (the ``symmetric group'').
\item And the map $A^n: (x,\ve) \in D_n \mapsto A_{(x,\ve)}^n \in S(V(C_n))$ is indeed a group homomorphism (using map-composition for the symmetric group), since $A_{(x,\ve) \circ (x',\ve')} = A_{(x,\ve)} \circ A_{(x',\ve')}$ (using $\circ$ on the right-hand side for map-composition) and $A_{(0,1)} = \id_{V(C_n)}$.

  The homomorphism $A^n$ is always injective (is a monomorphism), since only $A_{(0,1)}^n$ is the identity on $V(C_n)$ (using $n \ge 3$).

  Since $\abs{D_3} = 2 \cdot 3 = 6 = 3 ! = \abs{S(V(C_3))}$, for $n = 3$ the monomorphism $A^n$ is also an isomorphism (but not for $n > 3$).
\item All $A_{(x,\ve)}^n$ are automorphisms of $C_n$, since for $u, v \in V(C_n)$:
  \begin{multline*}
    \set{A(u),A(v)} \in E(C_n) \Lra A(v) - A(u) = \pm 1 \Lra\\
    (x + \ve v) - (x + \ve u) = \pm 1 \Lra v - u = \pm 1 \Lra \set{u,v} \in E(C_n).
  \end{multline*}
\end{enumerate}

\begin{lem}\label{lem:isoDnAut(Cn)}
  The map $A^n: D_n \ra \auto(C_n)$ is an isomorphism of groups for every $n \ge 3$.
\end{lem}
\begin{prf}
It remains to show that for every $\vp \in \auto(C_n)$ there exists $(x,\ve) \in D_n$ with $\vp = A_{(x,\ve)}^n$.
Let $x := \vp(0)$.
The neighbour $1$ of $0$ in $C_n$ must be mapped to one of the two neighbours of $\vp(0)$, namely to $\vp(0) - 1$ or $\vp(0) + 1$.
Let $\ve \in \set{\pm 1}$ such that $\vp(1) = \vp(0) + \ve$.
Now we have $\vp = A_{(x,\ve)}^n$; we prove $\fa\, v \in \NNZ : \vp(v) = A_{(x,\ve)}(v) = x + \ve v$ by induction:
For $v \in \set{0,1}$ this holds by definition, so assume $v \ge 2$.
$v$ is a neighbour of $v-1$, and thus $\vp(v) \in \set{\vp(v-1) \pm \ve}$.
By Induction Hypothesis holds $\vp(v-1) = x + \ve (v-1)$, and $\vp(v-2) = x + \ve (v-2)$.
If $\vp(v) = \vp(v-1) - \ve$, then $\vp(v) = \vp(v-2)$, which is impossible, and thus $\vp(v) = \vp(v-1) + \ve = x + \ve v$.
\Qed
\end{prf}

Let $\vec{C}_n$ for $n \ge 3$ be the directed version of $C_n$, that is $V(\vec{C}_n) = V(C_n) = \ZZ_n$, while
\begin{displaymath}
  E(\vec{C}_n) = \set{(a,b) \in \ZZ_n^2 : b - a = 1};
\end{displaymath}
note that $\auto(\vec{C}_n)$ is a subgroup of $\auto(C_n)$.
And let $Z_n := \ZZ_n \times \set{1}$ and $Z_n' := \ZZ_n \times \set{-1}$ ($Z_n$ is the set of rotations, $Z_n'$ the set of reflections).
So $Z_n$ is a subgroup of $D_n$, isomorphic to $\ZZ_n$ as an additive (commutative) group.
\begin{corol}\label{cor:isoZn}
  The map $A^n: D_n \ra \auto(C_n)$ restricted to $Z_n \subset D_n$ is an isomorphism of groups from $Z_n$ to $\auto(\vec{C}_n)$ for every $n \ge 3$.
\end{corol}
\begin{prf}
Elements of $Z_n$ yield automorphisms of $\vec{C}_n$, no element of $Z_n'$ does.
\Qed
\end{prf}
\begin{corol}\label{cor:isoZn'}
  The map $A^n: D_n \ra \auto(C_n)$ restricted to $Z_n' \subset D_n$ is a bijection from $Z_n'$ to $\isos(\vec{C}_n, \trans{\vec{C}_n})$ for every $n \ge 3$.
\end{corol}
\begin{prf}
$A_{(0,-1)}$ is an isomorphism from $\vec{C}_n$ to $\trans{\vec{C}_n}$ (running through the vertices in opposite direction, respecting the reverted arcs).
And $Z_n' = \set{(0,-1) \circ (x,1) : x \in \ZZ_n}$, so the assertion follows by Corollary \ref{cor:isoZn} and Lemma \ref{lem:isosstructs}.
\Qed
\end{prf}

For $n \ge 2$ define the path graph $P_n$ with $n$ vertices (and $n-1$ edges) by having $V(P_n) := V(C_n)$ and
\begin{displaymath}
  E(P_n) := E(C_n) \sm \set{\set{0,-1}} = \set{\set{a,a+1} : a \in \ZZ_n \sm \set{-1}}.
\end{displaymath}
Recall that the map $A^n_{(-1,-1)}$ maps $0 \mapsto n-1$, $1 \mapsto n-2$ and so on, and we call this map $s_{-1}: \ZZ_n \ra \ZZ_n$, $s_{-1}(x) := -x-1$, which is a bijection and defined for all $n \ge 2$.
\begin{corol}\label{cor:autoPn}
  The map $A^n: D_n \ra \auto(C_n)$ restricted to the subgroup $U := \set{(0,1),(-1,-1)} \subset D_n$ is an isomorphism of groups from $U$ to $\auto(P_n)$ for every $n \ge 3$.
  And for all $n \ge 2$ we have $\auto(P_n) = \set{\id, s_{-1}}$ (with $\id \ne s_{-1}$).
\end{corol}
\begin{prf}
For all $n \ge 2$, the permutation $s_{-1}$ of $\ZZ_n$ is indeed an automorphism of $P_n$, since for $a \ne -1$ holds $s_{-1}(\set{a,a+1}) = \set{-a-1, -(a+1)-1} = \set{-a-2, -a-1}$, where $-a-2 = -1 \Lra a=-1$.
We have $s_{-1}(0) = -1$.
Now every automorphism $f$ of $P_n$ must have $f(0) = 0$, in which case $f = \id$, or $f(0) = -1$, in which case $f = s_{-1}$, since $0, -1$ are the two vertices of degree $1$ in $P_n$, while every other vertex has degree $2$.
\Qed
\end{prf}

The directed path graph $\vec{P}_n$ for $n \ge 2$ is defined by $V(\vec{P}_n) := V(\vec{C}_n)$ and
\begin{displaymath}
E(\vec{P}_n) := E(\vec{C}_n) \sm \set{(-1,0)} = \set{(a,a+1) : a \in \ZZ_n \sm \set{-1}}.
\end{displaymath}
\begin{corol}\label{cor:isosdirpath}
  For $n \ge 2$ holds $\auto(\vec{P}_n) = \set{\id}$ and $\isos(\vec{P}_n, \trans{\vec{P}_n}) = \set{s_{-1}}$.
\end{corol}
\begin{prf}
$\vec{P}_n$ has exactly one vertex of out-degree one, namely $0$, and thus the identity is the only automorphism.
$s^{-1}$ maps the vertex of out-degree one of $\vec{P}_n$ to the vertex of out-degree one of $\trans{\vec{P}_n}$, and thus is the single isomorphism from $\vec{P}_n$ to $\trans{\vec{P}_n}$.
\Qed
\end{prf}

\section{Fixed points and near-fixed points}
\label{sec:fixedpointsnear}

We determine now the fixed points of the automorphisms of $C_n$:
\begin{lem}\label{lem:fixedpointsautoCn}
  Consider $n \ge 3$ and the automorphisms of $C_n$.
  \begin{enumerate}
  \item[1.] No nontrivial (non-identical) rotation has a fixed point.
  \item[2.] For every vertex $v$ there is exactly one reflection ($A^n_{(2 v, -1)}$) fixing $v$.
  \item[3.] So every vertex $v$ has exactly two automorphisms fixing $v$, the identity and one reflection.
  \item[4.] For odd $n$ every reflection has exactly one fixed point (namely $(x,-1)$ has the fixed point $2^{-1} x$).
  \item[5.] For even $n$:
    \begin{enumerate}
    \item[(a)] Half of the reflections have no fixed point, and the other half has exactly two fixed points (namely the reflection $(x,-1)$ has no fixed point for odd $x$, while for even $x$ we have the fixed points $\frac x2, \frac x2 + \frac n2$).
    \item[(b)] Now consider ``near-fixed points'', that is, solutions of $\sigma(v) - v = 1$ for reflections $\sigma = (x,-1)$.
      The sum of the number of fixed points and near-fixed points is always exactly two: So we have exactly two near-fixed points for the case of no fixed point (that is for odd $x$, and the near-fixed points are $\frac{x-1}2$ and $\frac{x-1}2 + \frac n2$), and no near-fixed points for the case of two fixed points.
    \end{enumerate}
  \end{enumerate}
\end{lem}
\begin{prf}
We have to consider the solutions of the equation $x * v = v$ for $x \in D_n$ and $v \in V(C_n)$.
For rotations we have $(x,1) * v = v \Lra x + v = v \Lra x = 0$, and thus no nontrivial (non-identical) rotation has a fixed point.
So it remains to consider reflections; we have $(x,-1) * v = v \Lra x - v = v \Lra x = 2 v$.
Thus for every vertex $v$ there is exactly one reflection (namely $(2 v, -1)$) fixing $v$.
Considering finally the fixed points of $(x,-1)$ (that is, solving the equation $x = 2 v$ for $v$), we have two cases by Lemma \ref{lem:lineqZn} (with $a = 2$):
\begin{enumerate}
\item If $n$ is odd (so $\gcd(a,n) = 1$ in Lemma \ref{lem:lineqZn}, and there is exactly one solution), then $2$ is multiplicatively invertible, that is, there is an element denoted $2^{-1} \in \ZZ_n$ with $2^{-1} + 2^{-1} = 1$ (in $\ZZ_n$).
  And indeed, for fixed $n$ we have $2^{-1} = \ceil{n/2} \in \ZZ_n$, for example for $n = 3$ we have $2^{-1} = 2$, while for $n = 5$ we have $2^{-1} = 3$.

  Thus $x = 2 v \Lra v = 2^{-1} x$.
  So here every reflection $(x,-1)$ has exactly one fixed point (namely $2^{-1} x$).
\item Finally assume $n$ is even (so $\gcd(a,n) = 2$ in Lemma \ref{lem:lineqZn}; note that here $2$ is not multiplicatively invertible --- division by $2$ below happens in $\ZZ$).

  We are considering the near-fixed-point case at the same time, yielding the equation $(x,-1) * v - v = \eta$ for $\eta \in \set{0, 1}$ (here $\eta = 0$ is the fixed-point case), which is equivalent to $2 v = x - \eta$ (so $b = x - \eta$ in Lemma \ref{lem:lineqZn}).
  \begin{enumerate}
  \item If $x$ is odd, then $(x,-1)$ has no fixed point, while having two near-fixed points, namely $v = \frac{x-1}2$ and $v = \frac{x-1}2 + \frac n2$.
  \item If $x$ is even, then $(x,-1)$ has (exactly) two fixed points, namely $v = \frac x2$ and $v = \frac x2 + \frac n2$, while having no near-fixed point.
    \Qed
  \end{enumerate}
\end{enumerate}
\end{prf}

\section{The action on tuples}
\label{sec:operationtuples}

In Definition \ref{def:bracelets} we referred to the natural action of Dihedral groups on tuples.
The basic idea is to consider the indices of an $m$-tuple as the vertices of the cycle graph $\cycleg_n$ (wrapping around at the end), and consider the action of $D_m$ on the vertices (according to Section \ref{sec:isosbasic}).

The exact definition is as follows.
We represent the elements of $\NNZ^m$ as maps $t: \ZZ_m \ra \NNZ$, with the set of all such tuples-as-maps as $\NNZ^{\ZZ_m}$.
The action of $(x,v) \in D_m$ on the indices $i \in \ZZ_m$ is the same as the action on the vertices in Section \ref{sec:isosbasic}, that is $(x,\ve) * i = x + \ve i \in \ZZ_m$.
The action $D_m \times \NNZ^{\ZZ_m} \ra \NNZ^{\ZZ_m}$ (by permutations of the indices) fulfils $t(i) = ((x,\ve) * t)((x,\ve) * i)$.
\begin{defi}\label{def:actDnTuples}
  For $(x,\ve) \in D_m$ and $t \in \NNZ^{\ZZ_m}$ let
  \begin{displaymath}
    ((x,\ve) * t)(i) := t((x,\ve)^{-1} * i)
  \end{displaymath}
  for $i \in \ZZ_m$.
\end{defi}
We have an action of a group, that is, $1 * t = t$ and $g * (g' * t) = (g \circ g') * t$.
(The latter follows from $(g * (g' * t))(i) = (g' * t)(g^{-1} * i) = t(g'^{-1} * (g^{-1} * i)) = t((g'^{-1} \circ g^{-1}) * i) = t((g \circ g')^{-1} * i) = (g \circ g')(t)(i)$.)

\section{Subgroups of Dihedral groups}
\label{sec:subgroupsDn}

By \cite[Theorem 3.1]{Conrad2019Dihedral2}, up to isomorphism the subgroups of $D_n$ ($n \ge 3$) are exactly as follows:
\begin{enumerate}
\item $\ZZ_2$.
\item $\ZZ_2 \times \ZZ_2$ (Klein four group) if $2 \teilt n$.
\item $\ZZ_d$ for every $d \teilt n$ (with $1 \le d \le n$, and $d \ne 2$ to keep cases disjoint).
\item $D_d$ for $d \teilt n$ and $d \ge 3$.
\end{enumerate}

\end{appendix}

\bibliographystyle{plainurl}

\newcommand{\noopsort}[1]{}

\end{document}